\documentclass[seceq]{ptptex}
\usepackage{wrapft}
\usepackage{graphicx}
\newcommand{\bra}[1]{\langle {#1} |}     %%
\newcommand{\ket}[1]{| {#1} \rangle}     %%
\newcommand{\rbra}[1]{( {#1} |}     %%
\newcommand{\rket}[1]{| {#1} )}     %%
\newcommand{\rdket}[1]{|\!| {#1} )}     %%
\newcommand{\rrket}[1]{| {#1} )\!)}     %%
\newcommand{\dket}[1]{|\!| {#1} \rangle}     %%
\newcommand{\maru}[1]{\stackrel{\tiny\circ} {#1}} %%
\newcommand{\wtilde}[1]{\widetilde{#1}} %%
\newcommand{\lsim}{{\stackrel{<}{\raisebox{-1ex}{$\sim$}}}}

\newcommand{\ovl}[1]{\overline{#1}}
\def\beq{\begin{eqnarray}}
\def\eeq{\end{eqnarray}}
\def\bsub{\begin{subequations}}
\def\esub{\end{subequations}}
\def\b{\begin{equation}}
\def\bs{\begin{split}}
\def\es{\end{split}}
\def\e{\end{equation}}
\renewcommand{\theequation}{\arabic{section}.\arabic{equation}}

\title{A Pseudo $su(1,1)$-Algebraic Deformation of the Cooper-Pair 
in the $su(2)$-Algebraic Many-Fermion Model}

\author{%       %Use \sc for the family name
Yasuhiko {\sc Tsue},$^{1}$  
Constan\c{c}a {\sc Provid\^encia},$^{2}$, 
Jo\~ao da {\sc Provid\^encia}$^{2}$ and\\ 
Masatoshi {\sc Yamamura}$^{3}$
}

\inst{%         %Affiliation, neglected when [addenda] or [errata]
$^{1}$Physics Division, Faculty of Science, Kochi University, Kochi 780-8520, Japan\\
$^{2}$Departamento de F\'{i}sica, Universidade de Coimbra, 3004-516 Coimbra, 
Portugal\\
$^{3}$Department of Pure and Applied Physics, 
Faculty of Engineering Science, Kansai University, Suita 564-8680, Japan
}

\abst{%
A pseudo $su(1,1)$-algebra is formulated as a possible deformation of the Cooper-pair in the 
$su(2)$-algebraic many-fermion system. 
With the aid of this algebra, it is possible to describe behavior of individual fermions 
which are generated as the result of interaction with the external environment. 
The form presented in this paper is a generalization of a certain simple case developed recently by 
the present authors. 
Basic idea follows the $su(1,1)$-algebra in the Schwinger boson representation for 
treating energy transfer between the harmonic oscillator and the external environment. 
Hamiltonian is given under the idea of the phase space doubling in the thermo-field dynamics 
formalism and the time-dependent variational method is applied to this Hamiltonian. 
Its trial state is constructed in the frame deformed from the BCS-Bogoliubov approach to 
the superconductivity. 
Several numerical results are shown. 
}

\begin{document}

\maketitle

\section{Introduction}

It may be hardly necessary to mention, but the BCS-Bogoliubov approach to the superconductivity has 
made a central contribution to study of nuclear structure theory. 
The orthogonal set in this approach is determined through two steps. 
At first step, the state $\rket{\phi_B}$ given in the following plays the leading part: 
\bsub\label{1-1}
\beq
& &\rket{\phi_B}=\frac{1}{\sqrt{\Gamma}}\exp \left(z{\wtilde S}_+\right)\rket{0}\ , 
\label{1-1a}\\
& &\Gamma=(1+|z|^2)^{2\Omega_0}\ . 
\label{1-1b} 
\eeq
\esub
Here, $\Gamma$, $z$, ${\wtilde S}_+$ and $\rket{0}$ denote normalization constant, complex parameter, 
the Cooper-pair creation and the fermion vacuum, respectively.
Including ${\wtilde S}_-$ and ${\wtilde S}_0$, the set $({\wtilde S}_{\pm,0})$ forms 
the $su(2)$-algebra. 
Clearly, $\rket{\phi_B}$ is the state with zero seniority. 
But, it is not eigenstate of the fermion-number operator and plays a role of the 
quasiparticle vacuum. 
At second step, the states with nonzero seniority are constructed by operating the quasiparticles on 
$\rket{\phi_B}$ in appropriate manner. 
On the other hand, the Cooper-pair can be treated by the conventional technique of the 
$su(2)$-algebra. 
The orthogonal set in this approach is determined also through two steps. 
First is to construct the minimum weight state $\rket{m}$, which does not contain any Cooper-pair: 
\beq\label{1-2}
{\wtilde S}_-\rket{m}=0\ . 
\eeq
Therefore, $\rket{m}$ is not necessary the state with zero seniority. 
Second is to construct the states orthogonal to $\rket{m}$ by operating ${\wtilde S}_+$ 
in appropriate manner. 
The above mention tells us that, for the two approaches, the orthogonal set is constructed in opposite orders. 
Therefore, without any argument, it may be not concluded that they are equivalent to each other.

In response to the above-mentioned situation, the present authors, recently, proposed a certain idea \cite{1}. 
In this idea, the quasiparticle in the conservation of the fermion-number, which is called the 
``quasiparticle", was introduced. 
Through the medium of this operator, it was shown that both are equivalent to each other in a certain sense. 
Further, in the paper following Ref.\citen{1}, the present authors discussed another role of the 
``quasiparticle": 
It leads to an idea of deformation of the Cooper-pair \cite{2}. 
Hereafter, this paper will be referred to as (A). 
Any state $\rket{\phi}$ with zero seniority including $\rket{\phi_B}$ obeys the condition (A.13), which is 
strongly related to the ``quasiparticle". 
This condition does not lead to fix the form of $\rket{\phi}$ automatically and, then, new condition 
additional to the condition (A.13) is required. 
If the condition (A.17) is added, we obtain $\rket{\phi_B}$. 
In (A), we treated the case of the condition (A.18) in detail. 
In this case, $\rket{\phi}$ is obtained in the form 
\bsub\label{1-3}
\beq
& &\rket{\phi}=\frac{1}{\sqrt{\Gamma}}\exp\left(z{\wtilde {\cal T}}_+\right)\rket{0}\ , 
\label{1-3a}\\
& &\Gamma=\sum_{n=0}^{2\Omega_0}(|z|^2)^n\ . 
\label{1-3b}
\eeq
\esub
Here, ${\wtilde {\cal T}}_+$ is an operator factorized in the product of ${\wtilde S}_+$ and a 
certain operator. 
The definition of ${\wtilde {\cal T}}_+$ including ${\wtilde {\cal T}}_-$ and ${\wtilde {\cal T}}_0$ is given 
in the relation (A.36). 
The commutation relations among ${\wtilde {\cal T}}_{\pm,0}$, which are shown in the relation (A.39), 
suggest us that, in spite of considering of the $su(2)$-algebraic many-fermion model, 
the set $({\wtilde {\cal T}}_{\pm,0})$ resembles the 
$su(1,1)$-algebra in behavior. 
The form (\ref{1-3b}) is given in the relation (A.25).

It is well known that, with the use of two kinds of boson operators, the $su(2)$- and the $su(1,1)$-algebra, 
the generators of which are denoted as ${\hat S}_{\pm,0}$ and ${\hat T}_{\pm,0}$, respectively, can be formulated. 
They are called the Schwinger boson representations \cite{3}. 
For these two algebras, we prepare two boson-spaces: 
(1) the space constructed under a fixed magnitude of the $su(2)$-spin, $s(=0, 1/2, 1, \cdots , s_{\rm max})$ 
and (2) the space constructed under a fixed magnitude of the $su(1,1)$-spin, $t(=1/2, 1, 3/2, \cdots , \infty)$. 
Following the idea of the boson mapping \cite{4}, any operator in the space (1) can be mapped into 
the space (2). 
In the space (1), we can find the set $({\hat {\cal T}}_{\pm,0})$, which obeys 
\beq\label{1-4}
{\hat {\cal T}}_{\pm,0}\ \stackrel{\textrm{(mapped)}}{\longrightarrow} \ {\hat T}_{\pm,0} \ .
\eeq
Naturally, the set $({\hat {\cal T}}_{\pm,0})$ shows the $su(1,1)$-like behavior and it is called 
the pseudo $su(1,1)$-algebra by the present authors \cite{5}. 
In (A), we presented a concrete expression of $({\wtilde {\cal T}}_{\pm,0})$ which corresponds to 
$({\hat {\cal T}}_{\pm,0})$ with $t=1/2$. 
On the other hand, we know that the mixed-mode boson coherent state constructed by $({\hat T}_{\pm,0})$ 
enable us to describe the ``damped and amplified harmonic oscillation" in the frame of the 
conservative form. 
Through this description, we can understand the energy transfer between the harmonic oscillator and the external environment. 
Further, by regarding the mixed-mode boson coherent state as the statistically mixed state, 
thermal effects in time-evolution are described with some interesting results \cite{5,6}. 
Therefore, with the aid of the set $({\hat {\cal T}}_{\pm,0})$, it may be also possible to describe the behavior of 
boson under consideration. 
Its examples are found in the pairing and the Lipkin model in the Holstein-Primakoff type 
boson realization \cite{7}. 
The results were shown in Ref.\citen{5}. 
Primitive form of the above idea is the phase-space doubling introduced in the thermo field dynamics formalism \cite{8}. 
However, it is impossible in the framework of the set $({\hat {\cal T}}_{\pm,0})$ to investigate 
the behavior of individual fermions. 
The form given in (A) may be useful for this problem, but, as is clear from the form (\ref{1-3}), 
the case of the state $\rket{\phi}$ with nonzero seniority cannot be treated in the frame of (A).

This paper aims at two targets. 
First is to generalize the pseudo $su(1,1)$-algebra with zero seniority to the case with nonzero seniority. 
Second is to apply the generalized form to a concrete many-fermion system. 
The $su(2)$-algebra in many-fermion model is characterized by $s$ and $s_0$; for a given $s$, $s_0=-s, -s+1, \cdots ,s-1, s$. 
The $su(1,1)$-algebra in the Schwinger boson representation is characterized by $t$ and $t_0$; 
for a given $t$, $t_0=t, t+1, \cdots , \infty$. 
The pseudo $su(1,1)$-algebra in the Schwinger boson representation, which is abbreviated to 
$B_{ps}$-form, is a possible deformation of the $su(1,1)$-algebra and, therefore, 
it should be characterized at least by $(t, t_0)$. 
However, we are now considering the pseudo $su(1,1)$-algebra which is a possible deformation of the Cooper-pair 
in the $su(2)$-algebraic many-fermion model. 
Hereafter, we will abbreviate it to $F_{ps}$-form. 
One of main problems at the first target is how to import $(t,t_0)$ in the $B_{ps}$-form into the $F_{ps}$-form characterized 
by $(s, s_0)$. 
Following an idea developed in this paper, we have 
\beq\label{1-5}
{\wtilde S}_{\pm,0}\ \stackrel{\textrm{(deformed)}}{\longrightarrow} \ {\wtilde {\cal T}}_{\pm,0} \ .
\eeq
Of course, a form generalized from $\rket{\phi}$ shown in the relation (\ref{1-3}) can be 
presented. 
This form also contains the complex parameter $z$ and the normalization constant $\Gamma$, which 
is a function of $x=|z|^2$. 
Another problem at the first target is how to calculate $\Gamma$ for the range $0 \leq x < \infty$. 
As a possible application of $F_{ps}$-form, we adopt the following scheme: 
Under the time-dependent variational method for a given Hamiltonian expressed in terms of 
$({\wtilde {\cal T}}_{\pm,0})$, we investigate the time-evolution of the system. 
The trial state is $\rket{\phi}$ and, then, 
our problem is reduced to finding the time-dependence of $z$. 
For the above task, we must calculate the expectation values of ${\wtilde {\cal T}}_{\pm,0}$. 
Naturally, $\Gamma$ appears in the expectation values. 
However, $\Gamma$ is complicated polynomial for $x$ and it may be impossible to handle it in 
a lump for the whole range. 
If dividing the whole range into the two, $0\leq x \leq \gamma$ and $\gamma \leq x <\infty$, 
$\Gamma$ becomes approximate, but simple for each range, at very accuracy. 
Here, $\gamma$ denotes a certain constant.

For second target, we must prepare a model for the application. 
The model is non-interacting many-fermion system in one single-particle level, 
which we will call the intrinsic system. 
The reason why we investigate such a simple model comes from the $su(1,1)$-algebra in the 
Schwinger boson representation. 
As was already mentioned, this algebra helps us to describe the harmonic oscillator interacting with the external environment. 
If we follow the thermo-field dynamics formalism, we prepare new degree of freedom for an auxiliary 
harmonic oscillator for the environment, that is, the phase space doubling. 
Further, as the interaction between both degrees of freedom, the form which is proportional to 
$({\hat T}_+-{\hat T}_-)$ is adopted. 
Our present scheme follows the above. 
Our problem is to describe the above-mentioned intrinsic system interacting with the 
external environment. 
For this aim, we introduce an auxiliary many-fermion system and as the interaction between both systems, 
we adopt the form proportional to $({\wtilde {\cal T}}_+-{\wtilde {\cal T}}_-)$. 
To the above Hamiltonian, we apply the time-dependent variational method. 
The trial state is of the form generalized from $\rket{\phi}$ shown in the 
relation (\ref{1-3}) and the variational parameters are $z$ and $z^*$ contained 
in this state. 
Through the variation, we obtain certain differential equations for ${\dot z}$ and ${\dot z}^*$. 
By solving them in appropriate manners including approximation, 
we can arrive at a certain type of the time-evolution. 
According to the result, the intrinsic system shows rather complicated cyclic behavior. 
One cycle can be represented in terms of a chain of different functions for 
the time; linear-, sinh- and sin-type.
This point is essentially different from the result obtained in the $su(1,1)$-algebraic boson model 
which permits infinite boson number. 
This case does not show any cyclic behavior. 
The above mention may be quite natural, because the present model is a 
kind of the $su(2)$-algebraic fermion model in which the Pauli principle works.

In next section, after recapitulating the $su(1,1)$-algebraic boson model 
presented by Schwinger, a pseudo $su(1,1)$-algebra is formulated as a possible 
deformation of the Schwinger boson representation, in which the maximum weight state is 
introduced. 
In \S 3, a possible pseudo $su(1,1)$-algebra as a deformation of the Cooper-pair is formulated 
in the frame of the $su(2)$-algebraic many-fermion model. 
Section 4 is devoted to giving conditions under which the two pseudo $su(1,1)$-algebras are equivalent to 
each other mainly by paying attention to the quantum numbers for the orthogonal sets of both algebras. 
In \S 5, the generalization from $\rket{\phi}$ shown in the relation (\ref{1-3}) is presented. 
Explicit expressions of the normalization constant $\Gamma$ and the expectation value 
of the fermion-number operator $N$ are given. 
Since $\Gamma$ and $N$ are of the complicated forms, in \S 6, the approximate expressions are presented 
in each of the two regions. 
In \S 7, a simple many-fermion model obeying the pseudo $su(1,1)$-algebra is served for the application of the idea 
developed in \S\S 2 - 6. 
Section 8, 9 and 10 are devoted to discussing various properties of $\Gamma$, i.e., 
$N$ in the approximate forms given in \S 6. 
In \S 11, following the scheme mentioned in \S 7, some concrete results are presented and it is shown that 
one cycle consists of a chain of the three different functions for the time. 
Finally, in \S 12, some remarks including future problem are given.

\section{The $su(1,1)$-algebra in the Schwinger boson representation and its deformation 
--- Pseudo $su(1,1)$-algebra ---}

With the use of two kinds of boson operators $({\hat a}, {\hat a}^*$) and (${\hat b}, {\hat b}^*$), the 
Schwinger boson representation of the $su(1,1)$-algebra can be formulated. 
This algebra is composed of three operators which are denoted as ${\hat T}_{\pm,0}$. 
They obey the relations 
\beq
& &{\hat T}_0^*={\hat T}_0\ , \qquad {\hat T}_{\pm}^*={\hat T}_{\mp}\ , 
\label{2-1}\\
& &[\ {\hat T}_+\ , \ {\hat T}_-\ ]=-2{\hat T}_0\ , \qquad
[\ {\hat T}_0\ , \ {\hat T}_{\pm}\ ]=\pm{\hat T}_{\pm}\ . 
\label{2-2}
\eeq
The Casimir operator, which is denoted as ${\hat {\mib T}}^2$, and its property are given by 
\beq
& &{\hat {\mib T}}^2={\hat T}_0^2-\frac{1}{2}\left({\hat T}_-{\hat T}_++{\hat T}_+{\hat T}_-\right)
={\hat T}_0\left({\hat T}_0\mp 1\right)-{\hat T}_{\pm}{\hat T}_{\mp}\ , 
\label{2-3}\\
& &[ \ {\hat T}_{\pm,0}\ , \ {\hat {\mib T}}^2\ ]=0\ .
\label{2-4}
\eeq
The Schwinger boson representation is presented in the form 
\beq\label{2-5}
{\hat T}_+={\hat a}^*{\hat b}^*\ , \qquad
{\hat T}_-={\hat b}{\hat a}\ , \qquad
{\hat T}_0=\frac{1}{2}({\hat a}^*{\hat a}+{\hat b}^*{\hat b})+\frac{1}{2}\ . 
\eeq
The eigenstate of ${\hat {\mib T}}^2$ and ${\hat T}_0$ with the eigenvalues $t(t-1)$ and $t_0$, respectively, 
which is constructed on the minimum weight state $\ket{t}$, is expressed in terms of the 
following form: 
\beq\label{2-6}
\ket{t,t_0}=\sqrt{\frac{(2t-1)!}{(t_0-t)!(t_0+t-1)!}}\left({\hat T}_+\right)^{t_0-t}\ket{t}\ , \qquad
(\bra{t,t_0}t,t_0\rangle=1)
\eeq
Here, $t$ and $t_0$ obey
\beq\label{2-7}
t=1/2,\ 1,\ 3/2, \cdots ,\ \infty\ , \qquad
t_0=t,\ t+1,\ t+2, \cdots ,\ \infty\ .
\eeq
Of course, $\ket{t}$ is given in the form 
\beq\label{2-8}
\ket{t}=\left(\sqrt{(2t-1)!}\right)^{-1}({\hat b}^*)^{2t-1}\ket{0}\ . \qquad 
\left(\ket{t=1/2}=\ket{0}\right)
\eeq
The state $\ket{t}$ satisfies the relation 
\beq\label{2-9}
{\hat T}_-\ket{t}=0\ , \qquad 
{\hat T}_0\ket{t}=t\ket{t}\ . 
\eeq
Concerning the state $\ket{t}$, we must give a small comments. 
The state $(\sqrt{(2t-1)!})({\hat a}^*)^{2t-1}\ket{0}$ satisfies also the relation (\ref{2-9}) and it is 
orthogonal to $\ket{t}$. 
This indicates that we have two types for the minimum weight states, 
which should be discriminated by the quantum number additional to $t$. 
We omit this discrimination and in this paper we will adopt the form (\ref{2-8}). 
The above is an outline of the $su(1,1)$-algebra in the Schwinger boson representation.

Since we are treating boson system, there does not exist any upper limit for 
the values of $t$ and $t_0$. 
In other words, there do not exist the terminal states. 
It can be seen in the relation (\ref{2-7}). 
As a possible variation, we will consider the case where there exists the terminal state for $t_0$: 
\beq\label{2-10}
t_0=t,\ t+1, \cdots , \ t_m-1,\ t_m.
\eeq
The reason why we investigate the above case will be mentioned in 
\S 3 
in relation to the $su(2)$-algebraic many-fermion model. 
In the space specified by the relation (\ref{2-10}), 
we introduce three operators defined as 
\beq\label{2-11}
{\hat {\cal T}}_+={\hat T}_+\sqrt{\frac{t_m-{\hat T}_0}{t_m-{\hat T}_0+\epsilon}}\ , 
\qquad
{\hat {\cal T}}_-=\sqrt{\frac{t_m-{\hat T}_0}{t_m-{\hat T}_0+\epsilon}}{\hat T}_-\ , 
\qquad
{\hat {\cal T}}_0={\hat T}_0\ .
\eeq
Here, $\epsilon$ denotes an infinitesimal positive parameter, which plays a role for avoiding 
the vanishing denominator. 
Successive operation of ${\hat {\cal T}}_+$ gives us the following: 
\bsub\label{2-12}
\beq
& &{\hat {\cal T}}_+\cdot\left({\hat {\cal T}}_+\right)^{t_0-t}\ket{t}=
\left({\hat T}_+\right)^{t_0+1-t}\ket{t}\quad
{\rm for}\quad
t_0=t,\ t+1, \cdots ,\ t_m-2,\ t_m-1,\nonumber\\
& &
\label{2-12a}\\
& &{\hat {\cal T}}_+\cdot\left({\hat {\cal T}}_+\right)^{t_m-t}\ket{t}=0\ , 
\label{2-12b}
\eeq
\esub
\vspace{-0.5cm}
\beq\label{2-13}
& &{\hat {\cal T}}_+\cdot\left({\hat {\cal T}}_+\right)^{t_0-t}\ket{t}=
\left({\hat T}_+\right)^{t_0+1-t}\ket{t}\quad
{\rm for}\quad
t_0=t_m+1,\ t_m+2, \cdots .\qquad\qquad
\eeq
Therefore, the present boson space spanned by the orthogonal set (\ref{2-6}) is divided into two subspaces 
and we are interested in the subspace governed by the relation (\ref{2-12}), in which 
$({\hat {\cal T}}_+)^{t_m-t}\ket{t}$ is the terminal state. 
In this subspace, the commutation relations for ${\hat {\cal T}}_{\pm,0}$ are given in the form 
\beq
& &[\ {\hat {\cal T}}_+\ , \ {\hat {\cal T}}_-\ ]=-2{\hat {\cal T}}_0
+(t_m+t)(t_m-t+1)\ket{t,t_m}\bra{t,t_m}\ , 
\label{2-14}\\
& &[\ {\hat {\cal T}}_0\ , \ {\hat {\cal T}}_{\pm}\ ]=\pm{\hat {\cal T}}_{\pm}\ . 
\label{2-15}
\eeq
We have also the relation 
\beq\label{2-16}
{\hat {\mib {\cal T}}}^2
&=&
{\hat {\cal T}}_0^2-\frac{1}{2}\left({\hat {\cal T}}_-{\hat {\cal T}}_++{\hat {\cal T}}_+{\hat {\cal T}}_-\right)
\nonumber\\
&=&t(t-1)+\frac{1}{2}(t_m+t)(t_m-t+1)\ket{t,t_m}\bra{t,t_m}\ . 
\eeq
Again we note the following relation:
\beq\label{2-17}
\left({\hat {\cal T}}_+\right)^{t_0-t}\ket{t}
=\left({\hat T}_+\right)^{t_0-t}\ket{t}\quad 
{\rm for}\quad
t_0=t,\ t+1,\cdots ,\ t_m-1,\ t_m.
\eeq
The operation of ${\hat {\cal T}}_+$ in the present subspace is essentially the 
same as that of ${\hat T}_+$. 
We call the set $({\hat {\cal T}}_{\pm,0}$) the pseudo $su(1,1)$-algebra. 
It contains positive parameter $t_m$. 
For practical purpose, 
we must find the condition for fixing the value of $t_m$. 
The relation (\ref{2-12a}) suggests us that the terminal state we called may be permitted 
to call the maximum weight state.

\setcounter{equation}{0}

\section{A $su(2)$-algebraic many-fermion model -- Pseudo $su(1,1)$-algebra --}

In \S 2, we presented the pseudo $su(1,1)$-algebra as a possible deformation 
of the $su(1,1)$-algebra. 
In this section, we will formulate the pseudo $su(1,1)$-algebra in the $su(2)$-algebraic 
many-fermion model, which was promised in (A). 
First, we will give an outline of the present many-fermion model. 
The constituents are confined in $4\Omega_0$ single-particle states, where $\Omega_0$ 
denotes integer or half-integer. 
Since $4\Omega_0$ is an even-number, all single-particle states are divided into 
equal parts $P$ and ${\ovl P}$. 
Therefore, as a partner, each single-particle state belonging to $P$ can find a 
single-particle state in ${\ovl P}$. 
We express the partner of the state $\alpha$ belonging to $P$ as ${\bar \alpha}$ and 
fermion operators in $\alpha$ and ${\bar \alpha}$ are denoted 
as $({\tilde c}_{\alpha}, {\tilde c}_{\alpha}^*$) and $({\tilde c}_{\bar \alpha},{\tilde c}_{\bar \alpha}^*)$, 
respectively. 
As the generators ${\wtilde S}_{\pm,0}$, we adopt the following form: 
\beq\label{3-1}
& &{\wtilde S}_+=\sum_{\alpha}s_{\alpha}{\wtilde c}_{\alpha}^*{\tilde c}_{\bar \alpha}^*\ , \quad
{\wtilde S}_-=\sum_{\alpha}s_{\alpha}{\wtilde c}_{\bar \alpha}{\tilde c}_{\alpha}\ , \nonumber\\
& &
{\wtilde S}_0=\frac{1}{2}{\wtilde N}-\Omega_0\ , \qquad
{\wtilde N}=\sum_{\alpha}({\wtilde c}_{\alpha}^*{\tilde c}_{\alpha}
+{\tilde c}_{\bar \alpha}^*{\tilde c}_{\bar \alpha})\ . \quad
\eeq
The symbol $s_{\alpha}$ denotes real number satisfying $s_{\alpha}^2=1$. 
The sum $\sum_{\alpha}$ $(\sum_{\bar \alpha}$) is carried out in all single-particle states 
in $P$ (${\ovl P}$) and we have 
$\sum_{\alpha}1=2\Omega_0\ (\sum_{\bar \alpha}1=2\Omega_0$). 
The operators ${\wtilde S}_{\pm,0}$ forms the $su(2)$-algebra obeying the relations 
\beq
& &{\wtilde S}_0^*={\wtilde S}_0\ , \qquad
{\wtilde S}_{\pm}^*={\wtilde S}_{\mp}\ , 
\label{3-2}\\
& &[\ {\wtilde S}_+\ , \ {\wtilde S}_-\ ]=2{\wtilde S}_0\ , \qquad
[\ {\wtilde S}_0\ , \ {\wtilde S}_{\pm}\ ]=\pm{\wtilde S}_{\pm}\ .
\label{3-3}
\eeq
The Casimir operator, which is denoted as ${\wtilde {\mib S}}^2$ and its property are 
given by 
\beq
& &{\wtilde {\mib S}}^2={\wtilde S}_0^2+\frac{1}{2}\left(
{\wtilde S}_-{\wtilde S}_+ +{\wtilde S}_+{\wtilde S}_-\right)
={\wtilde S}_0\left({\wtilde S}_0\mp 1\right)+{\wtilde S}_{\pm}{\wtilde S}_{\mp}\ , 
\label{3-4}\\
& &[\ {\wtilde S}_{\pm,0}\ , \ {\wtilde {\mib S}}^2\ ]=0\ .
\label{3-5}
\eeq
The eigenstate of ${\wtilde {\mib S}}^2$ and ${\wtilde S}_0$ with the eigenvalues $s(s+1)$ and $s_0$, 
respectively, is expressed in the form 
\beq\label{3-6}
\rket{s,s_0}=\sqrt{\frac{(s-s_0)!}{(2s)!(s+s_0)!}}\left({\wtilde S}_+\right)^{s+s_0}\rket{s}\ . 
\quad
\left(\rbra{s,s_0}s,s_0)=1\right)
\eeq
Here, $s$ and $s_0$ obey 
\beq\label{3-7}
s=0,\ 1/2,\ 1,\cdots ,\ \Omega_0 \ , \qquad
s_0=-s,\ -s+1, \cdots ,\ s-1, \ s .
\eeq
The state $\rket{s}$ denotes the minimum weight state satisfying 
\beq\label{3-8}
{\wtilde S}_-\rket{s}=0\ , \qquad 
{\wtilde S}_0\rket{s}=-s\rket{s}\ . 
\eeq
Since $\rket{s}$ is given in many-fermion system, it depends on not only 
$s$ but also the quantum numbers additional to $s$ and, 
recently, we presented an idea how to construct $\rket{s}$ in an explicit form \cite{9}. 
Later, we will sketch it. 
Needless to say, the operator ${\wtilde S}_+\ ({\wtilde S}_-)$ plays a role of creation (annihilation) 
of the Cooper-pair.

As a possible deformation of ${\wtilde S}_{\pm,0}$, i.e., deformation of the Cooper-pair, we introduce three operators in the space 
spanned by the set (\ref{3-6}). 
They are expressed in the form 
\beq\label{3-9}
{\wtilde {\cal T}}_+={\wtilde S}_+\sqrt{\frac{s+{\wtilde S}_0+2t'}{s-{\wtilde S}_0+\epsilon}}\ , 
\quad
{\wtilde {\cal T}}_-=\sqrt{\frac{s+{\wtilde S}_0+2t'}{s-{\wtilde S}_0+\epsilon}}{\wtilde S}_-\ , 
\quad
{\wtilde {\cal T}}_0=s+{\wtilde S}_0+t'\ .\quad
\eeq
Here, $\epsilon$ denotes an infinitesimal positive parameter. 
The form (\ref{3-9}) contains positive parameter $t'$ and in (A), we considered the case 
$t'=1/2$ for $s=\Omega_0$. 
The commutation relations for ${\wtilde {\cal T}}_{\pm,0}$ are given in the form 
\beq
& &[\ {\wtilde {\cal T}}_+\ , \ {\wtilde {\cal T}}_-\ ]=-2{\wtilde {\cal T}}_0
+(2s+2t')(2s+1)\rket{s,s}\rbra{s,s}\ , 
\label{3-10}\\
& &[\ {\wtilde {\cal T}}_0\ , \ {\wtilde {\cal T}}_{\pm}\ ]=\pm{\wtilde {\cal T}}_{\pm}\ . 
\label{3-11}
\eeq
The operator ${\wtilde {\mib {\cal T}}}^2$ is expressed as 
\beq\label{3-12}
{\wtilde {\mib {\cal T}}}^2
&=&
{\wtilde {\cal T}}_0^2-\frac{1}{2}\left({\wtilde {\cal T}}_-{\wtilde {\cal T}}_++{\wtilde {\cal T}}_+{\wtilde {\cal T}}_-\right)
\nonumber\\
&=&t'(t'-1)+\frac{1}{2}(2s+2t')(2s+1)\rket{s,s}\rbra{s,s}\ . 
\eeq
From the comparison with the relation (\ref{3-10})-(\ref{3-12}) with the relations 
(\ref{2-14})-(\ref{2-16}), we can understand that the set 
$({\wtilde {\cal T}}_{\pm,0})$ forms also the pseudo $su(1,1)$-algebra. 
Successive operation of ${\wtilde {\cal T}}_+$ on the state $\rket{s}$ gives us 
\bsub\label{3-13}
\beq
& &{\wtilde {\cal T}}_+\cdot\left({\wtilde {\cal T}}_+\right)^{s+s_0}\rket{s}
=\left({\wtilde {\cal T}}_+\right)^{s+s_0+1}\rket{s}\quad
{\rm for}\quad
s_0=-s,\ -s+s,\cdots ,\ s-1\ , \quad
\label{3-13a}\\
& &{\wtilde {\cal T}}_+\cdot\left({\wtilde {\cal T}}_+\right)^{2s}\rket{s}=0\ . 
\label{3-13b}
\eeq
\esub
The relation (\ref{3-13b}) tells us that $({\wtilde {\cal T}}_+)^{2s}\rket{s}$ is the 
maximum weight state. 
Further, we have 
\beq\label{3-14}
\left({\wtilde {\cal T}}_+\right)^{s+s_0}\rket{s}
=\sqrt{\frac{(2t'-1+s+s_0)!}{(2t'-1)!}\frac{(s-s_0)!}{(2s)!}}\left({\wtilde S}_+\right)^{s+s_0}\rket{s}\ .
\eeq
The relation (\ref{3-14}) suggests us that in order to describe the $su(2)$-algebraic 
model,
it may be enough to treat the model in the orthogonal set 
$\{({\wtilde S}_+)^{s+s_0}\rket{s}\}$. 
In spite of this fact, we describe in the orthogonal set 
$\{({\wtilde {\cal T}}_+)^{s+s_0}\rket{s}\}$. 
The reason may be clear in \S 5. 
It must be also noted that $\{({\wtilde {\cal T}}_+)^{s+s_0}\rket{s}; s_0=-s, -s+1,\cdots , s\}$ 
corresponds to $\{({\hat {\cal T}}_+)^{t_0-t}\ket{t}; t_0=t,t+1,\cdots , t_m\}$, which is 
defined in the relation (\ref{2-12}).

\setcounter{equation}{0}

\section{Condition for the equivalence of two pseudo $su(1,1)$-algebras}

In last two sections, we derived the pseudo $su(1,1)$-algebra from the two algebraic models: 
(1) the $su(1,1)$-algebra in the Schwinger boson representation and (2) the $su(2)$-algebra 
in many-fermion system. 
As was mentioned in \S 1, we call the first and the second as $B_{ps}$- and $F_{ps}$-form, respectively. 
Three quantities $t$, $t_0$ and $t_m$ characterize $B_{ps}$-form. 
In these three, $t$ and $t_0$ indicate the quantum numbers for the $su(1,1)$-algebra 
itself and, especially, $t$ determines the irreducible representation. 
The quantity $t_m$ is an artificial parameter introduced from the outside 
for defining the maximum weight state of $B_{ps}$-form. 
On the other hand, $F_{ps}$-form is characterized by four quantities, $s$, $s_0$, $\Omega_0$ and $t'$. 
The quantities $s$ and $s_0$ indicate the quantum numbers for the $su(2)$-algebra itself 
and $s$ determines the irreducible representation. 
The existence of the maximum weight state is guaranteed by $\Omega_0$. 
The quantity $t'$ is an artificial parameter introduced for constructing $F_{ps}$-form.

Under the above mention, let us search the condition which makes $B_{ps}$- and $F_{ps}$-form equivalent to each other. 
For this aim, we require the following correspondence: 
\beq\label{4-1}
\dket{t,t_0}\sim \rdket{s,s_0}\ .
\eeq
Here, $\dket{t,t_0}$ and $\rdket{s,s_0}$ are given as 
\bsub\label{4-2}
\beq
& &\dket{t,t_0}=\left({\hat {\cal T}}_+\right)^{t_0-t}\ket{t}\ , \quad
(t_0=t,\ t+1,\cdots ,\ t_m-1,\ t_m)
\label{4-2a}\\
& &\rdket{s,s_0}=\left({\wtilde {\cal T}}_+\right)^{s+s_0}\rket{s}\ . \quad
(s_0=-s,\ -s+1,\cdots ,\ s-1,\ s)
\label{4-2b}
\eeq
\esub
If the correspondence (\ref{4-1}) is permitted, the number of the states (\ref{4-2a}) should be 
equal to that of the states (\ref{4-2b}):
\beq\label{4-3}
t_m-t+1=2s+1\ , \quad {\rm i.e.,}\quad t_m-t=2s\ .
\eeq
Since $\dket{t,t_m}$ corresponds to $\rdket{s,s}$, the relations (\ref{2-14}) and (\ref{3-10}) should lead to 
\beq\label{4-4}
(t_m+t)(t_m-t+1)=(2s+2t')(2s+1)\ .
\eeq
Then, with the use of the relation (\ref{4-3}), we have
\beq\label{4-5}
t=t'\ .
\eeq
The eigenvalues of ${\hat {\cal T}}_0$ and ${\wtilde {\cal T}}_0$ for $\dket{t,t_0}$ and $\rdket{s,s_0}$ are 
given in the $t_0$ and $s+s_0+t'$, respectively, and they should be equal to each other:
\beq\label{4-6}
t_0=s+s_0+t'\ . 
\eeq
The cases $s_0=-s$ and $s_0=s$ correspond to the cases $t_0=t$ and $t_0=t_m$, respectively and 
they lead to $t_m=2s+t'$. 
They are consistent to the relations (\ref{4-5}) and (\ref{4-3}).

The above result is summarized as follows:
\beq\label{4-7}
t=t'\ , \qquad t_0=s+s_0+t'\ , \qquad t_m=2s+t'\ .
\eeq
We can see that $t$, $t_0$ and $t_m$ which characterize $B_{ps}$-form are expressed in terms of $s$, $s_0$ and 
$t'$ characterizing $F_{ps}$-form. 
However, usually, the $su(2)$-algebraic many-fermion model contains two quantum numbers 
except $\Omega_0$ which determines the framework of the model. 
As was already mentioned, $t'$ is introduced as an artificial parameter 
and $t$ determines the irreducible representation of the $su(1,1)$-algebra. 
Therefore, $t'$ may be a function of $\Omega_0$ and $s$, which determine the 
framework of the irreducible representation of the $su(2)$-algebra. 
As an example, in this paper, we will adopt the following form: 
\beq\label{4-8}
t'=\Omega_0+\frac{1}{2}-s\ (=t)\ , \quad {\rm i.e.,}\quad s+t=\Omega_0+\frac{1}{2}\ .
\eeq
If $t=1/2$, $s$ is equal to $\Omega_0$ and in (A), we investigated this case. 
The forms (\ref{4-7}) and (\ref{4-8}) give us the relation 
\beq\label{4-9}
t=t'=\Omega_0+\frac{1}{2}-s\ , \qquad t_0=\Omega_0+\frac{1}{2}+s_0\ , \qquad
t_m=\Omega_0+\frac{1}{2}+s\ .
\eeq

Final task of this section is to examine the validity of the relation (\ref{4-8}). 
For this examination, detailed structure of the state $\rket{s}$ must be investigated in relation to the 
state $\ket{t}$. 
Concerning the construction of the minimum weight state for the present $su(2)$-algebraic model, 
recently, the present authors presented an idea, with the aid of which the minimum weight state can be determined 
methodically \cite{9}. 
Following this idea, we will consider the present problem. 
First, we introduce the following $su(2)$-generators: 
\beq\label{4-10}
{\wtilde R}_+=\sum_{\alpha}{\tilde c}_{\alpha}^*{\tilde c}_{\bar \alpha}\ , \qquad
{\wtilde R}_-=\sum_{\alpha}{\tilde c}_{\bar \alpha}^*{\tilde c}_{\alpha}\ , \qquad
{\wtilde R}_0=\frac{1}{2}\sum_{\alpha}({\tilde c}_{\alpha}^*{\tilde c}_{\alpha}
-{\tilde c}_{\bar \alpha}^*{\tilde c}_{\bar \alpha})\ . 
\eeq
The generators ${\wtilde R}_{\pm,0}$ satisfy the relation 
\beq\label{4-11}
[\ {\rm any\ of}\ {\wtilde R}_{\pm,0}\ , \ {\rm any\ of}\ {\wtilde S}_{\pm,0}\ ]=0\ . 
\eeq
The relation (\ref{4-11}) suggests us that there exists the minimum weight state not only 
for $({\wtilde S}_{\pm,0})$ but also $({\wtilde R}_{\pm,0})$, which is 
denoted as $\rket{m_0}$:
\beq\label{4-12}
& &{\wtilde S}_-\rket{m_0}=0\ , \qquad 
{\wtilde R}_-\rket{m_0}=0\ , \nonumber\\
& &{\wtilde S}_0\rket{m_0}=-s\rket{m_0}\ , \qquad
{\wtilde R}_0\rket{m_0}=-r\rket{m_0}\ .
\eeq
Definition of ${\wtilde S}_-$, ${\wtilde R}_-$, ${\wtilde S}_0$ and ${\wtilde R}_0$ gives us the following form: 
\beq\label{4-13}
\rket{m_0}
=\left\{
\begin{array}{ll}
\rket{0}\ , & (r=0) \\
\displaystyle \prod_{i=1}^{2r}{\tilde c}_{{\bar \alpha}_i}^*\rket{0}\ . & (r=1/2,\ 1,\ 3/2,\cdots ,\ \Omega_0)
\end{array}
\right.
\eeq
It should be noted that $\rket{m_0}$ is composed of only the fermion creation operators belonging to 
${\ovl P}$ and symbolically we express $\rket{m_0}$ in the form 
\beq\label{4-14}
\rket{m_0}=({\tilde c}_{\ovl P}^*)^{2r}\rket{0}\ . 
\eeq
Here, ${\tilde c}_{\ovl P}^*$ and $2r$ denote any of ${\tilde c}_{\bar \alpha}^*$ and the number of ${\tilde c}_{\ovl P}^*$, 
respectively. 
The operation of ${\wtilde S}_0$ on $\rket{m_0}$ leads us to 
\beq\label{4-15}
{\wtilde S}_0\rket{m_0}
=\left(\frac{1}{2}\sum_{\alpha}({\tilde c}_{\alpha}^*{\tilde c}_{\alpha}+{\tilde c}_{\bar \alpha}^*{\tilde c}_{\bar \alpha})
-\Omega_0\right)\rket{m_0}
=-(\Omega_0-r)\rket{m_0}\ . 
\eeq
If $\rket{m_0}$ is adopted as $\rket{s}$, we have 
\beq\label{4-16}
s=\Omega_0-r\ . 
\eeq
We can see that $2r$ denotes the seniority number. 
Further, with the use of the raising operator, ${\wtilde R}_+$, and certain scalar operator for the 
$su(2)$-algebra $({\wtilde R}_{\pm,0})$, ${\wtilde {\cal P}}^*$, the minimum weight state 
$\rket{m}$ is obtained in the form 
$\rket{m}={\wtilde {\cal P}}^*\cdot({\wtilde R}_+)^{r+r_0}\rket{m_0}$. 
The above is our idea presented in Ref.\citen{2}

In \S 7, we will investigate the present pseudo $su(1,1)$-algebra under the idea of the 
phase space doubling in the thermo-field dynamics 
formalism. 
For this aim, it is enough to adopt $\rket{m_0}$ as the minimum weight state for the $su(2)$-algebra 
$({\wtilde S}_{\pm,0})$. 
In other words, if we adopt the form $\rket{m}={\wtilde {\cal P}}^*({\wtilde R}_+)^{r+r_0}\rket{m_0}$, 
the present pseudo $su(1,1)$-algebra becomes powerless for the idea of the phase space doubling. 
Under the above argument, let us consider the correspondence of $\ket{t}$ with $\rket{s}$. 
As for $\rket{s}$, we adopt the form $\rrket{r}=({\tilde c}_{\ovl P}^*)^{2r}\rket{0}\ (r=\Omega_0-s)$:
\beq\label{4-17}
\ket{t}=({\hat b}^*)^{2t-1}\ket{0}\ \sim\ 
\rket{s}=({\tilde c}_{\ovl P}^*)^{2r}\rket{0}\ . 
\eeq
Of course, the following correspondence may be permitted: 
\beq\label{4-18}
\ket{0}\ \sim\ \rket{0}\ . 
\eeq
Concerning $\ket{t}$ and $\rket{s}$ $(=\rrket{r})$, we have 
\beq\label{4-19}
{\hat b}^*\ket{t}=\ket{t+1/2}\ , \qquad {\tilde c}_{\ovl P}^*\rrket{r}=\rrket{r+1/2}\ . 
\eeq
Therefore, the following correspondence is obtained: 
\beq\label{4-20}
({\hat b}^*)^\nu\ket{0}\ \sim\ ({\tilde c}_{\ovl P}^*)^{\nu}\rket{0}\quad {\rm for}\quad
\nu=0,\ 1,\ 2,\cdots \ .
\eeq
Thus, the relation (\ref{4-17}) leads us to 
\beq\label{4-21}
2t-1=2r\ , \quad {\rm i.e.,}\quad 2t-1=2(\Omega_0-s)\ . 
\eeq
The above is nothing but the relation (\ref{4-8}). 
The operators ${\wtilde T}_{\pm,0}$ can be summarized in the form 
\beq\label{4-22}
& &{\wtilde {\cal T}}_+={\wtilde S}_+\sqrt{\frac{\Omega_0+\frac{1}{2}+t+{\wtilde S}_0}{\Omega_0+\frac{1}{2}-t-{\wtilde S}_0+\epsilon}}\ , 
\qquad
{\wtilde {\cal T}}_-=\sqrt{\frac{\Omega_0+\frac{1}{2}+t+{\wtilde S}_0}{\Omega_0+\frac{1}{2}-t-{\wtilde S}_0+\epsilon}}{\wtilde S}_-\ , 
\nonumber\\
& &{\wtilde {\cal T}}_0=\Omega_0+\frac{1}{2}+{\wtilde S}_0\ . 
\eeq
Thus, we could finish the task.

We know that the Cooper-pair in the BCS-Bogoliubov theory can be described 
by ${\wtilde S}_{\pm}$ and the set $({\wtilde S}_{\pm,0})$ forms the $su(2)$-algebra. 
On the other hand, ${\wtilde {\cal T}}_{\pm}$ can be regarded as a possible deformation of the Cooper-pair, 
which belongs still to the category of the $su(2)$-algebra. 
If we notice that the relation (\ref{2-11}) represents a possible deformation of the $su(1,1)$-algebra, our 
algebra, which we call the pseudo $su(1,1)$-algebra, may be expected to be useful 
for treating physical problem different from the superconductivity and its related problem.

\setcounter{equation}{0}

\section{A possible fermion-number non-conserving state in the $su(2)$-algebraic model}

In (A), we investigated the fermion-number non-conserving state shown in the form 
\beq\label{5-1}
\rket{\phi}=\frac{1}{\sqrt{\Gamma}}\exp\left(z{\wtilde {\cal T}}_+\right)\rket{\Omega_0}
\quad {\rm for}\quad t=1/2,\ {\rm i.e.,}\ s=\Omega_0\ .
\eeq
Here, $\Gamma$ and $z$ denote the normalization $((\phi\rket{\phi}=1)$ and complex parameter,
respectively. 
The state (\ref{5-1}) is an example of the deformation of the BCS-Bogoliubov state. 
In this section, we will develop its generalization to the case $t>1/2$, i.e., $s<\Omega_0$: 
\beq\label{5-2}
\rket{\phi}=\frac{1}{\sqrt{\Gamma}}\exp\left(z{\wtilde {\cal T}}_+\right)\rket{s}\ . 
\eeq
The state (\ref{5-2}) can be expanded to 
\beq\label{5-3}
\rket{\phi}=\frac{1}{\sqrt{\Gamma}}\sum_{n=0}^{2s}\frac{z^n}{n!}\left({\wtilde {\cal T}}_+\right)^n\rket{s}\ . 
\eeq
For the convenience of the treatment, we formulate in the $B_{ps}$-frame. 
Then, $\rket{\phi}$ corresponds to $\ket{\phi}$ given as 
\beq\label{5-4}
\ket{\phi}=\frac{1}{\sqrt{\Gamma}}\sum_{n=0}^{t_m-t}\frac{z^n}{n!}\left({\hat T}_+\right)^{n}\ket{t}\ . \quad 
(\bra{\phi}\phi\rangle=1)
\eeq
With the use of the relation (\ref{4-9}), $2s$ and $(t_m-t)$ can be expressed in the relation 
\beq\label{5-5}
2s=t_m-t=2\Omega_0-(2t-1)\ .
\eeq
%With the use of a new variable $x\ (=|z|^2)$, $\Gamma$ for the state (\ref{5-4}) is 
%given in the form 
The normalization constant $\Gamma$ can be expressed as a function of a new variable $x$ ($=|z|^2)$ in the form 
\beq\label{5-6}
{\rm (i)}\quad 
\Gamma=\Gamma_t(x)=\sum_{n=0}^{2\Omega_0-(2t-1)}x^n
\left(
\begin{array}{c}
2t-1+n \\
2t-1
\end{array}
\right) 
=1+2tx+\cdots \ . 
\quad (0\leq x < \infty)
\nonumber\\
& &
\eeq
Here, $\left(
\begin{array}{c}
2t-1+n \\
2t-1
\end{array}
\right)$
denotes the binomial coefficient and for deriving the above form the orthogonal set (\ref{2-6}) is used. 
We will treat $\Gamma$ in various values of $t$ and, 
hereafter, $\Gamma$ 
is denoted as $\Gamma_t(x)$. 
The function $\Gamma_t(x)$ is a polynomial for $x$, the degree of which is $2\Omega_0-(2t-1)$ 
and all the coefficients of $x^n\ (n=1,2,\cdots , 2\Omega_0-(2t-1))$ are positive. 
Therefore, we have another expression: 
\beq\label{5-7}
{\rm (ii)}\quad
\Gamma_t(x)&=&
\left(\begin{array}{c}
2\Omega_0 \\
2t-1
\end{array}
\right)
x^{2\Omega_0-(2t-1)}\sum_{n=0}^{2\Omega_0-(2t-1)}\left(\frac{1}{x}\right)^n
\left(
\begin{array}{c}
2\Omega_0-n \\
2t-1
\end{array}\right)
\left(\begin{array}{c}
2\Omega_0 \\
2t-1
\end{array}\right)^{-1}\nonumber\\
&=&
\left(\begin{array}{c}
2\Omega_0 \\
2t-1
\end{array}
\right)
x^{2\Omega_0-(2t-1)}
\left[1+\frac{1}{x}\cdot\left(\frac{2\Omega_0-(2t-1)}{2\Omega_0}\right)+\cdots \right]\ .
\eeq
Of course, the relations (\ref{5-6}) and (\ref{5-7}) are useful in the cases $x\sim 0$ and $x\rightarrow \infty$, respectively. 
First, we will discuss the relation to the $su(1,1)$-algebraic model. 
It is noted that $\Gamma_t(x)$ is rewritten to the form 
\beq\label{5-8}
{\rm (iii)}\quad
\Gamma_t(x)=\frac{1}{(1-x)^{2t}}\left[
1-x^{2\Omega_0+1}\sum_{n=0}^{2t-1}\left(\frac{1-x}{x}\right)^n
\left(
\begin{array}{c}
2\Omega_0+1\\
n
\end{array}
\right)\right]\ . \quad\qquad\qquad
\eeq
For this rewriting, we used the formula
\beq\label{5-9}
\Gamma_t(x)=\frac{1}{2t-1}\frac{d}{dx}\Gamma_{t-1/2}(x)\quad {\rm for}\quad t>1/2\ , 
\eeq
i.e., 
\beq
& &\Gamma_t(x)=\frac{1}{(2t-1)!}\left(\frac{d}{dx}\right)^{2t-1}\Gamma_{1/2}(x)\quad {\rm for}\quad t\geq 1/2\ , 
\label{5-10}\\
& &\Gamma_{1/2}(x)=\sum_{n=0}^{2\Omega_0}x^n=\frac{1-x^{2\Omega_0+1}}{1-x}\ . 
\eeq
If $2\Omega_0-(2t-1)\rightarrow \infty$, the expression (\ref{5-6}) is an infinite series which is convergent for $x<1$:
\beq\label{5-12}
\Gamma_t(x)=\frac{1}{(1-x)^{2t}}\ . \qquad (0\leq x <1)
\eeq
The form (\ref{5-12}) corresponds to the case of the $su(1,1)$-algebraic model and 
at $x=1$, it diverges. 
However, the form (\ref{5-6}) is a finite series defined in the range $0\leq x <\infty$ and, of course, at $x=1$, 
it is finite. 
This can be shown explicitly in the form 
\beq\label{5-13}
{\rm (iv)}\qquad\quad
\Gamma_t(x)=x^{2\Omega_0-(2t-1)}\sum_{n=0}^{2\Omega_0-(2t-1)}\left(\frac{1-x}{x}\right)^n
\left(
\begin{array}{c}
2\Omega_0+1\\
2t+n
\end{array}
\right)\ . \qquad\qquad
\eeq
The form (\ref{5-13}) can be derived from the relation (\ref{5-8}) through the relation 
\beq\label{5-14}
\sum_{n=0}^{2t-1}\left(\frac{1-x}{x}\right)^n
\left(
\begin{array}{c}
2\Omega_0+1\\
n
\end{array}
\right)
=
\frac{1}{x^{2\Omega_0+1}}-\sum_{n=2t}^{2\Omega_0+1}\left(\frac{1-x}{x}\right)^n
\left(
\begin{array}{c}
2\Omega_0+1\\
n
\end{array}
\right) \ . \quad
\eeq
The relation (\ref{5-13}) gives us the finite value at $x=1$: 
\beq\label{5-15}
\Gamma_t(x=1)=\left(
\begin{array}{c}
2\Omega_0+1\\
2t
\end{array}
\right) . 
\eeq
We showed four expressions for $\Gamma_t(x)$. 
It may be necessary to put each expression to its proper use. 
Through the state $\ket{\phi}$ (or $\rket{\phi}$), we can learn the difference between the 
$su(1,1)$- and the pseudo $su(1,1)$-algebraic model.

Next, we will treat the expectation value of the fermion number operator ${\wtilde N}$ for the 
state $\rket{\phi}$, which also depends on $t$ and $x$. 
For this aim, the following relation is useful:
\beq\label{5-16}
{\wtilde {\cal T}}_0=s+{\wtilde S}_0+t=\frac{\wtilde N}{2}+\frac{1}{2}\ , \quad{\rm i.e.,}\quad
{\wtilde N}=2{\wtilde {\cal T}}_0-1\ .
\eeq
Then, in the $B_{ps}$-form, we have the relation 
\beq
N&=&
2\bra{\phi}{\hat {\cal T}}_0\ket{\phi}-1=
2\bra{\phi}{\hat T}_0\ket{\phi}-1=2{\cal T}_0-1\nonumber\\
&=&2t-1+2\Lambda_t(x)\ , 
\label{5-17}\\
\Lambda_t(x)&=&x\cdot\frac{\frac{d\Gamma_t(x)}{dx}}{\Gamma_t(x)}=
\frac{tx\Gamma_{t+1/2}(x)}{\Gamma_{t}(x)}\ . 
\label{5-18}
\eeq
For the four forms of $\Gamma_t(x)$, $\Lambda_t(x)$ can be expressed in the form 
\beq
& &{\rm (i)'}\quad 
\Lambda_t(x)=\frac{\displaystyle \sum_{n=1}^{2\Omega_0-(2t-1)}nx^n
\left(
\begin{array}{c}
2t-1+n \\ 2t-1 
\end{array}\right)}{\displaystyle 1+\sum_{n=1}^{2\Omega_0-(2t-1)}x^n
\left(
\begin{array}{c}
2t-1+n \\ 2t-1 
\end{array}\right)}\ , 
\label{5-19}\\
& &{\rm (ii)'}\quad
\Lambda_t(x)=2\Omega_0-(2t-1)\nonumber\\
& &\qquad\qquad\qquad
-
\frac{\displaystyle \sum_{n=1}^{2\Omega_0-(2t-1)}n\left(\frac{1}{x}\right)^n
\left(
\begin{array}{c}
2\Omega_0-n \\ 2t-1 
\end{array}\right)
\left(
\begin{array}{c}
2\Omega_0 \\ 2t-1 
\end{array}\right)^{-1}
}{\displaystyle 1+\sum_{n=1}^{2\Omega_0-(2t-1)}\left(\frac{1}{x}\right)^n
\left(
\begin{array}{c}
2\Omega_0-n \\ 2t-1 
\end{array}\right)
\left(
\begin{array}{c}
2\Omega_0 \\ 2t-1 
\end{array}\right)^{-1}
}\ , 
\label{5-20-new}\\
& &{\rm (iii)'}\quad
\Lambda_t(x)=\frac{\displaystyle \frac{2tx}{1-x}-((2\Omega_0+1)-(2t-1))x^{2\Omega_0+1}
\left(\frac{1-x}{x}\right)^{2t-1}
\left(
\begin{array}{c}
2\Omega_0+1 \\ 2t-1 
\end{array}\right)}
{\displaystyle 1-x^{2\Omega_0+1}\sum_{n=1}^{2t-1}
\left(\frac{1-x}{x}\right)^n
\left(
\begin{array}{c}
2\Omega_0+n \\ n 
\end{array}\right)}\ , \nonumber\\
& &
\label{5-20}\\
& &{\rm (iv)'}\quad
\Lambda_t(x)=\frac{2t}{2t+1}(2\Omega_0-(2t-1))\nonumber\\
& &\qquad\qquad\qquad
\times
\frac{\displaystyle 1+\sum_{n=1}^{2\Omega_0-2t}\left(\frac{1-x}{x}\right)^n
\left(
\begin{array}{c}
2\Omega_0+1 \\ 2t+1+n 
\end{array}\right)
\left(
\begin{array}{c}
2\Omega_0+1 \\ 2t+1 
\end{array}\right)^{-1}
}{\displaystyle 1+\sum_{n=1}^{2\Omega_0-(2t-1)}\left(\frac{1-x}{x}\right)^n
\left(
\begin{array}{c}
2\Omega_0+1 \\ 2t+n 
\end{array}\right)
\left(
\begin{array}{c}
2\Omega_0+1 \\ 2t 
\end{array}\right)^{-1}
}\ . 
\label{5-21}
\eeq
The forms (i)${}'$ and (ii)${}'$ are suitable for investigating the cases $x\sim 0$ and $x\rightarrow \infty$: 
\bsub\label{5-22}
\beq
& &\Lambda_t(x)=2t(x+\cdots )\ , \qquad (x\sim 0)
\label{5-22a}\\
& &\Lambda_t(x)=(2\Omega_0-(2t-1))\left(1-\frac{1}{2\Omega_0}\cdot\frac{1}{x}+\cdots \right)\ . \qquad
(x\rightarrow \infty)
\label{5-22b}
\eeq
\esub
The form (iv)${}'$ is suited to the case $x\sim 1$:
\beq\label{5-23}
\Lambda_t(x)=\frac{2t}{2t+1}(2\Omega_0-(2t-1))\left(1+\frac{\Omega_0+1}{(t+1)(2t+1)}(x-1)+\cdots \right)\ . 
\eeq
The form (iii)${}'$ is related to the $su(1,1)$-algebraic model:
\beq\label{5-24}
\Lambda_t(x)=\frac{2tx}{1-x}\ . \qquad
\left( (2\Omega_0+1)-(2t-1)\rightarrow \infty \right)
\eeq
The relation (\ref{5-17}) gives us the expectation value of ${\wtilde N}$ denoted by $N$. 
The typical three cases are as follows: 
\bsub\label{5-25}
\beq
& &N=2t-1\ , \qquad (x=0)
\label{5-25a}\\
& &N=2t-1+2\left(\frac{2t}{2t+1}\right)(2\Omega_0-(2t-1))\nonumber\\
& &\ \ \ \ 
=\left(\frac{2t}{2t+1}\right)\cdot 4\Omega_0
-\left(\frac{2t-1}{2t+1}\right)\cdot (2t-1)\ , \qquad
(x=1)
\label{5-25b}\\
& &N=2t-1+2(2\Omega_0-(2t-1))
=4\Omega_0-(2t-1)\ . 
\qquad (x\rightarrow \infty)
\label{5-25c}
\eeq
\esub
We can see that at $t=1/2$, the above result is reduced to that in (A). 
The relation (\ref{4-21}) tells us that the number $(2t-1)$ indicates the seniority number. 
Therefore, $(2t-1)$ fermions belonging to ${\ovl P}$ cannot contribute to the fermion-pair ${\wtilde S}_+$ $({\wtilde S}_-)$ 
and, then, $(2t-1)$ single-particle states are not available for the formation of the 
fermion-pair. 
In this sense, the result (\ref{5-25}) is quite natural. 
The result (\ref{5-25}) tells us that the case $x=1$ corresponds to the intermediate situation 
between the cases $x=0$ and $x\rightarrow \infty$. 
Figure 1 shows various cases for $t$ in the case $\Omega_0=19/2$. 
In the range $0\leq x \lsim 2$, the slopes are steep and after $x\sim 2$, the slopes become gentle. 
More precisely, as $t$ increases, the point where the slope becomes gentle approaches to $x=0$.
This feature can be read in the result (\ref{5-25}). 
%
%%%%%%%%%%%%%%%%%%%%%%%%%%%%%%%%%%%%%%%%%%%%%%%%%%%%%%%%%%%%%%%%%%%%%%
\begin{figure}[t]
\begin{center}
\includegraphics[height=4.8cm]{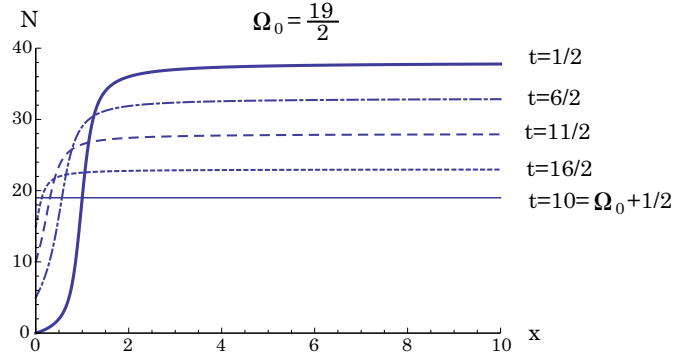}
\caption{
The figure shows $N$ as a function of $x$ with various $t$ for the case $\Omega_0=19/2$. 
The solid, dash-dotted, dashed and dotted curves represent the case $t=1/2$, 3, $11/2$ and 8, respectively. 
The thin line represents the case $t=\Omega_0+1/2\ (=10)$. 
}
\label{fig:1}
\end{center}
\end{figure}
%%%%%%%%%%%%%%%%%%%%%%%%%%%%%%%%%%%%%%%%%%%%%%%%%%%%%%%%%%%%%%%%%%%%%%%%
%

In the $B_{ps}$-form framework, the expectation value ${\cal T}_+=\rbra{\phi}{\wtilde {\cal T}}_+\rket{\phi}$ is 
given in the form 
\beq\label{5-26}
{\cal T}_+&=&\bra{\phi}{\hat {\cal T}}_+\ket{\phi}=\bra{\phi}{\hat T}_+\ket{\phi}\nonumber\\
&=&z^*\cdot\frac{1}{\Gamma}\sum_{n=0}^{2\Omega_0-(2t-1)}nx^{n-1}
\left(
\begin{array}{c}
2t-1+n \\ n
\end{array}
\right) \nonumber\\
&=&z^*\cdot \frac{1}{x}\Lambda_t(x)\ .
\eeq
It is noted that ${\cal T}_+$ is expressed in terms of the product of $z^*$ and 
the function of $x\ (=|z|^2)$, $\Lambda_t(x)/x$. 
In order to get a transparent understanding for ${\cal T}_+$, we introduce a new parameter 
$(y, y^*)$: 
\beq\label{5-27}
y=z\sqrt{\frac{\Lambda}{x}}=\frac{z}{|z|}\sqrt{\Lambda}\ , \qquad {\rm i.e.,}\qquad \Lambda=y^*y\ . 
\eeq
Then, ${\cal T}_+$ is expressed as 
\beq\label{5-28}
{\cal T}_+=y^*\sqrt{\frac{\Lambda}{x}}\ . 
\eeq
After lengthy calculation, we have the relation 
\beq
& &\frac{\Lambda}{x}=2t+y^*y-Y\ , 
\label{5-29}\\
& &Y=\frac{\displaystyle 2tx^{2\Omega_0+1}
\left[\left(\frac{1-x}{x}\right)^{2t}
\left(
\begin{array}{c}
2\Omega_0+1 \\ 2t 
\end{array}\right)
-\sum_{n=1}^{2t-1}
\left(\frac{1-x}{x}\right)^{n}
\left(
\begin{array}{c}
2\Omega_0+n \\ n 
\end{array}\right)\right]}
{\displaystyle 1-x^{2\Omega_0+1}\sum_{n=1}^{2t-1}
\left(\frac{1-x}{x}\right)^{n}
\left(
\begin{array}{c}
2\Omega_0+n \\ n 
\end{array}\right)} \ .\qquad
\label{5-30}
\eeq
Then, ${\cal T}_{\pm}$ can be expressed as 
\beq\label{5-31}
{\cal T}_+=y^*\cdot\sqrt{2t+y^*y-Y}\ , \qquad
{\cal T}_-=\sqrt{2t+y^*y-Y}\cdot y\ .
\eeq
The expectation value $N$ is expressed in the form 
\beq\label{5-32}
N=(2t-1)+2y^*y\ .
\eeq
With the use of the relation (\ref{5-17}), ${\cal T}_0$ is of the form 
\beq\label{5-33}
{\cal T}_0=t+y^*y\ .
\eeq
If $Y$ given in the relation (\ref{5-30}) can be neglected, the set $({\cal T}_{\pm,0})$ reduces 
to the classical counterpart of the set of the $su(1,1)$-generator ${\hat T}_{\pm,0}$, namely, 
it is the classical counterpart of the Holstein-Primakoff representation. 
It should be noted that $(y, y^*)$ is the 
canonical variable in the boson type. 
The above feature of the $su(1,1)$-algebra was discussed by the present authors with 
Kuriyama in detail\cite{5}.

\setcounter{equation}{0}

\section{Approximate expression for the expectation value of the fermion-number operator}

In \S 5, we gave the expectation value of ${\wtilde N}$ for $\rket{\phi}$. 
The result is too complicated to use it for practical purpose. 
Then, we must find the approximate expression which is fit for this purpose. 
As was already mentioned, roughly speaking, in the region where $x$ is sufficiently large, $N$ changes gently, 
but in the region $x\lsim 2$, especially $x\lsim 1$, it changes steeply. 
Therefore, it may be impossible to give an approximate expression of $N$ in terms of a well-behaved simple function of $x$ in the whole 
range $0\leq x <\infty$, but, if the range is limited, it may be possible. 
Judging from the behavior shown in Fig.1, it may be natural to divide the whole range into two: 
(1) $0\leq x \leq \gamma_t$ and (2) $\gamma_t\leq x <\infty$. 
Here, we conjecture that $\gamma_t$ is given in the form 
\beq\label{6-1}
\gamma_t=\frac{2\Omega_0-(2t-1)}{2\Omega_0}=1-\frac{2t-1}{2\Omega_0}\ (\leq 1)\ . 
\eeq
Later, we will give an interpretation of the relation (\ref{6-1}). 
We treat the ranges (1) and (2), separately.

First, we introduce the following function for the approximate expression of $\Gamma_t(x)$, 
which is denoted as $\Gamma_t^a(x)$:
\beq
& &\Gamma_t(x)=\left\{
\begin{array}{ll}
\displaystyle \left(\frac{1}{1-\alpha x}\right)^{\frac{2t}{\alpha}}\ \left(=\Gamma_t^{a_1}(x)\right) & 
{\rm for\ the\ range\ (1)} \\
\displaystyle 
\left(\begin{array}{c}
2\Omega_0 \\ 2t-1 
\end{array}
\right)
\left[x^{2\Omega_0}\left(\frac{1}{1-\frac{\beta}{x}}\right)^{\frac{1}{2\Omega_0\beta}}\right]^{2\Omega_0-(2t-1)}
\!\!\!\!\!\! \left(=\Gamma_t^{a_2}(x)\right) & 
{\rm for\ the\ range\ (2)}\ .  
\end{array}\right.\nonumber\\
& &
\label{6-2}
\eeq
Here, $\alpha$ and $\beta$ are real parameters which will be determined later. 
For the form (\ref{6-2}), we have the following relation: 
\bsub\label{6-3}
\beq
& &\Gamma_t^{a_1}(x)=1+2tx+\cdots \ , \qquad (\alpha x <1)
\label{6-3a}\\
& &\Gamma_t^{a_2}(x)=\left(
\begin{array}{c}
2\Omega_0 \\ 2t-1 
\end{array}\right)
x^{2\Omega_0-(2t-1)}\left(
1+\frac{2\Omega_0-(2t-1)}{2\Omega_0}\frac{1}{x}+\cdots \right)\ . \qquad
\left(\frac{\beta}{x}<1\right)\nonumber\\
& &\label{6-3b}
\eeq
\esub
The forms (\ref{6-2}) are reduced to the forms (\ref{5-6}) and (\ref{5-7}), 
if $x\sim 0$ and $x\rightarrow \infty$, respectively. 
From the above consideration, it may be understandable that $\Gamma_t^a(x)$ is a possible approximation of $\Gamma_t(x)$. 
The functions $(1-\alpha x)$ and $(1-\beta/x)$ should not have the points which make $1-\alpha x=0$ and $1-\beta /x=0$ in the 
ranges $0\leq x \leq \gamma_t$ and $\gamma_t \leq x < \infty$, respectively. 
These situations are realized under the condition
\beq\label{6-4}
\alpha < \frac{1}{\gamma_t}\ , \qquad \beta <\gamma_t\ . 
\eeq

%%%%%%%%%%%%%%%%%%%%%%%%%%%%%%%%%%%%%%%%%%%%%%%%%%%%%%%%%%%%%%%%%%%%%%
\begin{figure}[t]
\begin{center}
\includegraphics[height=5.2cm]{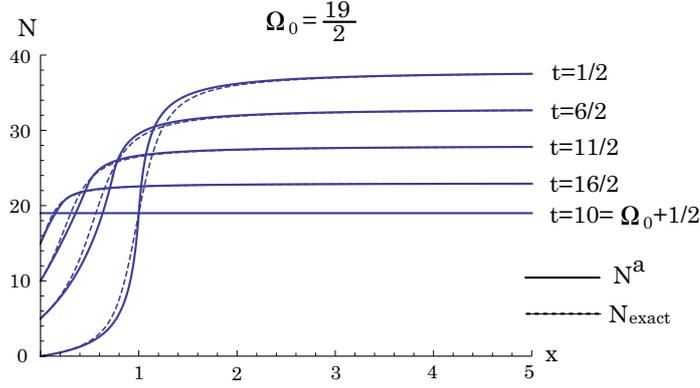}
\caption{
The figure shows $N^a$ as a function of $x$ with various $t$ for the case $\Omega_0=19/2$. 
The solid curves represent $N^a$ and, for comparison, the exact $N$ are depicted. 
It is noted that the horizontal scale is different from that of Fig.1.
}
\label{fig:2}
\end{center}
\end{figure}
%%%%%%%%%%%%%%%%%%%%%%%%%%%%%%%%%%%%%%%%%%%%%%%%%%%%%%%%%%%%%%%%%%%%%%%%

Through the relation (\ref{5-18}), we define the approximate form of $\Lambda_t(x)$ as follows: 
\beq\label{6-5}
& &\Lambda_t^{a}(x)=\frac{x\frac{d\Gamma_t^a(x)}{dx}}{\Gamma_t^a(x)}
=\left\{
\begin{array}{ll}
\displaystyle \frac{2tx}{1-\alpha x} & {\rm for}\quad 0\leq x \leq \gamma_t\ , 
\\
\displaystyle (2\Omega_0-(2t-1))\left(1-\frac{1}{2\Omega_0(x-\beta)}\right) & 
{\rm for}\quad \gamma_t\leq x < \infty \ .
\end{array}\right.\ \ 
%\nonumber\\
%& &
\eeq
We require the condition that the functions (\ref{6-5}) should connect with each other smoothly at $x=\gamma_t$: 
\bsub\label{6-6}
\beq
& &\frac{2t\gamma_t}{1-\alpha \gamma_t}=(2\Omega_0-(2t-1))\left(1-\frac{1}{2\Omega_0(\gamma_t-\beta)}\right)\ , 
\label{6-6a}\\
& &\frac{2t}{(1-\alpha \gamma_t)^2}=\frac{2\Omega_0-(2t-1)}{2\Omega_0(\gamma_t-\beta)^2}\ . 
\label{6-6b}
\eeq
\esub
The condition (\ref{6-6}) determines $\alpha$ and $\beta$ in the form 
\bsub\label{6-7}
\beq
& &\alpha=\frac{1}{\gamma_t}\left(1-\frac{1}{2\Omega_0}\sqrt{\frac{2t}{\gamma_t}}-\frac{2t}{2\Omega_0}\right)\ (=\alpha_t)\ , 
\label{6-7a}\\
& &\beta=\gamma_t\left(1-\frac{1}{2\Omega_0}\sqrt{\frac{2t}{\gamma_t}}-\frac{1}{2\Omega_0 \gamma_t}\right)\ (=\beta_t)\ . 
\label{6-7b}
\eeq
\esub
We can see that $\alpha$ and $\beta$ depend on $t$ and, then, hereafter, we 
express $\alpha$ and $\beta$ such as $\alpha_t$ and $\beta_t$. 
Clearly, they satisfy the condition (\ref{6-4}). 
The approximate expression of $N$, $N^a$, is given as 
\beq\label{6-8}
N^a=2t-1+2\Lambda_t^a(x)\ . 
\eeq
Figure 2 shows several concrete cases, together with $N$ shown in the relation (\ref{5-17}). 
We can see that the agreement is rather good. 
Next, we discuss the typical three cases $x=0$, 1 and $x\rightarrow \infty$. 
The cases $x=0$ and $x\rightarrow \infty$ agree with the exact results shown in the relations (\ref{5-25a}) and (\ref{5-25c}), 
because these two cases are constructed so as to reproduce the exact results. 
The case $x=1$ is expressed in the form 
\beq\label{6-9}
N^a=2t-1+2\left(1-\frac{1}{(2t+1)+\left(\sqrt{2t\left(1-\frac{2t-1}{2\Omega_0}\right)}-1\right)}\right)
(2\Omega_0-(2t-1))\ . \quad
\eeq
The exact result (\ref{5-25b}) can be expressed as 
\beq\label{6-10}
N=2t-1+2\left(1-\frac{1}{2t+1}\right)(2\Omega_0-(2t-1))\ . 
\eeq
The cases $2t=1$ and $2\Omega_0$ agree with the exact results, but in the other cases, disagreement with the exact one is not 
so much as imagined.

Let us discuss the quantity $\gamma_t$ which was introduced in the opening paragraph of this section. 
First, for $t=1/2$, we note the following relation: 
\beq\label{6-11}
\Lambda_{1/2}(x)+\Lambda_{1/2}\left(\frac{1}{x}\right)=2\Omega_0\ . 
\eeq
With the use of the formulae (i)${}'$ and (ii)${}'$, we can prove this relation. 
In (A), also we gave the relation (\ref{6-11}). 
This relation tells us that if $\Lambda_{1/2}$ for $0\leq x \leq 1$ is given, we are able to obtain $\Lambda_{1/2}$ 
for $1\leq x <\infty$ and vice versa. 
From the above argument, the range $0\leq x < \infty$ is divided by $x=1$; 
(1) $0\leq x \leq 1$ and (2) $1\leq x <\infty$. 
In the case $2\Omega_0-(2t-1)=0$, i.e., $t=\Omega_0+1/2$, $\Lambda_{t=\Omega_0+1/2}=0$ and the range 
$0\leq x<\infty$ is formally divided by $x=0$; 
(1) $x=0$ and (2) $0\leq x < \infty$. 
Combining the above two extreme cases with the behavior of $N\ (=2t-1+2\Lambda)$ shown in Fig.1, 
we conjectured that the range $0\leq x <\infty$ is divided by $x=\gamma_t=(2\Omega_0-(2t-1))/(2\Omega_0)$: 
(1) $0\leq x \leq \gamma_t$ and (2) $\gamma_t \leq x<\infty$. 
The parameter $\gamma_t$ is the ratio of the number of the single-particle states in ${\ovl P}$ which 
can contribute to the fermion-pair formation to the total number of the single-particle states in ${\ovl P}$. 
Therefore, if $\gamma_t$ is near to 1, the possibility for the fermion-pair formation is large and vice versa. 
The above is the interpretation of the conjecture for $\gamma_t$.

In the framework of our approximation, we generalized the relation (\ref{6-11}), which can be rewritten as 

\beq\label{6-12}
\Lambda_{1/2}(x)=2\Omega_0-\Lambda_{1/2}\left(\frac{1}{x}\right)\ .
\eeq
If $1\leq x <\infty$, we have $0\leq 1/x \leq 1$, i.e., $x\cdot (1/x)=1$. 
We generalize the relation (\ref{6-12}) to the case of arbitrary value of $t$. 
If $\gamma_t\leq x <\infty$, $\gamma_t^2/x$ obeys the inequality 
$0\leq \gamma_t^2/x \leq \gamma_t$, i.e., 
$x\cdot (\gamma_t^2/x)=\gamma_t^2$. 
Of course, if $t=1/2$, we have $x\cdot(1/x)=1$. 
Then, the relation (\ref{6-5}) for $0\leq x\leq \gamma_t$ gives 
\beq\label{6-13}
\Lambda_t^{a_1}\left(\frac{\gamma_t^2}{x}\right)
=\frac{2t\cdot\frac{\gamma_t^2}{x}}{1-\alpha_t\cdot\frac{\gamma_t^2}{x}}\ , \quad{\rm i.e.,}\quad
\frac{\gamma_t^2}{x}=\frac{\Lambda_t^{a_1}\left(\frac{\gamma_t^2}{x}\right)}{\alpha_t\Lambda_t^{a_1}\left(\frac{\gamma_t^2}{x}\right)+2t}\ . 
\eeq
The relation (\ref{6-13}) leads to 
\beq\label{6-14}
x=\gamma_t^2\left(\alpha_t+\frac{2t}{\Lambda_t^{a_1}\left(\frac{\gamma_t^2}{x}\right)}\right)\ . \qquad
(\gamma_t\leq x<\infty)
\eeq
Therefore,
the relation (\ref{6-5}) for $\gamma_t\leq x<\infty$ can be rewritten as 
\beq\label{6-15}
\Lambda_t^{a_2}(x)&=&
2\Omega_0 \gamma_t\left(1-\frac{1}{2\Omega_0(x-\beta_t)}\right)
\nonumber\\
&=&2\Omega_0 \gamma_t\Biggl(
1-\frac{1}{2\Omega_0\Bigl[\gamma_t^2\Bigl(\alpha_t+\frac{2t}{\Lambda_t^{a_1}\left(\frac{\gamma_t^2}{x}\right)}\Bigl)
-\beta_t\Bigl]}\Biggl)\ . 
\eeq
With the use of the explicit expressions of $\alpha_t$ and $\beta_t$ given in the relation (\ref{6-7}), 
we have the following: 
\beq\label{6-16}
\Lambda_t^{a_2}(x)=2\Omega_0 \gamma_t\left(
1-\frac{\Lambda_t^{a_1}\left(\frac{\gamma_t^2}{x}\right)}{(1-2t\gamma_t)\Lambda_t^{a_1}\left(\frac{\gamma_t^2}{x}\right)+2t\cdot 2\Omega_0}
\right)\ . 
\eeq
If $\Lambda_t^{a_1}$ is given, $\Lambda_t^{a_2}$ is obtained by the relation (\ref{6-16}). 
In the case $t=1/2$, $\Lambda_{1/2}^{a_2}(x)$ is expressed as 
\beq\label{6-17}
\Lambda_{1/2}^{a_2}(x)=2\Omega_0-\Lambda_{1/2}^{a_1}\left(\frac{1}{x}\right)\ . 
\eeq
Also, in the case $t=\Omega_0+1/2$, i.e., $2\Omega_0-(2t-1)=0$, we have 
\beq\label{6-18}
\Lambda_{t=\Omega_0+1/2}^{a_2}(x)=0\ . 
\eeq
In the exact case for $t>1/2$, numerically, the relation corresponding to the relation (\ref{6-16}) may be presented, but, 
in analytical form, it may be impossible.

Finally, we will investigate the parameters $\alpha_t$ and $\beta_t$ given in the 
relations (\ref{6-7a}) and (\ref{6-7b}), respectively. 
Both relations are rewritten as 
\bsub\label{6-19}
\beq
& &\alpha_t=1-\frac{1}{2\Omega_0-(2t-1)}\left(
\sqrt{\frac{2\Omega_0\cdot 2t}{2\Omega_0-(2t-1)}}+1\right)\ , 
\label{6-19a}\\
& &\beta_t=1-\frac{1}{2\Omega_0}\left(
2t+\sqrt{\frac{2t}{2\Omega_0}(2\Omega_0-(2t-1))}\right)\ . 
\label{6-19b}
\eeq
\esub
The above expressions tell us 
\beq\label{6-20}
\alpha_t < 1\ , \qquad \beta_t < 1\ . 
\eeq
In the $su(1,1)$-algebraic model, we have $\alpha_t=1$, which is realized in the 
case with $\Omega_0\rightarrow \infty$ and finite value of $t$. 
However, in our present model, $\Omega_0$ and $t$ are finite and $\alpha_t$ should obey the condition 
(\ref{6-20}). 
We do not know any model related to $\beta_t$ and, then, any comparison is impossible. 
Since $\alpha_t$ is decreasing for $2t$, the maximum value of $\alpha_t$ is given as 
\beq\label{6-21}
\alpha_{1/2}=1-\frac{1}{\Omega_0}\ , \qquad
\gamma_{1/2}=1\ . 
\eeq
At the point $2t=2t^0$, which will be later discussed, $\alpha_t$ vanishes $(\alpha_{t^0}=0$). 
After $\alpha_{t^0}=0$, $\alpha_t$ can change to $-\infty$: 
\beq\label{6-22}
& &\alpha_{\Omega_0}=-2\Omega_0\quad (\gamma_{\Omega_0}=\frac{1}{2\Omega_0})\ , \qquad
\alpha_{\Omega_0+1/2}\rightarrow -\infty\quad (\gamma_{\Omega_0+1/2}=0)\ . 
%\alpha_{\Omega_0+1/2}\rightarrow -\infty\ , \qquad
%x_{\Omega_0+1/2}=0\ .
\eeq
The quantity $\Lambda_t^{a_1}(x)$ in the range $\alpha_{1/2}>\alpha_t >0$ is of the type 
similar to that of the $su(1,1)$-algebraic model: 
$\Lambda_t^{a_1}(x)=2tx/(1-|\alpha_t| x)$. 
At $\alpha_{t^0}=0$, $\Lambda_t^{a_1}(x)=2t^0x$ and in the range $0>\alpha_t >-\infty$, 
$\Lambda_t^{a_1}(x)=2tx/(1+|\alpha_t|x)$. 
If $2\Omega_0$ and $2t$ can change continuously, $\alpha_t=0$ itself has own meaning. 
But, they are integers and we treat $\alpha_t=0$ as an auxiliary condition. 
It leads us to a certain cubic equation for $2t$ with one real solution and it is given as 
\bsub\label{6-23}
\beq\label{6-23a}
& &2t^0=2\Omega_0+\frac{1}{3}-(2\Omega_0)^{\frac{2}{3}}\left[
\sqrt[3]{\frac{1}{2}(A+B)}-\sqrt[3]{\frac{1}{2}(A-B)}\right]\ , 
\nonumber\\
& &A=\sqrt{\left(1-\frac{5}{54\Omega_0}\right)\left(1+\frac{1}{2\Omega_0}\right)}\ , \qquad
B=1+\frac{1}{6\Omega_0}-\frac{1}{54\Omega_0^2}\ .
\eeq
The expression (\ref{6-23a}) is approximated in the form 
\beq\label{6-23b}
2t^0=2\Omega_0-(2\Omega_0)^{\frac{2}{3}}+\frac{1}{3}(2\Omega_0)^{\frac{1}{3}}+\frac{1}{3}
-\frac{10}{81}(2\Omega_0)^{-\frac{1}{3}}\ . 
\eeq
\esub
As is conjectured in the relation (\ref{6-23}), $2t^0$ cannot be expected to be integer. 
Therefore, two integers $2t^+$ and $2t^-$ ($t^+<t^-$) which are the nearest to $2t^0$ 
must be searched: 
$\alpha_t>0$ for $1\leq 2t\leq 2t^+$ and $\alpha_t<0$ for $2t^-\leq 2t\leq 2\Omega_0+1$. 
For this searching, the relation (\ref{6-23}) is useful. 
For example, in the case $2\Omega_0=19$, the relations (\ref{6-23a}) and (\ref{6-23b}) give us 
$2t^0 \sim 13.0235$ and 13.0562, respectively and, therefore, 
$2t^+=13$ and $2t^-=14$. 
For them, we have $\alpha_{t^+}=8.5460\times 10^{-3}\ (>0)$ and 
$\alpha_{t^-}=-0.2764\ (<0)$. 
The treatment of $\beta_t$ is rather simple. 
As is clear from the relation (\ref{6-19b}), the maximum value of $\beta_t$ is also given in the case $t=1/2$: 
\beq\label{6-24}
\beta_{1/2}=1-\frac{1}{\Omega_0}\ . 
\eeq
Then, gradually decreases, at the point $2t=2\Omega_0-2$, $\beta_{\Omega_0-1}$ is 
given as 
\beq\label{6-25}
\beta_{\Omega_0-1}=\frac{1}{\Omega_0}\left(1-\frac{1}{2}\sqrt{3\left(1-\frac{1}{2\Omega_0}\right)}\right)
\ (>0)\ . 
\eeq
At the point $2t=2\Omega_0-1$, $\beta_{\Omega_0-1/2}$ is given as 
\beq\label{6-26}
\beta_{\Omega_0-1/2}=-\frac{1}{2\Omega_0}\left(\sqrt{2-\frac{1}{\Omega_0}}-1\right)\ (<0)\ . 
\eeq
At the terminal points $2t=2\Omega_0$ and $2\Omega_0+1$, we have 
\beq\label{6-27}
\beta_{\Omega_0}=\beta_{\Omega_0+1/2}=-\frac{1}{2\Omega_0}\ . 
\eeq
We can see that the sign of $\beta_t$ changes between $2t=2\Omega_0-2$ and $2\Omega_0-1$. 
The point which satisfies $\beta_t=0$ is given at $2t=2t^0{}'$ shown as 
\beq\label{6-28}
2t^0{}'=2\Omega_0-
\frac{2\Omega_0-1+2\Omega_0\sqrt{5+\frac{1}{\Omega_0}+\frac{1}{4\Omega_0^2}}}{2(2\Omega_0+1)}\ .
\eeq

\setcounter{equation}{0}

\section{A simple example of many-fermion model obeying the pseudo $su(1,1)$-algebra}

%In this section, we will apply the idea developed until the present to a certain simple many-fermion system. 

In next three sections, we intend to discuss an example of the application of the 
idea developed until the present. 
This section will be devoted to presenting a simple many-fermion model aimed at the application. 
As an illustrative example of our idea, first, we give a short summary of the ``damped and amplified oscillator." 
The starting Hamiltonian is the simplest, i.e., the harmonic oscillator: 
\beq\label{7-1}
{\hat H}_b=\omega{\hat b}^*{\hat b}\ . \qquad (\omega\ ;\ {\rm frequency})
\eeq
Here, $({\hat b}, {\hat b}^*)$ denotes boson operator. 
As an auxiliary degree of freedom for the ``damping and amplifying," new boson $({\hat a}, {\hat a}^*)$ 
is introduced. 
The Hamiltonian for $({\hat a}, {\hat a}^*)$ is also the harmonic oscillator type: 
\beq\label{7-2}
{\hat H}_a=\omega{\hat a}^*{\hat a}\ . 
\eeq
Further, as for the interaction between both degrees of freedom, the following form is adopted:
\beq\label{7-3}
{\hat V}_{ab}=-i\gamma({\hat a}^*{\hat b}^*-{\hat b}{\hat a})\ . \qquad (\gamma \ ; \ {\rm constant})
\eeq
The idea presented in Ref.\citen{6} is to adopt the Hamiltonian 
\beq\label{7-4}
{\hat H}={\hat H}_b-{\hat H}_a+{\hat V}_{ba}\ . 
\eeq
By treating ${\hat H}$ appropriately, we can describe the ``damped and amplified oscillation" 
in the conservative form. 
It should be noted that, for the Hamiltonian (\ref{7-4}), the form 
$({\hat H}_b+{\hat H}_a+{\hat V}_{ba})$ is not adopted. 
It shows that the Hamiltonian (\ref{7-4}) is not the energy of the entire system, but the 
generator for time-evolution. 
This is a significant feature of this approach. 
With the use of ${\hat T}_{\pm}$ defined in the relation (\ref{2-5}), ${\hat H}$ can be 
expressed as 
\beq\label{7-5}
{\hat H}=2\omega\left({\hat T}-\frac{1}{2}\right)-i\gamma\left({\hat T}_+-{\hat T}_-\right)\ .
\eeq
Here, ${\hat T}$ is defined as 
\bsub\label{7-6}
\beq
& &{\hat T}=-\frac{1}{2}({\hat a}^*{\hat a}-{\hat b}^*{\hat b})+\frac{1}{2}\ , 
\label{7-6a}\\
& &{\hat {\mib T}}^2={\hat T}\left({\hat T}-1\right)\ , \qquad
[\ {\hat T}_{\pm,0}\ , \ {\hat T}\ ]=0\ . 
\label{7-6b}
\eeq
\esub
By using the mixed-mode coherent states for the $su(1,1)$-algebra, the present authors, 
with Kuriyama, have investigated extensively the Hamiltonian (\ref{7-5}) and its variations \cite{5}.

The above illustrative example teaches us the following: 
In order to treat the system such as the ``damped and amplified oscillator" in an isolated system, 
so-called phase space doubling is required. 
The idea of the phase space doubling occupies main part of the thermo-field dynamics formalism \cite{8}. 
Then, the original intrinsic oscillator expressed in terms of the boson $({\hat b}, {\hat b}^*)$ and 
the ``external environment" expressed in terms of the boson $({\hat a}, {\hat a}^*)$ appear. 
The interaction between both systems is introduced. 
We will apply the above consideration to a simple many-fermion system.

We make the following translation into the fermion system: 
\beq
& &({\hat b}, {\hat b}^*)\rightarrow ({\tilde c}_{\bar \alpha}, {\tilde c}_{\bar \alpha}^*)\ , \qquad
({\hat a}, {\hat a}^*)\rightarrow ({\tilde c}_{\alpha}, {\tilde c}_{\alpha}^*)\ , 
\label{7-7}\\
& &{\hat T}-\frac{1}{2}\rightarrow {\wtilde {\cal T}}=-\frac{1}{2}\sum_{\alpha}
({\tilde c}_{\alpha}^*{\tilde c}_{\alpha}-{\tilde c}_{\bar \alpha}^*{\tilde c}_{\bar \alpha})\ (=-{\wtilde R}_0)\ , \qquad
{\hat T}_{\pm}\rightarrow {\wtilde {\cal T}}_{\pm}\ , 
\label{7-8}\\
& &\omega\ ({\rm frequency}) \rightarrow \varepsilon\ (\textrm{single-particle energy})\ . 
\label{7-9}
\eeq
Here, ${\wtilde {\cal T}}$ is introduced in the relation (\ref{4-10}) and the relation (\ref{4-11}) 
suggests us the relation $[\ {\wtilde {\cal T}}\ , \ {\wtilde {\cal T}}_{\pm}\ ]=0$. 
Under the above translation, our Hamiltonian is expressed in the form 
\beq\label{7-10}
{\wtilde H}=2\varepsilon {\wtilde {\cal T}}-i\gamma\left({\wtilde {\cal T}}_+-{\wtilde {\cal T}}_-\right)\ . 
\eeq
It may be clear that we have the translation 
\beq\label{7-11}
{\hat H}_b\rightarrow {\wtilde H}_{\ovl P}=\varepsilon{\wtilde N}_{\ovl P}\ , \quad
{\wtilde N}_{\ovl P}=\sum_{\alpha}{\tilde c}_{\bar \alpha}^*{\tilde c}_{\bar \alpha}\ , \qquad
{\hat H}_a\rightarrow {\wtilde H}_P=\varepsilon{\wtilde N}_P\ , \quad 
{\wtilde N}_P=\sum_{\alpha}{\tilde c}_{\alpha}^*{\tilde c}_{\alpha}\ . 
\eeq
The original intrinsic Hamiltonian ${\wtilde H}_{\ovl P}$ may be the simplest in many-fermion systems and our aim is to 
describe this system in the ``external environment."
The Hamiltonian (\ref{7-10}) was set up under an idea analogous to that in the case (\ref{7-5}). 
However, it may be permitted to regard the Hamiltonian (\ref{7-10}) as the energy of the 
entire system. 
Concerning this point, we will discuss the possibility in \S 11. 
It may be important to see that the conventional pairing Hamiltonian and the present one are expressed in terms of the $su(2)$-generators, 
${\wtilde S}_{\pm,0}$, but differently from the former, the latter does not commute with 
the total fermion-number operator. 
In this sense, the use of the state (\ref{1-1}) for the variational treatment in the pairing 
Hamiltonian is justified by the symmetry breaking. 
On the other hand, the use of the state (\ref{5-1}) (or (\ref{5-2})) may be natural as a possible trial state for 
the variation without any comment such as the symmetry breaking.

Our basic idea is to describe the Hamiltonian (\ref{7-10}) in the framework of the 
time-dependent variational method: 
\beq\label{7-12}
\delta\int \rbra{\phi}i\partial_{\tau}-{\wtilde H}\rket{\phi}d\tau=0\ . 
\eeq
Here, the state $\rket{\phi}$ is used for the trial state of the variation. 
In order to avoid the confusion between the time variable and the quantum number $t$, we will 
use $\tau$ for the time variable. 
For the relation (\ref{7-12}), the following are useful: 
\beq
& &\delta \int \rbra{\phi}i\partial_{\tau}\rket{\phi}d\tau=
\frac{i}{2}\int\left(\delta z^*\cdot {\dot z}-\delta z\cdot {\dot z}^*\right)
\left(\frac{\partial {\cal T}_+}{\partial z^*}+\frac{\partial {\cal T}_-}{\partial z}\right)d\tau\ , 
\label{7-13}\\
& &\delta \int \rbra{\phi}{\wtilde H}\rket{\phi}d\tau=
\int\left(\delta z^*\frac{\partial {\cal H}}{\partial z^*}
+\delta z\frac{\partial {\cal H}}{\partial z}\right)d\tau\ . 
\label{7-14}
\eeq
Here, ${\cal T}_{\pm}$ is given in the relation (\ref{5-26}) and ${\cal H}$ is defined as 
\beq\label{7-15}
{\cal H}={\cal H}(z^*,z)=\rbra{\phi}{\wtilde H}{\rket{\phi}}\ . 
\eeq
Then, the relations (\ref{7-12})-(\ref{7-14}) give us 
\beq\label{7-16}
i{\dot z}\frac{1}{2}
\left(\frac{\partial {\cal T}_+}{\partial z^*}+\frac{\partial {\cal T}_-}{\partial z}\right)
=\frac{\partial {\cal H}}{\partial z^*}\ , \qquad
-i{\dot z}^*\frac{1}{2}
\left(\frac{\partial {\cal T}_+}{\partial z^*}+\frac{\partial {\cal T}_-}{\partial z}\right)
=\frac{\partial {\cal H}}{\partial z}\ . 
\eeq
For the relation (\ref{7-16}), ${\cal H}$ is adopted in the following form: 
\beq\label{7-17}
{\cal H}=\varepsilon(2t-1)-i\gamma\left({\cal T}_+-{\cal T}_-\right)
=\varepsilon(2t-1)-\gamma\cdot i(z^*-z)\frac{\Lambda_t(x)}{x}\ .
\eeq
Under the Hamiltonian (\ref{7-17}), the relation (\ref{7-16}) is reduced to the differential equation 
\beq\label{7-18}
{\dot z}=-\gamma\left[
1-\frac{z^2}{x}\left(1-\frac{\Lambda_t(x)}{x\Lambda'_t(x)}\right)
\right] \ , \qquad
{\dot z}^*=-\gamma\left[
1-\frac{z^{*2}}{x}\left(1-\frac{\Lambda_t(x)}{x\Lambda'_t(x)}\right)\right] \ .\qquad
\eeq
Here, $\Lambda_t'(x)$ denotes the derivative of $\Lambda_t(x)$ for $x$.

The relation (\ref{7-18}) forms our basic framework for describing the time-evolution. 
In order to give the physical interpretation of the relation (\ref{7-18}), we examine the case $\Lambda_t^{a_1}(x)$. 
In this case, the relation (\ref{7-18}) becomes 
\beq\label{7-19}
{\dot z}=-\gamma(1-\alpha_t z^2)\ , \qquad
{\dot z}^*=-\gamma(1-\alpha_t z^*{}^2)\ . 
\eeq
If $z$ is expressed as $z=u+iv$ ($u$, $v$: real), we have 
\beq\label{7-20}
{\dot u}=-\gamma(1-\alpha_t u^2+\alpha_t v^2)\ , \qquad
{\dot v}=2\gamma\alpha_t uv\ .
\eeq
If we eliminate $v$ from the relation (\ref{7-20}), the following equation is 
derived: 
\beq\label{7-21}
{\ddot u}=4\gamma^2\alpha_t^2 u\left(\frac{1}{\alpha_t}-u^2\right)+6\alpha_t u{\dot u}\ . 
\eeq
If the relation (\ref{7-21}) is interpreted in Newton mechanics, a mass point with 
mass$=1$ moves in the one-dimensional space under the external force $4\gamma^2\alpha_t^2 u(1/\alpha_t -u^2)$ 
and the velocity-dependent force $6\alpha_t u{\dot u}$. 
The force $4\gamma^2\alpha_t^2 u(1/\alpha_t -u^2)$ is expressed in terms of the potential energy $V(u)$: 
\beq\label{7-22}
4\gamma^2\alpha_t^2 u\left(\frac{1}{\alpha_t}-u^2\right)
=-\frac{dV(u)}{du} \ , \qquad
V(u)=-4\gamma^2\left(\frac{1}{2}\alpha_t u^2
-\frac{1}{4}\alpha_t^2 u^4\right)\ . 
\eeq
The cases $\alpha_t>0$ and $\alpha_t<0$ correspond to double well-like and single well-like potentials, respectively 
and the case $\alpha_t=0$ to no external force. 
Existence of the velocity-dependent force suggests that our model enable us to describe the 
dissipation phenomena in many-fermion system. 
If we can solve the equation of motion (\ref{7-21}), $u$ can be determined as a function of $\tau$ and 
the second of the relation (\ref{7-20}) gives us the following form:
\beq\label{7-23}
v(\tau)=e^{2\gamma\alpha_t\int^{\tau}u(\tau')d\tau'}\ . 
\eeq
Here, we omitted the initial condition for $u$ and $v$. 
Thus, we are able to obtain $u(\tau)$ and $v(\tau)$ and, then, $x$ is determined as a function of $\tau$: 
\beq\label{7-24}
x(\tau)=u(\tau)^2+v(\tau)^2\ . 
\eeq
The case $\Lambda_t^{a_2}(x)$ is not so simple as the case 
$\Lambda_t^{a_1}(x)$, because, in classical mechanics, we cannot find any simple example 
analogous to this case. 
The above is an outline of our model discussed in next sections.

Finally, we give the expectation values of ${\wtilde N}_{\ovl P}$ and ${\wtilde N}_{P}$ for $\rket{\phi}$, 
$N_{\ovl P}$ and $N_P$: 
\bsub\label{7-25}
\beq
& &N_{\ovl P}=2t-1+\Lambda_t(x)\ , 
\label{7-25a}\\
& &N_P=\Lambda_t(x)\ . 
\label{7-25b}
\eeq
\esub
The expectation value of ${\wtilde N}(={\wtilde N}_{\ovl P}+{\wtilde N}_P)$ is given as 
$N=(2t-1+\Lambda_t(x))+\Lambda_t(x)$ and it is nothing but the result (\ref{5-17}).

\setcounter{equation}{0}

\section{Various properties of $\Lambda_t^{a_1}(x)$ for describing its 
time-dependence}

Let us investigate various properties of $\Lambda_t^{a_1}(x)$. 
First, we notice that the present system is of two-dimension and, then, there exist two constants 
of motion. 
One is the quantum number $t$ and the second, which will be denoted as $\kappa$, is 
given through the relation 
\beq\label{8-1}
i(z^*-z)\frac{\Lambda_t(x)}{x}=2\kappa\ .
\eeq
It may be self-evident, because ${\cal H}$ itself shown in the relation (\ref{7-17}) is a 
constant of motion. 
If $z$ is expressed in the form $z=u+iv$, we have 
\beq\label{8-2}
i(z^*-z)=2v\ .
\eeq
The relation (\ref{8-1}) leads to 
\beq\label{8-3}
v=\frac{\kappa}{y}\ , \qquad y=\frac{\Lambda_t(x)}{x}\ . \qquad
(x=|z|^2=u^2+v^2\ , \quad y\geq 0)
\eeq
Inversely, $x$ can be expressed as a function of $y$: 
\beq\label{8-4}
x=f_t(y)\ .
\eeq
Then, $u$ can be given in the form 
\beq\label{8-5}
u=\pm\sqrt{x-v^2}=\pm \frac{1}{y}\sqrt{y^2 f_t(y)-\kappa^2}\ . 
\eeq
The sign $+$ or $-$ may be appropriately chosen. 
Later, we will discuss this problem. 
As is clear in the above argument, $(u,v)$ and also $x$ are functions of $y$. 
Since $u^2 \geq 0$, the relation (\ref{8-5}) gives us the inequality 
\beq\label{8-6}
y^2 f_t(y) \geq \kappa^2 \ .
\eeq
The inequality (\ref{8-6}) suggests us that the value of $y$ cannot vary freely. 
We will apply the above scheme to the cases $\Lambda_t^{a_1}(x)$.

For the case $\Lambda_t(x)/x=\Lambda_t^{a_1}(x)/x=2t/(1-\alpha_t x)$, we have 
\beq\label{8-7}
x=f_t(y)=\frac{1}{\alpha_t}\left(1-\frac{2t}{y}\right)\ ,\quad {\rm i.e.,}\quad
\Lambda_t^{a_1}(x)=\frac{1}{\alpha_t}(y-2t)\ . 
\eeq
The relation (\ref{8-7}) is applicable in the range $0\leq x \leq \gamma_t$. 
In \S 10, this point will be discussed. 
Then, the inequality (\ref{8-6}) is reduced to 
\beq\label{8-8}
\frac{1}{\alpha_t}\cdot y(y-2t)\geq \kappa^2\ ,
\qquad\qquad\qquad\qquad\qquad\qquad
\eeq
i.e., 
\beq
& &
w=\left\{
\begin{array}{ll}
+y(y-2t)\geq \rho^2 \ , &\qquad (\alpha_t >0) \\
-y(y-2t)\geq \rho^2\ , & \qquad (\alpha_t <0) 
\end{array}
\right.
\label{8-9}\\
& &
\rho=\sqrt{|\alpha_t|}\kappa\ . 
\label{8-10}
\eeq
The behavior of the relation (\ref{8-9}) is depicted in Fig. 3. 
%
%
%%%%%%%%%%%%%%%%%%%%%%%%%%%%%%%%%%%%%%%%%%%%%%%%%%%%%%%%%%%%%%%%%%%%%%
\begin{figure}[t]
\begin{center}
\includegraphics[height=4.8cm]{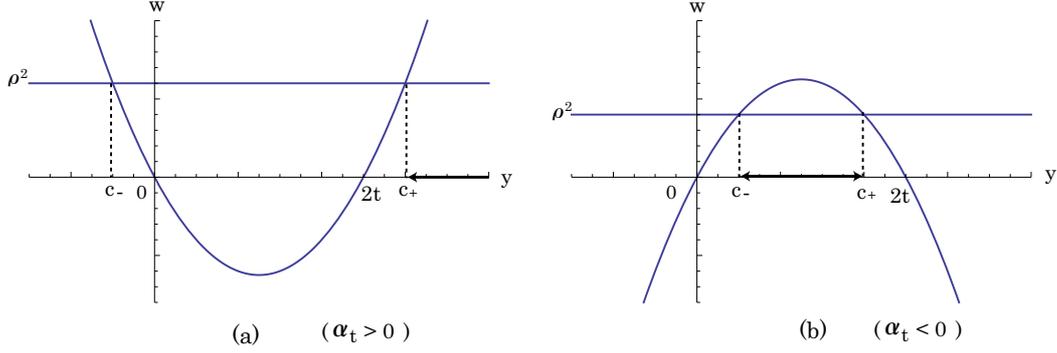}
\caption{
The figures shows the inequality in (\ref{8-9}) schematically with $\Omega_0=19/2$: (a) $t=5/2$, then 
$\alpha_t\approx 0.766\ (>0)$ (b) $t=15/2$, then $\alpha_t \approx -0.710 \ (<0)$.  
}
\label{fig:3}
\end{center}
\end{figure}
%%%%%%%%%%%%%%%%%%%%%%%%%%%%%%%%%%%%%%%%%%%%%%%%%%%%%%%%%%%%%%%%%%%%%%%%
%
%
In Fig.3, we can find out the following restriction to $y$:  
\bsub\label{8-11}
\beq
& &y\geq c_+\ , \quad {\rm i.e.}\quad y\geq \frac{1}{2}(c_+ +c_-)+\frac{1}{2}(c_+-c_-)\ , \quad (\alpha_t>0) 
\label{8-11a}\\
& &c_- \leq y \leq c_+\ , \quad {\rm i.e.,}\quad 
\frac{1}{2}(c_+ +c_-)-\frac{1}{2}(c_+-c_-)\leq y \leq
\frac{1}{2}(c_+ +c_-)+\frac{1}{2}(c_+-c_-) \ . \nonumber\\
& &\qquad\qquad\qquad\qquad\qquad\qquad\qquad\qquad\qquad\qquad\qquad\qquad\qquad\qquad
(\alpha_t<0)
\label{8-11b}
\eeq
\esub
Here, $c_{\pm}$ denote solutions of quadratic equation 
\bsub\label{8-12}
\beq
& &y^2-2ty-\rho^2=0 \ , \qquad \ (\alpha_t>0)
\label{8-12a}\\
& &-y^2+2ty-\rho^2=0 \ . \qquad (\alpha_t<0)
\label{8-12b}
\eeq
\esub
The above equation is obtained by equating both sides of the inequality (\ref{8-9}). 
Therefore, with the use of new variable $\chi$, $y$ can be parametrized in the form  
\bsub\label{8-13}
\beq
& &y=\frac{1}{2}(c_+ +c_-)+\frac{1}{2}(c_+-c_-)\cosh \chi \ ,  \qquad (\alpha_t>0) 
\label{8-13a}\\
& &y= 
\frac{1}{2}(c_+ +c_-)+\frac{1}{2}(c_+-c_-)\cos \chi \ .  \qquad\ \  (\alpha_t<0) 
\label{8-13b}
\eeq
\esub
In \S 10, we will discriminate between the former (\ref{8-13a}) 
and the later (\ref{8-13b}) in terms of the 
notations $y_+$ and $y_-$. 
With the use of Eq.(\ref{8-13}), $u$ and $v$ can be expressed as follows: 
\beq
& &
u=\left\{
\begin{array}{l}
\pm \frac{1}{\sqrt{|\alpha_t|}}\cdot\frac{1}{y}\sqrt{(y-c_+)(y-c_-)}
=\pm\frac{1}{\sqrt{|\alpha_t|}}\cdot\frac{1}{y}\cdot\frac{1}{2}(c_+-c_-)|\sinh \chi|\ ,\  
(\alpha_t>0) \\
\pm \frac{1}{\sqrt{|\alpha_t|}}\cdot\frac{1}{y}\sqrt{(c_+ -y)(y-c_-)}
=\pm\frac{1}{\sqrt{|\alpha_t|}}\cdot\frac{1}{y}\cdot\frac{1}{2}(c_+-c_-)|\sin \chi|\ , \ \  
(\alpha_t < 0) 
\end{array}
\right.\qquad
\label{8-14}\\
& &v=\frac{1}{\sqrt{|\alpha_t|}}\cdot \frac{\rho}{y}\ . \qquad (\alpha_t \neq 0)
\label{8-15}
\eeq
By substituting Eq.(\ref{8-13}) into the relation (\ref{8-7}), $x$ can be expressed in terms of 
$\cosh \chi$ and $\cos \chi$. 
Then, we can express $\Lambda_t^{a_1}(x)=2tx/(1-\alpha_t x)$ as a function of $\cosh \chi$ and $\cos \chi$.

Since Eq.(\ref{8-12}) gives us the solutions $c_{\pm}=t\pm\sqrt{t^2+\rho^2}$ for $\alpha_t>0$ and 
$c_{\pm}=t\pm\sqrt{t^2-\rho^2}$ for $\alpha_t <0$, the relation (\ref{8-13}) can be expressed as 
\beq\label{8-16}
y=\left\{
\begin{array}{ll}
t+\sqrt{t^2+\rho^2}\cosh \chi\ , &\quad (\alpha_t>0) \\
t+\sqrt{t^2-\rho^2}\cos \chi \ , & \quad (\alpha_t <0)
\end{array}
\right.
\eeq
Then, $u$ and $v$ are obtained in the form 
\beq\label{8-17}
& &u=\left\{\!\!
\begin{array}{l}
\displaystyle 
\pm\frac{1}{\sqrt{|\alpha_t|}}\cdot\frac{\sqrt{t^2+\rho^2}|\sinh \chi|}{t+\sqrt{t^2+\rho^2}\cosh \chi} \\
\displaystyle 
\pm\frac{1}{\sqrt{|\alpha_t|}}\cdot\frac{\sqrt{t^2-\rho^2}|\sin \chi|}{t+\sqrt{t^2-\rho^2}\cos \chi} 
\end{array}
\right.\!\! , \ \ 
v=\left\{\!\!
\begin{array}{ll}
\displaystyle 
\frac{1}{\sqrt{|\alpha_t|}}\cdot\frac{\rho}{t+\sqrt{t^2+\rho^2}\cosh \chi}\ , & (\alpha_t>0) \\
\displaystyle 
\frac{1}{\sqrt{|\alpha_t|}}\cdot\frac{\rho}{t+\sqrt{t^2-\rho^2}\cos \chi}\ . & (\alpha_t<0) 
\end{array}
\right.
\nonumber\\
& &
\eeq
We can express $z$ as the function of $\rho$ and $\chi$. 
The quantity $x$ is obtained in the form 
\bsub\label{8-18}
\beq
& &x= 
\frac{1}{\alpha_t}\cdot\frac{\sqrt{t^2+\rho^2}\cosh \chi-t}{\sqrt{t^2+\rho^2}\cosh \chi+t} \ ,  \qquad (\alpha_t>0)
\label{8-18a}\\
& &x= 
\frac{1}{\alpha_t}\cdot\frac{\sqrt{t^2-\rho^2}\cos \chi-t}{\sqrt{t^2-\rho^2}\cos \chi+t} \ .  \ \ \qquad (\alpha_t<0) 
\label{8-18b}
\eeq
\esub
Thus, we have the following form for $\Lambda_t^{a_1}(x)$: 
\bsub\label{8-19}
\beq
& &\Lambda_t^{a_1}(x)=
\frac{1}{\alpha_t}(\sqrt{t^2+\rho^2}\cosh \chi -t)\ , \qquad (\alpha_t>0) 
\label{8-19a}\\
& &\Lambda_t^{a_1}(x)=
\frac{1}{\alpha_t}(\sqrt{t^2-\rho^2}\cos \chi -t)\ , \ \qquad (\alpha_t<0)
\label{8-19b}
\eeq
\esub
It may be necessary for determining the time-dependence of $\Lambda_t^{a_1}(x)$ to investigate the 
behavior of $\chi$ for the time.

The starting variables for describing the present model are $u$ and $v$. 
As is shown in the relations (\ref{8-14}) and (\ref{8-15}), the new variables are $\rho$ and $\chi$. 
Depending on $\alpha_t>0$ and $\alpha_t<0$, the connection to $(u, v)$ is 
different from each other. 
However, $\rho$ is a constant of motion and ${\cal H}$ can be expressed as 
\beq\label{8-20}
{\cal H}=\varepsilon (2t-1)-\frac{2\gamma}{\sqrt{|\alpha_t|}}\cdot\rho\ , \qquad {\dot \rho}=0\ . 
\eeq
Therefore, the time-dependence of $(u, v)$ is given through $y$ which is a function of $\chi$. 
First, we notice the relation 
\beq\label{8-21}
{\dot x}={\dot z}^*z+z^*{\dot z}=-2\gamma(1-\alpha_t x)\cdot u\ . 
\eeq
Here, we used relation (\ref{7-19}). 
With the use of the relation (\ref{8-21}), we have ${\dot y}$ in the following form:
\beq\label{8-22}
{\dot y}=\frac{d}{dx}\left(\frac{\Lambda_t^{a_1}(x)}{x}\right)\cdot{\dot x}=-2\gamma\alpha_t y\cdot u\ , 
\eeq
i.e.,
\beq\label{8-23}
{\dot y}=
\left\{
\begin{array}{ll}
\mp 2\gamma\sqrt{|\alpha_t|}\cdot\frac{1}{2}(c_+-c_-)|\sinh \chi| \ , & \quad (\alpha_t>0) \\
\mp 2\gamma\sqrt{|\alpha_t|}\cdot\frac{1}{2}(c_+-c_-)|\sin \chi| \ . & \quad (\alpha_t<0)
\end{array}
\right.
\eeq
On the other hand, the relation (\ref{8-13}) gives us ${\dot y}$ in the form 
\beq\label{8-24}
{\dot y}=
\left\{
\begin{array}{ll}
+\frac{1}{2}(c_+-c_-)\sinh \chi\cdot {\dot \chi} \ , & \quad (\alpha_t>0) \\
-\frac{1}{2}(c_+-c_-)\sin \chi\cdot{\dot \chi} \ . & \quad (\alpha_t<0)
\end{array}
\right.
\eeq
Combining the relations (\ref{8-23}) and (\ref{8-24}), ${\dot \chi}$ is obtained: 
\bsub\label{8-25}
\beq
& &
{\dot \chi}=\mp 2\gamma\sqrt{|\alpha_t|}\cdot\frac{|\sinh \chi|}{\sinh \chi}
=\left\{
\begin{array}{ll}
\mp 2\gamma\sqrt{|\alpha_t|} & \quad {\rm for}\ \ \chi>0\ , \\
\pm 2\gamma\sqrt{|\alpha_t|} & \quad {\rm for}\ \ \chi<0\ ,
\end{array}
\right.
\quad 
(\alpha_t>0)
\label{8-25a}\\
& &
{\dot \chi}=\pm 2\gamma\sqrt{|\alpha_t|}\cdot\frac{|\sin \chi|}{\sin \chi}
=\left\{
\begin{array}{ll}
\pm 2\gamma\sqrt{|\alpha_t|} & \quad {\rm for}\ \ \chi>0\ , \\
\mp 2\gamma\sqrt{|\alpha_t|} & \quad {\rm for}\ \ \chi<0\ ,
\end{array}
\right.
\quad 
(\alpha_t<0)
\label{8-25b}
\eeq
\esub
Our final aim is to present the time-dependence of $\Lambda_t^{a_1}(x)$, which is also a 
function of $y$. 
The quantity $y$ contains $\cosh \chi$ or $\cos \chi$. 
As can be seen in the relation (\ref{8-25}), $\chi$ is given in the following two cases: 
\bsub\label{8-26}
\beq
& &{\rm (i)}\quad \chi=\chi_+=+2\gamma\sqrt{|\alpha_t|}\cdot \tau +\chi_+^0\ , 
\label{8-26a}\\
& &{\rm (ii)}\quad \chi=\chi_-=-2\gamma\sqrt{|\alpha_t|}\cdot \tau -\chi_-^0\ . 
\label{8-26b}
\eeq
\esub
Here, $\pm\chi_{\pm}^0$ denote the initial values of $\chi_{\pm}$ ($\tau=0$). 
Then, we have 
\beq\label{8-27}
\cosh \chi_+=\cosh (2\gamma\sqrt{|\alpha_t|}\cdot\tau+\chi_+^0)\ , \qquad
\cosh \chi_-=\cosh (2\gamma\sqrt{|\alpha_t|}\cdot\tau+\chi_-^0)\ . 
\eeq
If $\chi_+^0=\chi_-^0$, the case (ii) is nothing but the case (i). 
The case $\cos \chi$ is also in the same situation as the above. 
The above argument suggests us that it may be enough to adopt the case (i): 
\beq\label{8-28}
{\dot \chi}=+2\gamma\sqrt{|\alpha_t|}\ , \quad{\rm i.e.,}\quad
\chi=2\gamma\sqrt{|\alpha_t|}\cdot \tau+\chi^0\ . \quad
(\chi^0 \ : \ {\rm constant})
\eeq
The above argument gives us the time-dependence of $\Lambda_t^{a_1}(x)$. 
Further, this procedure suggests the following form for $u$: 
\bsub\label{8-29}
\beq
& &
u=\frac{1}{\sqrt{|\alpha_t|}}\cdot\frac{\sqrt{t^2+\rho^2}\sinh \chi}{t+\sqrt{t^2+\rho^2}\cosh \chi}\ , \qquad
(\alpha_t>0)\ , 
\label{8-29a}\\
& &
u=\frac{1}{\sqrt{|\alpha_t|}}\cdot\frac{\sqrt{t^2-\rho^2}\sin \chi}{t+\sqrt{t^2-\rho^2}\cos \chi}\ . \qquad\ \ 
(\alpha_t<0)
\label{8-29b}
\eeq
\esub

\setcounter{equation}{0}

\section{Various properties of $\Lambda_t^{a_2}(x)$ for describing its 
time-dependence --- general arguments ---}

The aim of this section is to formulate the case $\Lambda_t^{a_2}(x)$. 
In order to make the discussion in parallel to the case $\Lambda_t^{a_1}(x)$, it may be inconvenient for 
formulating the case $\Lambda_t^{a_2}(x)$ to use the variables $z$, $z^*$ and $x$ used 
in the case $\Lambda_t^{a_1}(x)$. 
These three variables are denoted by $z'$, $z'{}^*$ and $x'$,
respectively:  
\beq\label{9-1}
z\rightarrow z'\ , \qquad 
z^*\rightarrow z'{}^*\ , \qquad x\rightarrow x'\ , \qquad 
{\rm i.e.,}\qquad x'=|z'|^2\ . 
\eeq
In the new notations for the variables, the relations (\ref{7-17}) and (\ref{7-18}) are 
expressed as 
\beq
& &{\cal H}=\varepsilon(2t-1)-\gamma\cdot i(z'{}^*-z')\frac{\Lambda_t^{a_2}(x')}{x'}\ , 
\label{9-2}\\
& &{\dot z}'=-\gamma\left[
1-\frac{z'{}^2}{x'}\left(1-\frac{\Lambda_t^{a_2}(x')}{x'\Lambda_t^{a_2}{}'(x')}\right)\right] , 
\quad
{\dot z}'{}^*=-\gamma\left[
1-\frac{z'{}^{*2}}{x'}\left(1-\frac{\Lambda_t^{a_2}(x')}{x'\Lambda_t^{a_2}{}'(x')}\right)\right] . 
\label{9-3}
\eeq
We redefine $z$, $z^*$ and $x$ in the following form:
\beq\label{9-4}
z=\frac{\gamma_t}{z'{}^*}\ , \qquad z^*=\frac{\gamma_t}{z'}\ , \qquad x=|z|^2=\frac{\gamma_t^2}{x'}\ . 
\eeq
In the new variables, we have 
\beq\label{9-5}
\gamma_t \leq x' < \infty \ \ \longrightarrow\ \  
0\leq x \leq \gamma_t \ .
\eeq
It may be important to see that the range for $x$ is the same as that in the case 
$\Lambda_t^{a_1}(x)$. 

%
%
%%%%%%%%%%%%%%%%%%%%%%%%%%%%%%%%%%%%%%%%%%%%%%%%%%%%%%%%%%%%%%%%%%%%%%
\begin{figure}[t]
\begin{center}
\includegraphics[height=6cm]{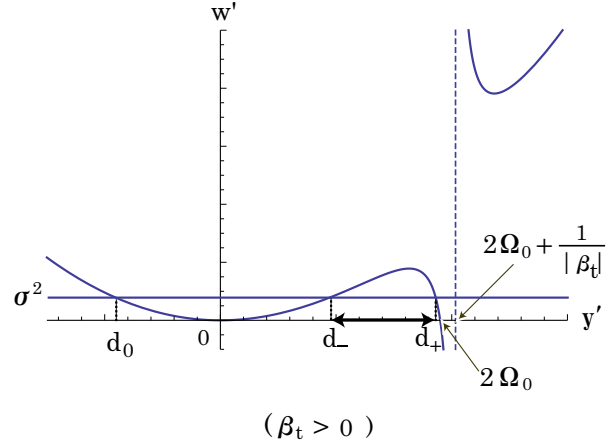}
\caption{
The behavior of the relation (\ref{9-13}) for $\beta_t>0$ is depicted. 
Here, $\Omega_0=19/2$ and $t=5/2$ are adopted which lead to $\beta_t=(14-15/\sqrt{57})/19 \approx 0.632\ (>0)$.
}
\label{fig:4}
\end{center}
\end{figure}
%%%%%%%%%%%%%%%%%%%%%%%%%%%%%%%%%%%%%%%%%%%%%%%%%%%%%%%%%%%%%%%%%%%%%%%%
%
%

With the use of the new variables, ${\cal H}$ can be rewritten to 
\beq
& &{\cal H}=\varepsilon (2t-1)-\gamma\cdot i(z^*-z)\frac{\Lambda_t^{a_2}(\frac{\gamma_t^2}{x})}{\gamma_t}\ , 
\label{9-6}\\
& &\frac{\Lambda_t^{a_2}\left(\frac{\gamma_t^2}{x}\right)}{\gamma_t}=2\Omega_0
\left(\frac{\gamma_t^2-\left(\beta_t+\frac{1}{2\Omega_0}\right)x}{\gamma_t^2-\beta_t x}\right)\ . 
\label{9-7}
\eeq
The relations (\ref{8-1}) - (\ref{8-3}) are reduced to 
\beq
& &i(z^*-z)\frac{\Lambda_t^{a_2}\left(\frac{\gamma_t^2}{x}\right)}{\gamma_t}=2\kappa\ , 
\label{9-8}\\
& &i(z^*-z)=2v\ , 
\label{9-9}\\
& &v=\frac{\kappa}{y'}\ , \qquad
y'=\frac{\Lambda_t^{a_2}\left(\frac{\gamma_t^2}{x}\right)}{\gamma_t}\ . 
\label{9-10}
\eeq
The function $x=f_t(y')$ in the present case is given by 
\beq\label{9-11}
x=f_t(y')=\frac{\gamma_t^2(2\Omega_0-y')}{\beta_t(2\Omega_0-y')+1}\ . 
\eeq
We can treat $u$ in the present case under the same idea as that of the case $\Lambda_t^{a_1}(x)$. 
Since $u^2=x-v^2 \geq 0$, we have the following inequality: 
\beq\label{9-12}
\frac{\gamma_t^2}{\beta_t}\cdot\frac{y'{}^2(y'-2\Omega_0)}{y'-\left(2\Omega_0+\frac{1}{\beta_t}\right)} \geq \kappa^2\ , 
\eeq
i.e., 
\beq
& &w'=\left\{
\begin{array}{ll}
\displaystyle +\frac{y'{}^2(y'-2\Omega_0)}{y'-\left(2\Omega_0+\frac{1}{|\beta_t|}\right)} \geq \sigma^2\ , &\quad (\beta_t >0) \\
\displaystyle -\frac{y'{}^2(y'-2\Omega_0)}{y'-\left(2\Omega_0-\frac{1}{|\beta_t|}\right)} \geq \sigma^2\ , &\quad (\beta_t <0)
\end{array}
\right. 
\label{9-13}\\
& &\sigma=\frac{\sqrt{|\beta_t|}}{\gamma_t}\kappa\ , \qquad {\dot \sigma}=0\ . 
\label{9-14}
\eeq
The case $\gamma_t=0$ appears in $2t-1=2\Omega_0$ and, later, we will consider this case. 
The behavior of the relation (\ref{9-13}) for $\beta_t>0$ is depicted in Fig.4. 
In Fig.4, we can find out the relation 
\beq\label{9-15}
d_- \leq y' \leq d_t\ , \quad {\rm i.e.,}\quad
\frac{1}{2}(d_++d_-)-\frac{1}{2}(d_+-d_-) \leq y' \leq 
\frac{1}{2}(d_++d_-)+\frac{1}{2}(d_+-d_-)\ . 
\eeq
Therefore, the same idea as that shown in the relation (\ref{8-13}) for $\alpha_t<0$ can be adopted: 
\bsub\label{9-16}
\beq
y'=\frac{1}{2}(d_++d_-)+\frac{1}{2}(d_+-d_-)\cos \psi\ . \qquad (\beta_t>0)
\label{9-16a}
\eeq
%
%
%%%%%%%%%%%%%%%%%%%%%%%%%%%%%%%%%%%%%%%%%%%%%%%%%%%%%%%%%%%%%%%%%%%%%%
\begin{figure}[t]
\begin{center}
\includegraphics[height=5.0cm]{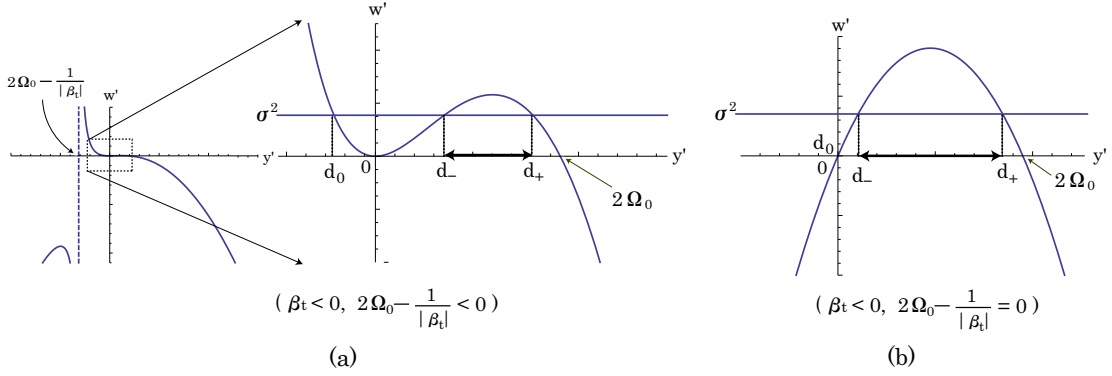}
\caption{
The behavior of the relation (\ref{9-13}) for $\beta_t<0$ is depicted 
in the case (a) $2\Omega_0-1/|\beta_t|<0$ and (b) $2\Omega_0-1/|\beta_t|=0$, separately.
Here, in (a), $\Omega_0=19/2$ and $t=18/2$ are adopted which lead to $\beta_t=(1-6/\sqrt{19})/19 \approx -0.0198\ (<0)$ 
and $2\Omega_0-1/|\beta_t|\approx -31.466\ (<0)$. 
In (b), $\Omega_0=19/2$ and $t=19/2$ are adopted which lead to $\beta_t=-1/19\ (<0)$ 
and $2\Omega_0-1/|\beta_t|=0$. 
}
\label{fig:5}
\end{center}
\end{figure}
%%%%%%%%%%%%%%%%%%%%%%%%%%%%%%%%%%%%%%%%%%%%%%%%%%%%%%%%%%%%%%%%%%%%%%%%
%
%
Here, $\psi$ denotes new parameter and, later, the explicit forms of $d_{\pm,0}$ will be shown. 
In order to treat the $\beta_t<0$, some comments are necessary. 
As was shown in the relations (\ref{6-26}) and (\ref{6-27}), the present case 
$\beta_t<0$ appears only in the three cases: 
\beq
& &{\rm (i)}\quad 
t=\Omega_0-\frac{1}{2}\qquad 
\left(\beta_{\Omega_0-1/2}=-\frac{1}{2\Omega_0}\left(\frac{1-\frac{1}{\Omega_0}}{\sqrt{2-\frac{1}{\Omega_0}}+1}
\right)\ , \quad \gamma_{\Omega_0-1/2}=\frac{1}{\Omega_0}\right)\ , 
\nonumber\\
& &{\rm (ii)}\quad 
t=\Omega_0\qquad 
\left(\beta_{\Omega_0}=-\frac{1}{2\Omega_0}\ , \quad \gamma_{\Omega_0}=\frac{1}{2\Omega_0}\right)\ , 
\nonumber\\
& &{\rm (iii)}\quad 
t=\Omega_0+\frac{1}{2}\qquad 
\left(\beta_{\Omega_0+1/2}=-\frac{1}{2\Omega_0}\ , \quad \gamma_{\Omega_0+1/2}=0\right)\ . 
\nonumber
\eeq
Later, we will contact with the case (iii) separately. 
The cases (i) and (ii) give us $2\Omega_0-1/|\beta_t|<0$ and 
$2\Omega_0-1/|\beta_t|=0$, respectively. 
The behavior of the relation (\ref{9-13}) for $\beta_t<0$ is depicted in Fig.5(a) and (b), separately. 
We can see that the parametrization of the above case is of the same as that shown in the relation (\ref{9-16a}): 
\beq\label{9-16b}
y'=\frac{1}{2}(d_++d_-)+\frac{1}{2}(d_+-d_-)\cos \psi\ . \qquad (\beta_t<0)
\eeq
\esub
In \S 10, we will discriminate between the former (\ref{9-16a}) and the later (\ref{9-16b}) in terms of 
the notations $y'_+$ and $y'_-$.

By equating both sides of the relation (\ref{9-13}), we derive the following cubic equation: 
\bsub\label{9-17}
\beq
& &y'{}^3-2\Omega_0 y'{}^2-\sigma^2y'+\sigma^2\left(2\Omega_0+\frac{1}{|\beta_t|}\right)=0\ , \qquad (\beta_t>0)
\label{9-17a}\\
& &-y'{}^3+2\Omega_0 y'{}^2-\sigma^2y'+\sigma^2\left(2\Omega_0-\frac{1}{|\beta_t|}\right)=0\ . \qquad (\beta_t<0)
\label{9-17b}
\eeq
\esub
Three real solutions of Eq.(\ref{9-17}) give us $d_{\pm,0}$:
\bsub\label{9-18}
\beq
& &d_+=\frac{2}{3}\left(\Omega_0+\sqrt{4\Omega_0^2+3\sigma^2}\cos\frac{\theta}{3}\right)\ , \nonumber\\
& &d_-=\frac{2}{3}\left(\Omega_0-\sqrt{4\Omega_0^2+3\sigma^2}\cos\left(\frac{\theta+\pi}{3}\right)\right)\ , \nonumber\\
& &d_0=\frac{2}{3}\left(\Omega_0-\sqrt{4\Omega_0^2+3\sigma^2}\cos\left(\frac{\theta-\pi}{3}\right)\right)\ , \qquad 
(\beta_t>0)
\label{9-18a}\\
& &d_+=\frac{2}{3}\left(\Omega_0+\sqrt{4\Omega_0^2-3\sigma^2}\cos\frac{\theta}{3}\right)\ , \nonumber\\
& &d_-=\frac{2}{3}\left(\Omega_0-\sqrt{4\Omega_0^2-3\sigma^2}\cos\left(\frac{\theta+\pi}{3}\right)\right)\ , \nonumber\\
& &d_0=\frac{2}{3}\left(\Omega_0-\sqrt{4\Omega_0^2-3\sigma^2}\cos\left(\frac{\theta-\pi}{3}\right)\right)\ . \qquad 
(\beta_t<0)
\label{9-18b}
\eeq
\esub
Here, $\theta$ ($0\leq \theta \leq \pi$) denotes a parameter which satisfies 
\bsub\label{9-19}
\beq
& &\cos \theta =\frac{8\Omega_0^3-9\left(\frac{3}{2|\beta_t|}+2\Omega_0\right)\sigma^2}
{\left(\sqrt{4\Omega_0^2+3\sigma^2}\right)^3}\qquad {\rm for}\quad \beta_t>0\ , 
\label{9-19a}\\
& &\cos \theta =\frac{8\Omega_0^3-9\left(\frac{3}{2|\beta_t|}-2\Omega_0\right)\sigma^2}
{\left(\sqrt{4\Omega_0^2-3\sigma^2}\right)^3}\qquad {\rm for}\quad \beta_t<0\ . 
\label{9-19b}\\
& &\frac{3}{2|\beta_t|}-2\Omega_0 > 0 \qquad {\rm for}\quad \beta_t<0\ .  
\label{9-19c}
\eeq
\esub
The relations (\ref{6-26}) and (\ref{6-27}) support the inequality (\ref{9-19c}). 
The above three quantities $d_{\pm,0}$ satisfy 
\beq\label{9-20}
d_++d_-+d_0=2\Omega_0\ .
\eeq
With the use of the relation (\ref{9-18}), we have the following expression:
\bsub\label{9-21}
\beq
& &\frac{1}{2}(d_++d_-)=\frac{1}{3}\left(2\Omega_0+\sqrt{4\Omega_0^2+3\sigma^2}\cos \left(\frac{\pi-\theta}{3}\right)\right)\ , 
\nonumber\\
& &\frac{1}{2}(d_+-d_-)=\frac{1}{\sqrt{3}}\sqrt{4\Omega_0^2+3\sigma^2}\sin \left(\frac{\pi-\theta}{3}\right)\ , 
\qquad (\beta_t>0)
\label{9-21a}\\
& &\frac{1}{2}(d_++d_-)=\frac{1}{3}\left(2\Omega_0+\sqrt{4\Omega_0^2-3\sigma^2}\cos \left(\frac{\pi-\theta}{3}\right)\right)\ , 
\nonumber\\
& &\frac{1}{2}(d_+-d_-)=\frac{1}{\sqrt{3}}
\sqrt{4\Omega_0^2-3\sigma^2}\sin \left(\frac{\pi-\theta}{3}\right)\ . 
\qquad (\beta_t<0)
\label{9-21b}
\eeq
\esub
By substituting the above result (\ref{9-21}) into the relation (\ref{9-16}), we are able to obtain 
$y'$ as a function of $\cos \psi$.

With the use of $y'$, we have the expressions for $x'$ and $\Lambda_t^{a_2}(x')$ in the form 
\beq
& &x'=\frac{\gamma_t^2}{x}=\beta_t+\frac{1}{2\Omega_0-y'}\ , 
\label{9-22}\\
& &\Lambda_t^{a_2}(x')=\Lambda_t^{a_2}\left(\frac{\gamma_t^2}{x}\right)=\gamma_t y'\ . 
\label{9-23}
\eeq
The relation (\ref{9-22}) is applicable in the range $\gamma_t \leq x' < \infty$. 
In \S 10, this point will be discussed. 
%The relation (\ref{9-16}) determines $y'$. 
In the relation (\ref{9-16}), $y'$ is given as a function of $\cos\psi$. 
Therefore, if the time-dependence of $\psi$ is determined, 
we are able to have 
the time-dependence of $\Lambda_t^{a_2}(x')$. 
Concerning to this point, we can see that in the case $\gamma_t=0$, $\Lambda_t^{a_2}(x')$ vanishes. 
It is quite natural and the reason is simple: 
The case $\gamma_t=0$ gives us the relation $2t-1=2\Omega_0$ and the relations (\ref{4-21}) and (\ref{5-17}) 
suggest us that this case corresponds to the maximum seniority number, that is, 
there does not exist the possibility of the creation of the Cooper-pair.

Let us investigate the time-dependence of $\psi$. 
The basic idea is the same as that in the case $\Lambda_t^{a_1}(x)$. 
First, we notice that the relation (\ref{9-3}) can be rewritten as follows: 
\beq\label{9-24}
{\dot z}=\frac{\gamma}{\gamma_t}\left[z^2-x\left(1-\frac{\Lambda_t^{a_2}(x')}{x'\Lambda_t^{a_2}{}'(x')}\right)\right]\ , 
\quad
{\dot z}^*=\frac{\gamma}{\gamma_t}\left[z^{*2}-x\left(1-\frac{\Lambda_t^{a_2}(x')}{x'\Lambda_t^{a_2}{}'(x')}\right)\right]\ . 
\eeq
Here, of course, $x'=\gamma_t^2/x$. 
Then, we can calculate ${\dot x}$: 
\beq\label{9-25}
{\dot x}={\dot z}^* z+z^*{\dot z}
=2\gamma\cdot\frac{x^2}{\gamma_t^3}\cdot\frac{\Lambda_t^{a_2}(x')}{\Lambda_t^{a_2}{}'(x')}\cdot u\ . 
\eeq
In a way similar to the case $\Lambda_t^{a_1}(x)$, $u$ is obtained in the form 
\beq\label{9-26}
u=\left\{
\begin{array}{ll}
\displaystyle \pm\frac{\gamma_t}{\sqrt{|\beta_t|}}\cdot\frac{1}{y'}
\sqrt{(d_+-y')(y'-d_-)}\cdot\sqrt{\frac{y'-d_0}{(2\Omega_0+\frac{1}{|\beta_t|})-y'}}\ , & \qquad 
(\beta_t>0) \\
\displaystyle \pm\frac{\gamma_t}{\sqrt{|\beta_t|}}\cdot\frac{1}{y'}
\sqrt{(d_+-y')(y'-d_-)}\cdot\sqrt{\frac{y'-d_0}{y'-(2\Omega_0-\frac{1}{|\beta_t|})}}\ . & \qquad 
(\beta_t<0) 
\end{array}
\right.
\eeq
Here, it is noted that in the case $2\Omega_0-1/|\beta_t|=0$, $d_0=0$ and it corresponds to 
the situation shown in Fig.5(b). 
The relations (\ref{9-10}) and (\ref{9-25}) lead to the following form for ${\dot y}'$: 
\beq\label{9-27}
{\dot y}'=-\frac{2\gamma}{\gamma_t}y'\cdot u\ . 
\eeq
By substituting the quantity $u$ shown in the relation (\ref{9-26}), ${\dot y}'$ is 
obtained. 
On the other hand, the result (\ref{9-16}) gives us 
\beq\label{9-28}
{\dot y}'=-\frac{1}{2}(d_+-d_-)\sin \psi \cdot {\dot \psi}\ . 
\eeq
Equating the expressions (\ref{9-27}) and (\ref{9-28}) and treating the double sign $\pm$ in the 
same idea as that in the case $\Lambda_t^{a_1}(x)$, we obtain ${\dot \psi}$ in the following form: 
\beq\label{9-29}
{\dot \psi}=\left\{
\begin{array}{ll}
\displaystyle 
\frac{2\gamma}{\sqrt{|\beta_t|}}\sqrt{\frac{(\Omega_0-\frac{3}{2}d_0)+\frac{1}{2}(d_+-d_-)\cos \psi}
{\frac{1}{|\beta_t|}+(\Omega_0+\frac{1}{2}d_0)-\frac{1}{2}(d_+-d_-)\cos \psi}}\ , &\quad (\beta_t>0) \\
\displaystyle 
\frac{2\gamma}{\sqrt{|\beta_t|}}\sqrt{\frac{(\Omega_0-\frac{3}{2}d_0)+\frac{1}{2}(d_+-d_-)\cos \psi}
{\frac{1}{|\beta_t|}-(\Omega_0+\frac{1}{2}d_0)+\frac{1}{2}(d_+-d_-)\cos \psi}}\ . &\quad (\beta_t<0) 
\end{array}
\right.
\eeq
Here, we used the relation (\ref{9-20}). 
By solving the differential equation (\ref{9-29}), we obtain $\psi$ as a function of $\tau$. 
But, in spite of simple form, 
the exact solution may be impossible to obtain in analytical form except the following two cases: 
(1) $d_+=d_-$ for $\beta_t > 0$ and $\beta_t < 0$ and 
(2) $\Omega_0-(3/2)\cdot d_0=1/|\beta_t|-(\Omega_0+d_0/2)$. 
The case (2) corresponds to Fig. 5(b). 
Therefore, we must search reasonable approximation for the solution.

\setcounter{equation}{0}

\section{Various properties of $\Lambda_t^{a_2}(x)$ for describing its 
time-dependence --- procedure for the application ---}

\begin{figure}[t]
\begin{center}
\includegraphics[height=4.5cm]{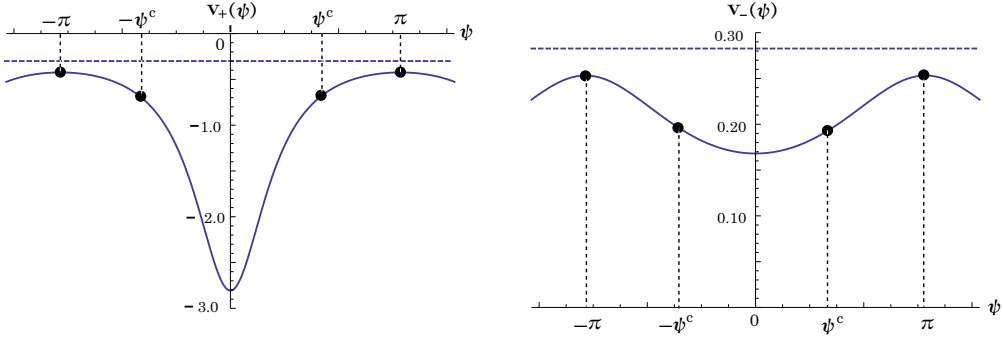}
\caption{
The figure shows $v_{\pm}(\psi)$ as function of $\psi$ in the range 
$-\pi \leq \psi \leq \pi $. 
The horizontal dotted line corresponds to $E_{\pm}/(A_{\pm}\pm B)$ $(E_{\pm}=\mp 1)$.
}
\label{fig:6add}
\end{center}
\end{figure}

Let us consider a guide to the approximation for the differential equation (\ref{9-29}). 
We start in rewriting this equation: 
\bsub\label{9-30}
\beq
& &\frac{1}{2}{\cal J}{\dot \psi}^2+V_+(\psi)=E_+\ , \quad
V_+(\psi)=-\frac{A_++B}{A_+-\cos\psi}\ , \quad E_+=-1\ , \ (\beta_t>0)
\label{9-30a}\\
& &\frac{1}{2}{\cal J}{\dot \psi}^2+V_-(\psi)=E_-\ , \quad
V_-(\psi)=+\frac{A_--B}{A_-+\cos\psi}\ , \quad E_-=+1\ . \ (\beta_t<0)\qquad\qquad
\label{9-30b}
\eeq
\esub
Here, ${\cal J}$, $A_{\pm}$ and $B$ are defined as 
\bsub\label{9-31}
\beq
{\cal J}=\frac{|\beta_t|}{2\gamma^2}\ , \qquad
A_{\pm}=\frac{\frac{2}{|\beta_t|}\pm (2\Omega_0+d_0)}{d_+-d_-}\ , \qquad
B=\frac{2\Omega_0-3d_0}{d_+-d_-}\ . 
\label{9-31a}
\eeq
With the use of the relation (\ref{9-20}) and Figs.4 and 5, we can show that $A_{\pm}$ and $B$ obey the inequality 
\beq\label{9-31b}
A_{\pm}>1\ , \qquad B>1\ , \qquad A_->B\ . 
\eeq
\esub
The expression (\ref{9-30}) can be regarded as the total energy $E_{\pm}$ with the kinetic energy 
${\cal J}{\dot \psi}^2/2$ and the potential energy $V_{\pm}(\psi)$. 
With the aid of the inequality (\ref{9-31b}), we can prove the following relation: 
\beq\label{9-32}
V_{\pm}(\psi)<E_{\pm}\ . \quad (-1 \leq \cos\psi \leq 1)
\eeq
Therefore, if the angle variable $\psi$ changes continuously in the range $-\infty < \psi < \infty$, 
the present case can be understood in terms of the rotational motion 
with the moment of inertia ${\cal J}$ and the periodically changing angular velocity. 
However, in the present case, $\psi$ does not change continuously, because the original variable $x'$ 
changes in the range $\gamma_t \leq x' <\infty$ and the other $x$ in the range 
$0 \leq x \leq\gamma_t$. 
The relation (\ref{9-22}) suggests us that for certain angle denoted as $\psi^c$, the 
variable $\psi$ should obey the condition 
\beq\label{9-33}
\cos\psi^c \leq \cos \psi \leq 1\ . 
\eeq
This means that any result derived from the relation (\ref{9-30}) in the range 
$-1\leq \cos\psi \leq \cos\psi^c$ is meaningless. 
This range is treated in relation to the variable $\chi$ discussed in \S 8. 
In this sense, the quantity $\cos\psi^c$ plays an essential role in our 
treatment. 
Including the value of $\cos\psi^c$, the detail will be considered in \S 10. 
The above mention is illustrated in Fig.\ref{fig:6add} for the range $-\pi \lsim \psi \lsim \pi$. 
The longitudinal axis represents $v_{\pm}(\psi)$ defined as 
\bsub\label{9-34}
\beq
& &V_{\pm}(\psi)=(A_{\pm}\pm B)\cdot v_{\pm}(\psi)\ , 
\label{9-34a}\\
& &v_{\pm}(\psi)=\mp\frac{1}{A_{\pm}\mp \cos\psi}\ , \qquad 
v_{\pm}(-\psi)=v_{\pm}(\psi)\ . 
\label{9-34b} 
\eeq
\esub
Hereafter, we will treat the angle $\psi$ in the range
\beq\label{9-35}
-\pi \leq \psi \leq \pi\ .
\eeq

%
%
%%%%%%%%%%%%%%%%%%%%%%%%%%%%%%%%%%%%%%%%%%%%%%%%%%%%%%%%%%%%%%%%%%%%%%
%%%%%%%%%%%%%%%%%%%%%%%%%%%%%%%%%%%%%%%%%%%%%%%%%%%%%%%%%%%%%%%%%%%%%%%%
%

\begin{figure}[t]
\begin{center}
\includegraphics[height=5.0cm]{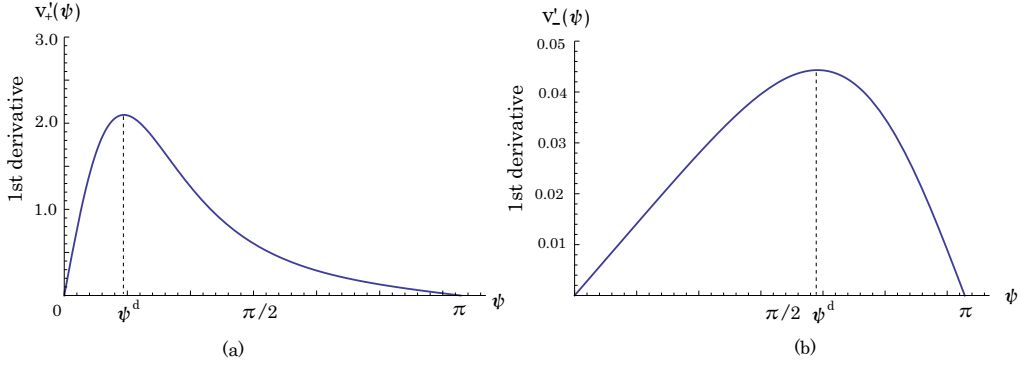}
\caption{
The figure shows $v'_{\pm}(\psi)$ as function of $\psi$ in the range 
$0 \leq \psi \leq \pi $.
}
\label{fig:7add}
\end{center}
\end{figure}

Let us present an idea for the approximation of $v_{\pm}(\psi)$, which 
will be adopted in this paper. 
Needless to say, we search a possible approximation in the range 
\beq\label{9-36}
-\psi^c \leq \psi \leq \psi^c\ , \qquad 
0 \leq \psi^c \leq \pi\ . 
\eeq
But, for the time being, forgetting the range (\ref{9-36}), $\psi$ is treated in 
the range (\ref{9-35}). 
Behavior of $v_{\pm}(\psi)$ is illustrated in Fig.\ref{fig:6add}. 
In order to understand this behavior more precisely, we take up the first and the second derivative of $v_{\pm}(\psi)$ for 
$\psi$, $v_{\pm}'(\psi)$ and $v''_{\pm}(\psi)$: 
\bsub\label{9-37}
\beq
& &v_{\pm}'(\psi)=\frac{\sin \psi}{(A_{\pm}\mp \cos\psi)^2}\ , 
\label{9-37a}\\
& &v_{\pm}''(\psi)=\pm \frac{\cos^2\psi \pm A_{\pm}\cos\psi -2}{(A_{\pm}\mp \cos \psi)^3}\ . 
\label{9-37b}
\eeq
\esub
Since $V_{\pm}(\psi)$ represents the potential energy, the force $F_{\pm}(\psi)$ acting on 
the present system is given in the form 
\beq\label{9-38}
F_{\pm}(\psi)=-\frac{dV_{\pm}(\psi)}{d\psi}=-(A_{\pm}\pm B)\cdot v_{\pm}'(\psi)\ . 
\eeq
Therefore, we can learn characteristics of $F_{\pm}(\psi)$ through $v_{\pm}'(\psi)$. 
Since $F_{\pm}(-\psi)$, i.e., $v_{\pm}'(-\psi)=-v_{\pm}'(\psi)$, it may be 
enough to investigate $v_{\pm}'(\psi)$ in the range $0\leq \psi \leq \pi$. 
Its behavior is shown in Fig.\ref{fig:7add}. 
%
%
%%%%%%%%%%%%%%%%%%%%%%%%%%%%%%%%%%%%%%%%%%%%%%%%%%%%%%%%%%%%%%%%%%%%%%
%%%%%%%%%%%%%%%%%%%%%%%%%%%%%%%%%%%%%%%%%%%%%%%%%%%%%%%%%%%%%%%%%%%%%%%%
%
The angle $\psi^d$ gives us the maximum value of $v_{\pm}'(\psi)$ and its 
value is determined by $v_{\pm}''(\psi)=0$: 
\beq\label{9-39}
\cos\psi^d=\pm\frac{1}{2}\left(\sqrt{A_{\pm}^2+8}-A_{\pm}\right)\ . 
\eeq
The flectional point of $v_{\pm}(\psi)$ is given by $\psi^d$ and we have the following: \\
\hspace{1cm}
(i) $v_{\pm}'(\psi)$ is increasing in the range $\psi < \psi^d$, \\
\hspace{1cm}
(ii) $v_{\pm}'(\psi)$ is decreasing in the range $\psi > \psi^d$. \\
The above two cases and the relation $F_{\pm}(-\psi)=F_{\pm}(\psi)$ 
teach us that the force under consideration is attractive for the point, the center of the 
force and as $|\psi|$ increases, the strength of the force increases in the case (i) and 
decreases in the case (ii). 
This indicates that the property of the force is transformed at $\psi=\psi^d$. 
The above is a distinctive feature of $F_{\pm}(\psi)$. 
With this feature in mind, we consider the approximation of $v_{\pm}(\psi)$ through $v_{\pm}'(\psi)$.

\begin{figure}[t]
\begin{center}
\includegraphics[height=5.0cm]{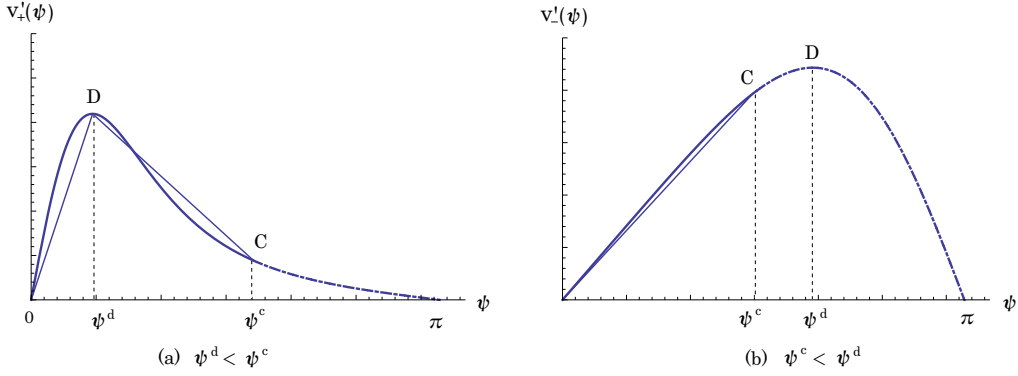}
\caption{
The figure shows $v'_{\pm}(\psi)$ as function of $\psi$ in the range 
$0 \leq \psi \leq \pi $. 
The solid curves and dot-dashed curves represent the exact results. 
The thin lines represent the approximate ones. 
(a) The parameters are taken as $\kappa=-15/2$, $t=5/2$ and $\Omega_0=19/2$. 
Here, $\alpha_t\approx 0.7656\ (>0) $ and $\beta_t\approx 0.6323\ (>0)$ are derived. 
Also, 
$A_+\approx 1.3564$, 
$\psi^c\approx 1.7123$ and $\psi^d\approx 0.4729$ are obtained. 
(b) The parameters are taken as $\kappa=-1$, $t=18/2$ and $\Omega_0=19/2$. 
In this parameter set, $\alpha_t\approx -6.0384\ (<0) $ and $\beta_t\approx -0.0198\ (<0)$ are derived. 
Here, 
$A_-\approx 4.9493$, 
$\psi^c\approx 1.4877$ and $\psi^d\approx 1.9558$ are obtained. 
}
\label{fig:8}
\end{center}
\end{figure}

In order to obtain the idea, first, we must introduce the angle $\psi^c$ into the above argument. 
In the case $v_+'(\psi)$, we have two possibilities, 
which are illustrated in Fig.\ref{fig:8}. 
As is clear from the relation (\ref{9-36}), the force $F_+(\psi)$ has its meaning for the solid curve OC. 
Our idea may be the simplest and it is summarized as follows: 
(a) In the case (a), the curves OD and DC are replaced with the straight lines OD and DC. 
(b) In the case (b), the curve OC is replaced with the straight line OC. 
The above scheme is also applicable to the case $v_-'(\psi)$. 
It should be noted that the above approximation preserves the distinctive feature of $F_{\pm}(\psi)$ 
already mentioned and the values of $v_{\pm}'(\psi)$ at $\psi=0$, $\psi^d$ and $\psi^c$. 
By adopting the symbol $v_{\pm}^a(\psi)$ for the approximate form of $v_{\pm}(\psi)$, 
the above idea is formulated as follows: 
\bsub\label{9-40}
\beq
{\rm (a)}
%\hspace{-0.3cm}
& &v_{\pm}^{a}{}'(\psi)=
-w_{\pm}^{cd}\cdot(\psi-\psi^c)
+v_{\pm}'(\psi^c)\ , \quad \left(w_{\pm}^{cd}=-\frac{v_{\pm}'(\psi^c)-v_{\pm}'(\psi^d)}{\psi^c-\psi^d}\right)
\quad {\rm (DC)}\nonumber\\
& &
\label{9-40a}\\
& &
v_{\pm}^{a}{}'(\psi)
=w_{\pm}^d\cdot\psi\ , \quad\qquad\qquad\qquad\qquad  \left(w_{\pm}^d=\frac{v_{\pm}'(\psi^d)}{\psi^d}\right)\qquad {\rm (OD)}
\label{9-40b} 
\eeq
\esub
%\vspace{-0.2cm}
\beq
& &
%\hspace{-1.4cm}
{\rm (b)}\quad
v_{\pm}^a{}'(\psi)=w_{\pm}^c\cdot\psi\ . \quad\left(w_{\pm}^c=\frac{v_{\pm}'(\psi^c)}{\psi^c}\right)\qquad\qquad\qquad\quad\qquad\quad{\rm (OC)}
\qquad
\label{9-41}
\eeq
%
%
%%%%%%%%%%%%%%%%%%%%%%%%%%%%%%%%%%%%%%%%%%%%%%%%%%%%%%%%%%%%%%%%%%%%%%
%%%%%%%%%%%%%%%%%%%%%%%%%%%%%%%%%%%%%%%%%%%%%%%%%%%%%%%%%%%%%%%%%%%%%%%%
%
Naturally, the relations (\ref{9-40}) and (\ref{9-41}) give us 
\beq\label{9-42}
v_{\pm}^a{}'(0)=v_{\pm}'(0)\ , \qquad
v_{\pm}^a{}'(\psi^d)=v_{\pm}'(\psi^d)\ , \qquad
v_{\pm}^a{}'(\psi^c)=v_{\pm}'(\psi^c)\ . 
\eeq
By integrating the relations (\ref{9-40}) and (\ref{9-41}), we are able to obtain the approximate form of 
$v_{\pm}(\psi)$, $v_{\pm}^a(\psi)$.

%
%
%%%%%%%%%%%%%%%%%%%%%%%%%%%%%%%%%%%%%%%%%%%%%%%%%%%%%%%%%%%%%%%%%%%%%%
\begin{figure}[t]
\begin{center}
\includegraphics[height=5.0cm]{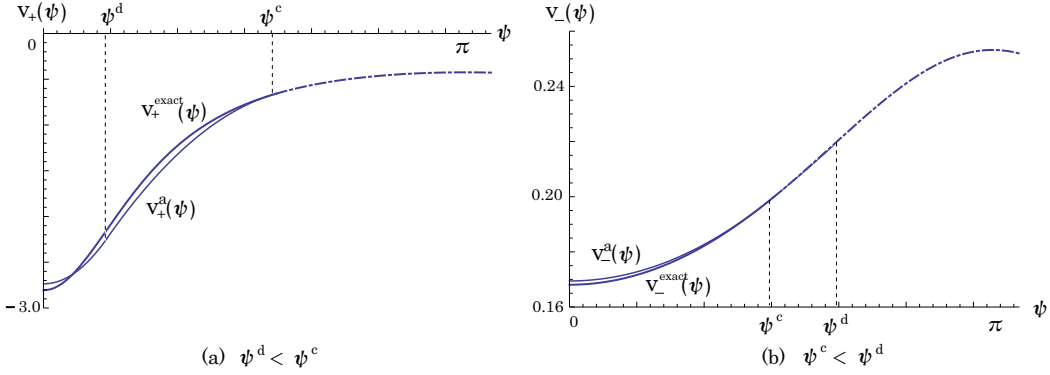}
\caption{
The figure shows $v_{\pm}(\psi)$ as function of $\psi$ in the range 
$0 \leq \psi \leq \pi $. 
The solid and dot-dashed curves represent the exact results. 
The thin curves represent the approximate ones. 
The parameters are the same ones used in Fig.\ref{fig:8}.
%(a) $\kappa=-15/2$, $t=5/2$ and $\Omega_0=19/2$, which lead to 
%$\alpha_t\approx 0.7656\ (>0) $ and $\beta_t\approx 0.6323\ (>0)$. 
%Also, $A_+\approx 1.3564$, $\psi^c\approx 1.7123$ and $\psi^d\approx 0.4729$ are obtained. 
%(b) $\kappa=-1$, $t=18/2$ and $\Omega_0=19/2$, which lead to 
%$\alpha_t\approx -6.0384\ (<0) $ and $\beta_t\approx -0.0198\ (<0)$. 
%Here, $A_-\approx 4.9493$, $\psi^c\approx 1.4877$ and $\psi^d\approx 1.9558$ are obtained. 
}
\label{fig:9}
\end{center}
\end{figure}
%%%%%%%%%%%%%%%%%%%%%%%%%%%%%%%%%%%%%%%%%%%%%%%%%%%%%%%%%%%%%%%%%%%%%%%%
%

For the integration, we require the condition 
\beq\label{9-43}
v_{\pm}^a(\psi^c)=v_{\pm}(\psi^c)\ . 
\eeq
As was already mentioned, the angle $\psi^c$ plays a role of the door way to 
the range treated by $\chi$. 
Therefore, for obtaining $v_{\pm}^a(\psi)$, consideration to the behavior of $v_{\pm}(\psi)$ 
at $\psi=\pm\psi^c$ and the neighbouring region should be prior to any other region. 
The above consideration suggests us the condition (\ref{9-43}). 
By integrating the relation of (\ref{9-40a}) under the condition 
(\ref{9-43}), we have 
the following expression for $v_{\pm}^a(\psi)$: 
\bsub\label{9-44}
\beq
v_{\pm}^a(\psi)=
-\frac{1}{2}w_{\pm}^{cd}\cdot (\psi-\psi^c)^2
+v_{\pm}'(\psi^c)\cdot(\psi-\psi^c)+v_{\pm}(\psi^c)\ . \ \ {\rm (DC)}
\label{9-44a}
\eeq
If we require that, at $\psi=\psi^d$, the value of $v_{\pm}^a(\psi)$ from the side OD agrees with 
that from the side DC for the relation of (\ref{9-40b}), we obtain the expression 
\beq
v_{\pm}^a(\psi)=\frac{1}{2}w_{\pm}^d\cdot(\psi^2-\psi^d{}^2)
-\frac{1}{2}(v_{\pm}'(\psi^c)+v_{\pm}'(\psi^d))\cdot(\psi^c-\psi^d)+v_{\pm}(\psi^c)\ . 
\ {\rm (OD)}\nonumber\\ 
& &\label{9-44b}
\eeq
\esub
The above requirement may be acceptable, because the present system 
conserves the energy. 
For the case (b), the condition (\ref{9-41}) gives us the following expression: 
\beq\label{9-45}
v_{\pm}^a(\psi)=\frac{1}{2}w_{\pm}^c(\psi^2-\psi^c{}^2)
+v_{\pm}(\psi^c)\ . \qquad {\rm (OC)}
\eeq
By replacing $\psi$ with $-\psi$, we have the expressions in the range $-\psi^c\leq \psi \leq 0$. 
Thus, we obtained the approximate expressions of $v_{\pm}(\psi)$ in our scheme. 
It should be noted that, owing to the approximation, we are forced to have 
$v_{\pm}^a(0)\neq v_{\pm}(0)$, and $v_{\pm}^a(\psi^d)\neq v_{\pm}(\psi^d)$. 
%The accuracy of the above approximation will be discussed in \S 10. 
In Fig.\ref{fig:9}, the solid and dot-dashed curves represent the exact $v_{\pm}(\psi)$. 
Under the idea formulated by (\ref{9-40})-(\ref{9-42}), the approximate $v_{\pm}^a(\psi)$ are 
obtained and are shown by thin curves. 
Figure \ref{fig:9} shows that the $v_{\pm}^a(\psi)$ presents a good approximation 
for the exact result in the range $-\psi^c \leq\psi \leq\psi^c$ under consideration.

Finally, we will sketch the approximate solution of $\psi$ as a function of $\tau$. 
The relation (\ref{9-30}) leads us to the following approximate expression for ${\dot \psi}$: 
\beq\label{9-46add}
{\dot \psi}=\sqrt{\frac{2}{\cal J}(E_{\pm}-(A_{\pm}\pm B)v_{\pm}^a(\psi))}\ . 
\eeq
For $v_{\pm}^a(\psi)$, the relations (\ref{9-44}) and (\ref{9-45}) must be used. 
As can be seen in the forms (\ref{9-44}) and (\ref{9-45}), the potential energy 
is expressed as a quadratic function of $\psi$. 
For the coefficients of $\psi^2$, we have the inequalities 
\beq\label{9-46}
%\frac{v_{\pm}'(\psi^c)-v_{\pm}'(\psi^d)}{\psi^c-\psi^d} 
w_{\pm}^{cd}> 0 \ , \qquad
%\frac{v_{\pm}'(\psi^d)}{\psi^d} 
w_{\pm}^d > 0 \ , \qquad
%\frac{v_{\pm}'(\psi^c)}{\psi^c} 
w_{\pm}^c > 0 \ . 
\eeq
Therefore, $\psi$ can be simply expressed in the form 
\bsub\label{9-48}
\beq
& &\psi(\tau)=\psi_h(\tau)={\cal A}_h\sinh (\omega_h\tau+\alpha_h)+{\cal B}_h\ , \qquad {\rm (CD)}
\label{9-48a}\\
& &\psi(\tau)=\psi_n(\tau)={\cal A}_n\sin (\omega_n\tau+\alpha_n)+{\cal B}_n\ . \qquad\ \  {\rm (OD),\ (OC)}
\label{9-48b}
\eeq
\esub
Then, our problem is reduced to determine the coefficients 
$({\cal A}_h, \omega_h,\alpha_h,{\cal B}_h)$ and $({\cal A}_n, \omega_n,\alpha_n,{\cal B}_n)$. 
In next section, some examples will be given.

\setcounter{equation}{0}

\section{Discussion}

One of the aims of this paper is to describe a simple many-fermion model obeying the pseudo $su(1,1)$-algebra 
in terms of the time-dependent variational method. 
In this description, the function $\Lambda_t(x)$ plays a central role. 
For its original form, we adopted an approximate form which consists of two 
parts: 
$\Lambda_t^{a_1}(x)$ for $0\leq x \leq \gamma_t$ and $\Lambda_t^{a_2}(x')$ for 
$\gamma_t \leq x' <\infty$. 
Treating both parts independently in \S\S 8 and 9, we derived various features induced by 
these two functions. 
Therefore, it is inevitable to investigate connection between the results 
derived from two forms. 
For this aim, it may be convenient to discuss the connection under four categories, although they are 
correlated with one another.

First is related to constants of motion. 
We already mentioned that $t$ is one of them, i.e., common to the two parts. 
The others are $\rho$ in the range $0\leq x \leq \gamma_t$ and $\sigma$ in the range 
$\gamma_t \leq x' < \infty$ shown in the relations (\ref{8-10}) and (\ref{9-14}), respectively. 
They are not independent to each other. 
By eliminating $\kappa$ in both relations, we have 
\bsub\label{10-1}
\beq
\sigma=\frac{1}{\gamma_t}\sqrt{\frac{|\beta_t|}{|\alpha_t|}}\cdot \rho\ , \quad {\rm i.e.,}\quad 
\sigma^2=\frac{1}{\gamma_t^2}\cdot \frac{|\beta_t|}{|\alpha_t|}\cdot \rho^2\ .
\label{10-1a}
\eeq
As is clear from Figs.3(b), 4 and 5, $\rho^2$ and $\sigma^2$ have the maximum values. 
On the other hand, Fig.3(a) shows that in this case, formally, $\rho^2$ is permitted 
to become $\infty$. 
But, the relation (\ref{10-1}) teaches us that in this case, also, there exists the 
maximum value, because $\sigma^2$ has the maximum value. 
For example, in the case $\beta_t > 0$, the maximum value of 
$\sigma^2$, $(\sigma^2)_{\rm max}$, is given by 
\beq
(\sigma^2)_{\rm max}=\frac{d_m^2(d_m-2\Omega_0)}{d_m-\left(2\Omega_0+\frac{1}{\beta_t}\right)}\ , 
\qquad
d_m=2\Omega_0\left(1-\frac{2}{3+\sqrt{9+16\Omega_0|\beta_t|}}\right)\ . 
\label{10-1b}
\eeq
\esub
Here, $d_m$ denotes the value of $y'$ which makes $w'$ in the 
relation (\ref{9-13}) the maximum.

The ranges covered by $x$ and $x'$ are $0\leq x \leq \gamma_t$ and $\gamma_t \leq x' < \infty$, respectively. 
Second is related to these ranges. 
The relation (\ref{8-7}) and (\ref{9-22}) lead to the following inequalities: 
\bsub\label{10-2}
\beq
& &0 \leq \frac{1}{\alpha_t}\left(1-\frac{2t}{y}\right) \leq \gamma_t\ , 
\label{10-2a}\\
& &\gamma_t \leq \beta_t+\frac{1}{2\Omega_0-y'} < \infty \ . 
\label{10-2b}
\eeq
\esub
The inequality (\ref{10-2a}) gives us 
\beq
& &2t \leq y_+ \leq 2\Omega_0\left(1-\frac{1}{1+\sqrt{2t\gamma_t}}\right)\ , 
\label{10-3}\\
& &2\Omega_0\left(1-\frac{1}{1+\sqrt{2t\gamma_t}}\right) \leq y_- \leq 2t\ . 
\label{10-4}
\eeq
Also, the inequality (\ref{10-2b}) gives us 
\beq\label{10-5}
2\Omega_0\left(1-\frac{1}{1+\sqrt{2t\gamma_t}}\right) \leq y_{\pm}' \leq 2t\ .
\eeq
For the derivation of the inequalities (\ref{10-4}) and (\ref{10-5}), we used the 
relation (\ref{6-7}). 
It should be noted that although $y_+$ contains $\cosh \chi$, it should be finite.

At the point $x=x'=\gamma_t$, $y$ connects to $y'$. 
As was shown in the relations (\ref{8-13}) and (\ref{9-16}), $y$ and $y'$ consist of $y_{\pm}$ and $y_{\pm}'$, 
respectively. 
Therefore, it is necessary to investigate, for example, if $y_+$ can connect to $y'$ or not. 
Third is concerned with the above. 
Formally, we can find four combinations between $y$ and $y'$: 
$(y_+, y_+')$, $(y_+, y'_-)$, $(y_-, y'_+)$ and $(y_-, y'_-)$. 
The relations from (\ref{6-19}) to (\ref{6-28}) with the interpretations for them lead us to the following three cases: 
\bsub\label{10-6}
\beq
& &{\rm (i)}\quad {\rm if}\quad \frac{1}{2}\leq t < t^0\ , \qquad 0 < \alpha_t \leq \beta_t\ , \quad ({\rm in\ the\ case}\ t=1/2\ , 
\alpha_t=\beta_t) 
\label{10-6a}\\
& &{\rm (ii)}\quad {\rm if}\quad t^0 < t < t^0{}'\ , \qquad \alpha_t < 0 < \beta_t\ , 
\label{10-6b}\\
& &{\rm (iii)}\quad {\rm if}\quad t^0{}' < t \leq \Omega_0+\frac{1}{2}\ , \qquad \alpha_t < \beta_t < 0\ . 
\label{10-6c}
\eeq
\esub
If the relation $\alpha_t < \beta_t$ is noticed, the above three cases may be understandable. 
The cases (i), (ii) and (iii) correspond to the combinations $(y_+, y'_+)$, $(y_-, y_+')$ and 
$(y_-, y_-')$, respectively. 
Therefore, $y_+$ cannot connect with $y_-'$.

For the above three combinations, we show the maximum values of the squares of the constants of motion $\kappa^2$ 
introduced in the relation (\ref{8-1}). 
The conditions $c_+-c_-=0$ and $d_+-d_-=0$ give us the maximum values 
$\kappa_m^{(i)2}$ $(i=1, 2, 3, 4)$ for the cases (1), (2), (3) and (4) related to Figs.3(a), 3(b), 4 and 5, respectively. 
With the use of these conditions, we obtain the following results: 
\bsub\label{10-7}
\beq
& &{\rm (1)}\quad \kappa_m^{(1)}{}^2 \rightarrow \infty\ , \qquad (y_+;\ {\rm Fig.3(a)})
\label{10-7a}\\
& &{\rm (2)}\quad \kappa_m^{(2)}{}^2 = \frac{t^2}{|\alpha_t|}\ , \qquad (y_-;\ {\rm Fig.3(b)})
\label{10-7b}\\
& &{\rm (3)}\quad \kappa_m^{(3)}{}^2 = \frac{\gamma_t^2}{|\beta_t|}\cdot\frac{16\Omega_0^3}{9(4\Omega_0+\frac{3}{|\beta_t|})}\ , 
\qquad (y_+';\ {\rm Fig.4})
\label{10-7c}\\
& &{\rm (4)}\quad \kappa_m^{(4)}{}^2 = \frac{\gamma_t^2}{|\beta_t|}\cdot\frac{4\Omega_0^2}{3}\ . 
\qquad (y_-';\ {\rm Fig.5})
\label{10-7d}
\eeq
\esub
For the combination $(y_+, y_+')$, we choose the smaller value of $\kappa_m^2$, i.e., 
$\kappa_m^{(3)2}$. 
For the combination $(y_-, y_+')$ and $(y_-, y_-')$, we choose the smaller values 
of $\kappa_m^2$ for each case; 
${\rm min}\{ \kappa_m^{(2)2}, \kappa_m^{(3)2}\}$ and ${\rm min}\{ \kappa_m^{(2)2}, \kappa_m^{(4)2}\}$, 
respectively.

%
%
%%%%%%%%%%%%%%%%%%%%%%%%%%%%%%%%%%%%%%%%%%%%%%%%%%%%%%%%%%%%%%%%%%%%%%
\begin{figure}[t]
\begin{center}
\includegraphics[height=5.0cm]{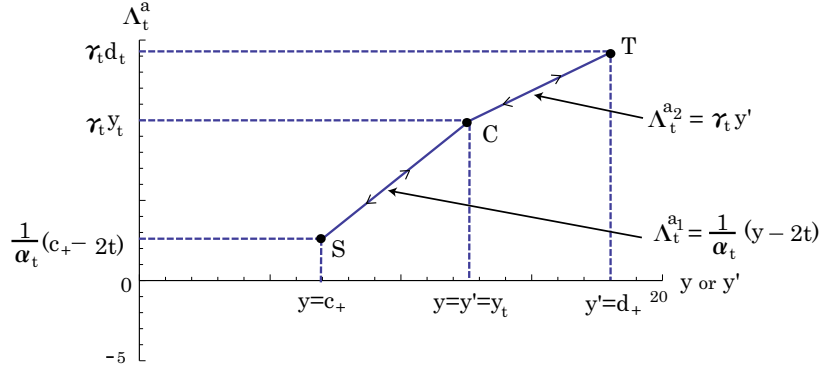}
\caption{
The path of the evolution is illustrated in the case of $\Omega_0=19/2$ and $t=5/2$. 
}
\label{fig:6}
\end{center}
\end{figure}
%%%%%%%%%%%%%%%%%%%%%%%%%%%%%%%%%%%%%%%%%%%%%%%%%%%%%%%%%%%%%%%%%%%%%%%%
%

Forth is related to giving the explicit expression of the connection. 
First, let us notice again that $\Lambda_t^{a_1}(x)$ and $\Lambda_t^{a_2}(x')$ 
should connect with each other smoothly at $x=x'=\gamma_t$: 
\beq\label{10-8}
\Lambda_t^{a_1}(\gamma_t)=\Lambda_t^{a_2}(\gamma_t)\ , \qquad
\Lambda_t^{a_1}{}'(\gamma_t)=\Lambda_t^{a_2}{}'(\gamma_t)\ .
\eeq
The explicit expression of the relation (\ref{10-8}) are presented in the relation (\ref{6-6}). 
The first of the relation (\ref{10-8}) and the definition of $y$ and $y'$ in (\ref{8-3}) and (\ref{9-10}) 
lead to 
\beq\label{10-9}
(y)_t=(y')_t\ (=y_t)\ . 
\eeq
Here, $(y)_t$ and $(y')_t$ denote the values of $y$ and $y'$ at the point $x=x'=\gamma_t$, respectively 
and, with the use of the relation (\ref{6-7}), $y_t$ is given by 
\beq\label{10-10}
y_t=2\Omega_0\left(1-\frac{1}{1+\sqrt{2t\gamma_t}}\right)\ . 
\eeq
The above is the connection between the results derived from the two forms.

Our final task is to investigate the time-evolution of $\Lambda_t$ in the approximate form, $\Lambda_t^a$. 
The path of the evolution is illustrated in Fig.\ref{fig:6}. 
The lines SC and CT correspond to $\Lambda_t^{a_1}=(y-2t)/\alpha_t$ and 
$\Lambda_t^{a_2}=\gamma_t y'$, respectively. 
Here, $y$ and $y'$ are given in the relations (\ref{8-13}) and (\ref{9-16}), respectively. 
They depend on two constants of motion, $t$ and $\kappa$. 
It is noted that $\Lambda_t^a$ has the minimum and the maximum values which correspond to 
$y=c_+$ and $y'=d_+$, respectively. 
At the point C, $\Lambda_t^a$ changes from $\Lambda_t^{a_1}$ to $\Lambda_t^{a_2}$, i.e., 
from $y$ to $y'$ and vice versa. 
The dependence of $\Lambda_t^a$ on the time $\tau$ 
may be periodic. 
One cycle consists of four paths  
(S$\rightarrow$C,\ C$\rightarrow$T,\ T$\rightarrow$C,\ C$\rightarrow$S), 
%(S$\rightarrow$C$\rightarrow$T$\rightarrow$C'(=C)$\rightarrow$S), 
which is shown in Fig.\ref{fig:6}.

%
%%%%%%%%%%%%%%%%%%%%%%%%%%%%%%%%%%%%%%%%%%%%%%%%%%%%%%%%%%%%%%%%%%%%%%
\begin{figure}[t]
\begin{center}
\includegraphics[height=5.5cm]{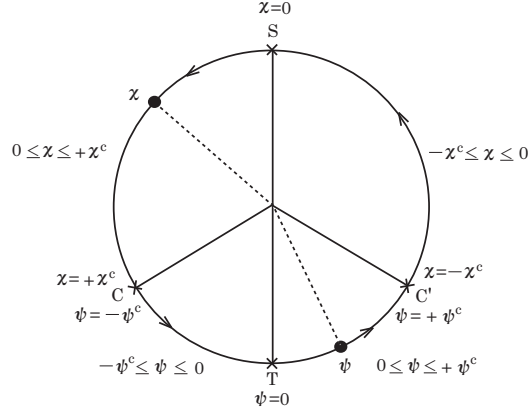}
\caption{
The values of $\chi$ and $\psi$ on each point are shown.  
}
\label{fig:7}
\end{center}
\end{figure}
%%%%%%%%%%%%%%%%%%%%%%%%%%%%%%%%%%%%%%%%%%%%%%%%%%%%%%%%%%%%%%%%%%%%%%%%
%

With the use of the relation (\ref{10-9}) with (\ref{8-13}) and (\ref{9-16}), we have the 
following relations: 
\beq
& &\cosh \chi^c=\frac{y_t-\frac{1}{2}(c_++c_-)}{\frac{1}{2}(c_+-c_-)}\quad
(\alpha_t>0)\ , \qquad
\cos \chi^c=\frac{y_t-\frac{1}{2}(c_++c_-)}{\frac{1}{2}(c_+-c_-)}\quad (\alpha_t<0)\ , \nonumber\\
& &
\label{10-15}\\
& &\cos\psi^c=\frac{y_t-\frac{1}{2}(d_++d_-)}{\frac{1}{2}(d_+-d_-)}\quad (\beta_t>0\ , \ \beta_t<0)\ . 
\label{10-16}
\eeq
Here, $\chi^c$ and $\psi^c$ denote the values of $\chi$ and $\psi$ at the point C, respectively. 
It is noticed that if $\chi^c$ and $\psi^c$ are positive, $-\chi^c$ and $-\psi^c$ also satisfy the 
relations (\ref{10-15}) and (\ref{10-16}), respectively. 
The angle $\psi^c$ is nothing but that introduced in \S 9. 
On the other hand, $\chi$ and $\psi$ are equal to 0 at the point S and the point T, respectively. 
The above consideration permits us to 
choose $\chi$ and $\psi$ including the signs of $\chi^c$ and $\psi^c$ in the form shown 
in Fig.\ref{fig:7}.
For the cycle (S$\rightarrow$C$\rightarrow$T$\rightarrow$C'(=C)$\rightarrow$S), it may be enough to regard ${\dot \chi}$ 
and ${\dot \psi}$ as positive, ${\dot \chi}>0$ and ${\dot \psi}>0$, at any position except 
the point S with ${\dot \chi}=0$ and the point T with ${\dot \psi}=0$. 
It is easily verified by $\sinh \chi=\sin \chi=\sin \psi=0$ for $\chi=\psi=0$.  
The time derivatives ${\dot \chi}$ and ${\dot \psi}$ are given 
in the relations (\ref{8-28}) and (\ref{9-46add}), respectively. 
We already mentioned that there are three cases for the combinations at the point 
C: $(y_+,y'_+)$, $(y_+,y'_-)$ and $(y_-,y'_-)$. 
If applying this rule to the present cycle, we obtain the following three cases: 
\beq
& &{\rm (i)}\ \ \ \alpha_t>0\quad ({\rm S}\rightarrow {\rm C})\ , \qquad
\beta_t>0\quad ({\rm C}\rightarrow {\rm T}\rightarrow {\rm C'(=C)})\ , \qquad
\alpha_t>0\quad ({\rm C'}\rightarrow {\rm S})\ , \nonumber\\
& &{\rm (ii)}\ \ \alpha_t>0\quad ({\rm S}\rightarrow {\rm C})\ , \qquad
\beta_t<0\quad ({\rm C}\rightarrow {\rm T}\rightarrow {\rm C'(=C)})\ , \qquad
\alpha_t>0\quad ({\rm C'}\rightarrow {\rm S})\ , \nonumber\\
& &{\rm (iii)}\ \alpha_t<0\quad ({\rm S}\rightarrow {\rm C})\ , \qquad
\beta_t<0\quad ({\rm C}\rightarrow {\rm T}\rightarrow {\rm C'(=C)})\ , \qquad
\alpha_t<0\quad ({\rm C'}\rightarrow {\rm S})\ . \nonumber
\eeq

We take up the case where the cycle starts from the point S, that is, 
the initial condition is given by 
$\chi^0=0$ at $\tau^0=0$. 
Here, $\chi^0$ and $\tau^0$ appear in the relation (\ref{8-28}). 
In order to demonstrate our idea, we present several results derived in the 
case (i) with $\psi^c > \psi^d$. 
In this case, there exists the flectional point D specified by 
$\psi=\pm \psi^d$ between C and T and also between T and C'. 
Then, the path (C$\rightarrow$T$\rightarrow$C'(=C)) is decomposed 
into the three: C$\rightarrow$D, D$\rightarrow$T and T$\rightarrow$D'(=D). 
We discriminate between D and D' by the condition 
($\psi_D=-\psi^d$, $\psi_{D'}=\psi^d$). 
General solutions of the paths (S$\rightarrow$C, C'$\rightarrow$S), 
(C$\rightarrow$D, D'$\rightarrow$C') and (D$\rightarrow$T, T$\rightarrow$D') 
are given by the relations (\ref{8-28}), (\ref{9-48a}) and (\ref{9-48b}), respectively. 
The parameters $(\omega_h, {\cal A}_h)$ and $(\omega_n, {\cal A}_n)$ contained in the relation 
(\ref{9-48}) are obtained in the form 
\bsub\label{10-13add}
\beq
& &\omega_h=\sqrt{\frac{A_++B}{\cal J}\cdot w_+^{cd}}\ , 
\label{10-13add-a}\\
& &{\cal A}_h=\sqrt{\frac{2}{w_+^{cd}}\left(\frac{E_+}{A_++B}-v_+(\psi^c)\right)
-\left(\frac{v_+'(\psi^c)}{w_+^{cd}}\right)^2}\ , \qquad\qquad\qquad\qquad\qquad\qquad
\label{10-13addb}
\eeq
\esub
\bsub\label{10-14add}
\beq
& &\omega_n=\sqrt{\frac{A_++B}{\cal J}\cdot w_+^{d}}\ , 
\label{10-14add-a}\\
& &{\cal A}_n=\sqrt{\frac{2}{w_+^{d}}\left(\frac{E_+}{A_++B}-v_+(\psi^c)\right)
+{\psi^{d}}^2+\frac{1}{w_+^d}\left(v_+'(\psi^c)+v_+'(\psi^d)\right)
\left(\psi^c-\psi^d\right)}\ . \nonumber\\
& &
\label{10-14add-b}
\eeq
\esub
Here, we used the relation (\ref{9-40}) or (\ref{9-30a}) with 
$V_+(\psi)=(A_++B)v_+^a(\psi)$. 
The other parameters $(\alpha_h, {\cal B}_h)$ and 
$(\alpha_n, {\cal B}_n)$ can be determined through the conditions 
governing each path.

\begin{figure}[t]
\begin{center}
\includegraphics[height=5.8cm]{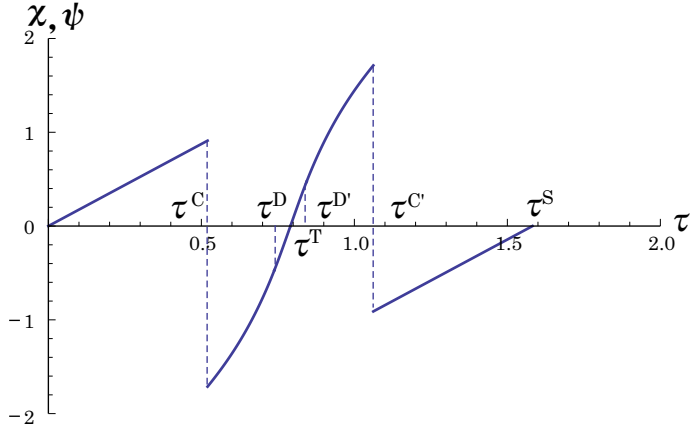}
\caption{
It is illustrated how $\chi(\tau)$ and $\psi(\tau)$ behave in one cycle. 
Here, the same parameters as those used in Fig.{\ref{fig:8}} (a) are adopted.   
}
\label{fig:12}
\end{center}
\end{figure}

Let $\tau^C$, $\tau^D$, $\tau^T$, $\tau^{D'}$, $\tau^{C'}$ and $\tau^S$ denote the arrival, i.e., 
departure times at the points C, D, T, D', C' and S, respectively, after the cycle starts from 
S at the time $\tau^0=0$. 
Then, these times obey the following condition: 
\beq\label{10-15add}
\begin{array}{ll}
(1)\ \chi(0)=0\ , \quad \chi(\tau^C)=\chi^c\ , & (2)\ \psi_h(\tau^C)=-\psi^c\ , \quad \psi_h(\tau^D)=-\psi^d\ , \\
(3)\ \psi_n(\tau^D)=-\psi^d\ , \quad \psi_n(\tau^T)=0\ , & (4)\ \psi_n(\tau^T)=0\ , \quad \psi_n(\tau^{D'})=\psi^d\ , \\
(5)\ \psi_h(\tau^{D'})=\psi^d\ , \quad \psi_h(\tau^{C'})=\psi^c\ , & (6)\ \chi(\tau^{C'})=-\chi^c\ , \quad \chi(\tau^S)=0\ . 
\end{array}
\eeq
Under the condition (\ref{10-15add}), we can determine $(\alpha_h, {\cal B}_h)$ and $(\alpha_n, {\cal B}_n)$ 
and $\chi(\tau)$ and $\psi(\tau)$ for each path are given in the following form: 
\bsub\label{10-16add}
\beq
& &(1)\ {\rm S}\rightarrow{\rm C}\ ; \ 
\chi(\tau)=2\gamma\sqrt{|\alpha_t|} \cdot\tau\ , \quad (0\leq \tau \leq \tau^C)
\label{10-16add-a}\\
& &(2)\ {\rm C}\rightarrow{\rm D}\ ; \ 
\psi(\tau)={\cal A}_h\sinh\left[\omega_h (\tau-\tau^C)+\sinh^{-1}\left(\frac{v_+'(\psi^c)}{{\cal A}_h w_+^{cd}}\right)\right]
-\frac{v_+'(\psi^c)}{w_+^{cd}}-\psi^c\ , \nonumber\\
& &\qquad\qquad\qquad\qquad\qquad\qquad\qquad\qquad\qquad\qquad\qquad\qquad (\tau^C\leq \tau \leq \tau^D)
\label{10-16add-b}\\
& &(3)\ {\rm D}\rightarrow{\rm T}\ ; \ 
\psi(\tau)={\cal A}_n\sin\left[\omega_n (\tau-\tau^D)-\sin^{-1}\left(\frac{\psi^d}{{\cal A}_n}\right)\right]\ , 
\quad (\tau^D \leq \tau \leq \tau^T)
\label{10-16aadd-c}\\
& &(4)\ {\rm T}\rightarrow{\rm D'}\ ; \ 
\psi(\tau)={\cal A}_n\sin\left[\omega_n (\tau-\tau^T)\right]\ , 
\quad (\tau^T \leq \tau \leq \tau^{D'})
\label{10-16aadd-d}\\
& &(5)\ {\rm D'}\rightarrow{\rm C'}\ ; \ 
\psi(\tau)={\cal A}_h\sinh\left[\omega_h (\tau-\tau^{D'})-\sinh^{-1}\left(\frac{1}{{\cal A}_h}\left(
\psi^c-\psi^d+\frac{v_+'(\psi^c)}{w_+^{cd}}\right)\right)\right]
\nonumber\\
& &\qquad\qquad\qquad\qquad\qquad
+\left(\psi^c-\psi^d+\frac{v_+'(\psi^c)}{w_+^{cd}}\right)+\psi^d\ , \quad
(\tau^{D'}\leq \tau \leq \tau^{C'})
\label{10-16add-e}\\
& &(6)\ {\rm C'}\rightarrow{\rm S}\ ; \ 
\chi(\tau)=2\gamma\sqrt{|\alpha_t|}\cdot(\tau-\tau^{C'})-\chi^c\ . \quad (\tau^{C'} \leq \tau \leq \tau^S)
\label{10-16add-f}
\eeq
\esub
In connection to the relations (\ref{7-21}) and (\ref{7-22}), it was 
suggested that our model enables us to describe the dissipation phenomena in 
many-fermion system. 
The paths (1) and (6) correspond to this suggestion. 
It indicates that this dissipation cannot be observed at any time. 
The condition (\ref{10-15add}) also gives us the time intervals for the paths: 
\bsub\label{10-17add}
\beq
\tau^S-\tau^{C'}&=&
\tau^C=\frac{\chi^c}{2\gamma\sqrt{|\alpha_t|}}\ , 
\label{10-17add-a}\\
\tau^{C'}-\tau^{D'}&=&
\tau^D-\tau^C\nonumber\\
&=&\frac{1}{\omega_h}\left\{
\sinh^{-1}\left[\frac{1}{{\cal A}_h}\left(\psi^c-\psi^d+\frac{v'_+(\psi^c)}{w_+^{cd}}\right)\right]-
\sinh^{-1}\left(\frac{v'_+(\psi^c)}{{\cal A}_h w_+^{cd}}\right)\right\}\ , \quad
\label{10-17add-b}\\
\tau^{D'}-\tau^T&=&\tau^T-\tau^D\nonumber\\
&=&\frac{1}{\omega_n}\sin^{-1}\left(\frac{\psi^d}{{\cal A}_n}\right)\ . 
\label{10-17add-c}
\eeq
\esub
The results (\ref{10-17add-a})$\sim$(\ref{10-17add-c}) give us $\tau^C$ etc. 
For example, we have 
\bsub\label{10-18add}
\beq
& &\tau^S=2\tau^T\ , 
\label{10-18add-a}\\
& &\tau^T=\frac{\chi^c}{2\gamma\sqrt{|\alpha_t|}}+\frac{1}{\omega_h}
\left\{
\sinh^{-1}\left[\frac{1}{{\cal A}_h}\left(\psi^c-\psi^d+\frac{v'_+(\psi^c)}{w_+^{cd}}\right)\right]-
\sinh^{-1}\left(\frac{v'_+(\psi^c)}{{\cal A}_h w_+^{cd}}\right)\right\}
\nonumber\\
& &\qquad\qquad
+\frac{1}{\omega_n}\sin^{-1}\left(\frac{\psi^d}{{\cal A}_n}\right)\ . 
\label{10-18add-b}
\eeq
\esub
The time $\tau^S$ is nothing but the period of the cycle. 
Figure \ref{fig:12} illustrates how $\chi(\tau)$ and $\psi(\tau)$ behave in one cycle. 
%
%%%%%%%%%%%%%%%%%%%%%%%%%%%%%%%%%%%%%%%%%%%%%%%%%%%%%%%%%%%%%%%%%%%%%%
%%%%%%%%%%%%%%%%%%%%%%%%%%%%%%%%%%%%%%%%%%%%%%%%%%%%%%%%%%%%%%%%%%%%%%%%
%
%
%%%%%%%%%%%%%%%%%%%%%%%%%%%%%%%%%%%%%%%%%%%%%%%%%%%%%%%%%%%%%%%%%%%%%%
\begin{figure}[t]
\begin{center}
\includegraphics[height=5.8cm]{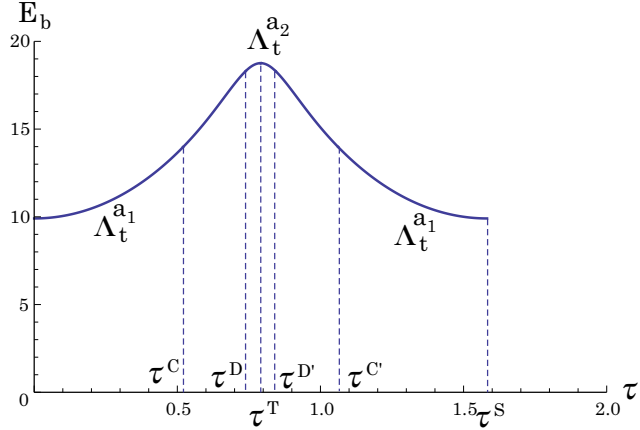}
\caption{
The approximate energy of the intrinsic system, $\langle {\wtilde H}_{\ovl P}\rangle=\varepsilon N_{\ovl P}$, 
is shown as a function of time $\tau$ under the same parameter set as those used in Fig.{\ref{fig:8}} (a). 
}
\label{fig:13}
\end{center}
\end{figure}
%%%%%%%%%%%%%%%%%%%%%%%%%%%%%%%%%%%%%%%%%%%%%%%%%%%%%%%%%%%%%%%%%%%%%%%%
%
Here, in the regions of $\tau\leq \tau^C$ and $\tau \geq \tau^{C'}$, $\chi(\tau)$ is shown and is a linear function 
with respect to time $\tau$. 
In the region of $\tau^C \leq \tau \leq \tau^{C'}$, $\psi(\tau)$ is used, and from $\tau^C$ to $\tau^T$, 
$\psi(\tau)$ is changed from $\sinh$-type to $\sin$-type function and vice versa from $\tau^T$ to $\tau^{C'}$. 
Our main interest is concerned with the energy of the intrinsic system 
expressed in the form 
\beq\label{11-12add}
E_b\equiv\rbra{\phi}{\wtilde H}_{\ovl P}\rket{\phi}
=\varepsilon N_{\ovl P}\ . 
\eeq
Here, ${\wtilde H}_{\ovl P}$ and $N_{\ovl P}$ are given in the relations 
(\ref{7-11}) and (\ref{7-25}), respectively and, therefore, it may be 
enough for the understanding of $E_b$ to consider $N_{\ovl P}$ ($\varepsilon=1$). 
In Fig.\ref{fig:13}, the result
%the energy of the intrinsic system, 
%$E_b\equiv \langle {\wtilde H}_{\ovl P}\rangle=\varepsilon N_{\ovl P}$ in (\ref{7-11}), 
is shown as a function of $\tau$ under the approximation developed in \S\S 8 - 10 with the same parameters as those used in Fig.\ref{fig:8} (a). 
In the region of $\tau\leq \tau^C$ and $\tau \geq \tau^{C'}$, $x\leq \gamma_t$ is satisfied where 
$\Lambda_t^{a_1}(x)$ is used as the approximation of $\Lambda_t(x)$. 
In the region of $\tau^C \leq \tau \leq \tau^{C'}$, $\Lambda_t^{a_2}(x)$ is used in the approximation of $\Lambda_t(x)$.  
It is seen that, in one cycle, 
the energy flows into the intrinsic system from external environment from the time 0 to $\tau^T$ and 
vice versa from $\tau^T$ to $\tau^S$.

We will discuss some problems related to the result shown in Fig.\ref{fig:13}. 
The energy $E_b$ shows periodical behavior for time $\tau$. 
Such behavior cannot be expected in the $su(1,1)$-algebraic model, in which, 
following the cosh-type change, $E_b$ increases or decreases. 
Since our model belongs to the $su(2)$-algebraic model, the periodical behavior appears. 
In Fig.\ref{fig:13}, we can see that there exist the minimum and the maximum value 
in $E_b$. 
We consider these two values for the case 
$(\alpha_t > 0 , \beta_t > 0)$ in a rather general form. 
The relation (\ref{8-13a}) teaches us that if $\chi=0$, $y$ becomes the minimum, i.e., 
$y_{\rm min}=c_+$. 
Then, with the aid of the expression (\ref{8-7}), we have 
\beq\label{11-13add}
\left(\Lambda_t^{a_1}\right)_{\rm min}
=\frac{1}{\alpha_t}(y_{\rm min}-2t)=\frac{1}{\alpha_t}(c_+-2t)
=\frac{\rho^2}{\alpha_t}\cdot\frac{1}{t+\sqrt{t^2+\rho^2}}\ . 
\eeq
On the other hand, the relation (\ref{9-16a}) gives us the maximum value of 
$y'$, if $\psi=0$, i.e., $y'_{\rm max}=d_+$. 
Then, the relation (\ref{9-23}) gives us 
\beq\label{11-14add}
\left(\Lambda_t^{a_2}\right)_{\rm max}
=\gamma_t y'_{\rm max}=\gamma_t d_+\ . 
\eeq
Of course, $d_+$ is a solution of the cubic equation (\ref{9-17a}) and we use a possible 
approximate solution : 
\beq\label{11-15add}
d_+=2\Omega_0\left(
1-\frac{\sigma^2}{2\Omega_0|\beta_t|}\cdot \frac{2}{4\Omega_0^2-\sigma^2+
\sqrt{(4\Omega_0^2-\sigma^2)^2-\frac{4\sigma^2(2\Omega_0+d_m)}{|\beta_t|}}}\right)\ . 
\eeq
Here, $d_m$ is given in the relation (\ref{10-1b}). 
In the cases $\sigma^2=0$ and $(\sigma^2)_{\rm max}$, 
the solution (\ref{11-15add}) is exact. 
With the aid of the relation (\ref{7-25a}), 
$(E_b)_{\rm min}$ and $(E_b)_{\rm max}$ are expressed as follows: 
\bsub\label{11-16add}
\beq
& &(E_b)_{\rm min}=2t-1+\frac{\rho^2}{\alpha_t}\cdot
\frac{1}{t+\sqrt{t^2+\rho^2}}\ , 
\label{11-16a}\\
& &(E_b)_{\rm max}=2\Omega_0-
\frac{\sigma^2}{|\beta_t|}\cdot
\frac{2\gamma_t}{4\Omega_0^2-\sigma^2+
\sqrt{(4\Omega_0^2-\sigma^2)^2-\frac{4\sigma^2(2\Omega_0+d_m)}{|\beta_t|}}}\ . 
\label{11-16b}
\eeq
\esub
In the case $\rho^2=0$, $(E_b)_{\rm min}=2t-1$ and as $\rho^2$ increases, 
$(E_b)_{\rm min}$ increases. 
Inversely, in the case $\sigma^2=0$, $(E_b)_{\rm max}=2\Omega_0$ and as 
$\sigma^2$ increases, $(E_b)_{\rm max}$ decreases. 
For the case $(2\Omega_0=19,\ 2t=5,\ 2\kappa=-15)$, we 
have 
$(E_b)_{\rm min}=9.9072$ and $(E_b)_{\rm max}=18.7569$, which is very near to 
$(E_b)_{\rm max}=18.7562$ calculated under the exact solution of the 
cubic equation (\ref{9-17a}). 
This result may support the validity of the approximate 
form (\ref{11-15add}).

Next, on the basis of the above argument, we investigate the trial state (\ref{5-3}) 
which leads us to $(E_b)_{\rm min}$ and $(E_b)_{\rm max}$. 
In the ranges $0 \leq x \leq \gamma_t$ and $\gamma_t \leq x' < \infty$, 
the state (\ref{5-3}) contains the parameters $z\ (x=|z|^2)$ and 
$z'\ (x'=|z'|^2)$, respectively. 
Here, the relation (\ref{9-1}) should be noted. 
In the range $0 \leq x \leq\gamma_t$, we note the relation (\ref{8-18a}) 
and, then, the value of $x$ at $\chi=0$ is the minimum: 
\beq\label{11-17add}
x_{\rm min}=\frac{1}{\alpha_t}\cdot \frac{\sqrt{t^2+\rho^2}-t}{\sqrt{t^2+\rho^2}+t}\ 
(=x_{\rm min}(\rho^2))\ . 
\eeq
The function $x_{\rm min}(\rho^2)$ is increasing for $\rho^2$ with 
$x_{\rm min}(\rho^2=0)=0$, i.e., $z=0$. 
Therefore, the state $\rket{\phi}$ corresponding to $z=0$ is the minimum weight 
state (\ref{4-13}), $\rket{m_0}$, which contains $(2t-1)$ fermions only in 
${\ovl P}\ (N_{\ovl P}=2t-1,\ N_P=0)$. 
In the range $\gamma_t \leq x' < \infty$, we note the relation (\ref{9-22}) 
and the value of $x'$ at $\psi=0$ is the maximum: 
\beq\label{11-18add}
x'_{\rm max}&=&
|\beta_t|+\frac{1}{2\Omega_0-y'_{\rm max}}=|\beta_t|+\frac{1}{2\Omega_0-d_+} \nonumber\\
&=&
\frac{|\beta_t|}{\sigma^2}\left(
4\Omega_0^2+\sqrt{(4\Omega_0^2-\sigma^2)^2-\frac{4\sigma^2(2\Omega_0+d_m)}{|\beta_t|}}
\right)\ (=x'_{\rm max}(\sigma^2))\ . 
\eeq
Here, we used the approximate expression (\ref{11-15add}) for $d_+$. 
The function $x'_{\rm max}(\sigma^2)$ is decreasing for $\sigma^2$ with 
$x'_{\rm max}(\sigma^2=0)\rightarrow \infty$, i.e., 
$z'\rightarrow \infty$. 
Therefore, the state $\rket{\phi}$ corresponding to $z'\rightarrow \infty$ is 
$\left({\wtilde {\cal T}}_+\right)^{2\Omega_0-(2t-1)}\rket{m_0}\ 
(2s=2\Omega_0-(2t-1))$ which contains the maximum number of fermion permitted by 
the seniority coupling scheme 
($N_{\ovl P}=2\Omega_0,\ N_P=2\Omega_0-(2t-1)$). 
From the above argument, it may be clear that as $\kappa^2$ increases from 
$\kappa^2=0$, $\rho^2$ and $\sigma^2$ also increase from $\rho^2=\sigma^2=0$ and 
$(E_b)_{\rm min}$ and $(E_b)_{\rm max}$ become larger than $(2t-1)$ 
and smaller than $2\Omega_0$, respectively. 
If at $\tau=0$, the cycle starts in the point S with $(E_b)_{\rm min}$, it passes the 
critical points C and D and at $\tau=\tau^T$ arrives at T. 
Although the points C and D are introduced under the approximation adopted 
in this paper, they play an essential role for treating the present model 
in a well-known simple mathematical form.

The above is a basic part which our simple many-fermion model produces 
under the pseudo $su(1,1)$-algebra.

\setcounter{equation}{0}

\section{Concluding remarks}

In this section, we will give some remarks on the Hamiltonian (\ref{7-10}). 
This Hamiltonian was set up under the correspondences (\ref{7-7})-(\ref{7-9}). 
The original boson Hamiltonian (\ref{7-5}) is a generator for time-evolution and does not represent the energy of the entire system. 
It aims at the description of the ``damped and amplified harmonic oscillator". 
By regarding the mixed-mode boson coherent state as a statistically mixed state, 
we can describe the harmonic oscillator at finite temperature, which will be 
shown in the relation (\ref{11-6}). 
In this sense, the above-mentioned description provides us a possible entrance to the problems 
related to finite temperature. 
The Hamiltonian (\ref{7-10}) can be regarded as the fermion version of the harmonic oscillator in the 
$su(1,1)$-algebra in the Schwinger boson representation. 
Nevertheless, it may be possible to treat the Hamiltonian (\ref{7-10}) as the energy of the entire system. 
It was already mentioned in \S 7. 
In order to confirm this conjecture, we reexamine the correspondences (\ref{7-7})-(\ref{7-9}).

Let us start in the relation (\ref{7-9}). 
The frequency $\omega$ is positive, but, the single-particle energy $\varepsilon$ is not always positive. 
Therefore, instead of the relation (\ref{7-11}), it is permissible to set up the following form: 
\beq\label{11-1}
{\wtilde H}_{\ovl P}=\varepsilon\sum_{\alpha}{\tilde c}_{\bar \alpha}^*{\tilde c}_{\bar \alpha}\ , \qquad
{\wtilde H}_{P}=\varepsilon\sum_{\alpha}{\tilde c}_{\alpha}^*{\tilde c}_{\alpha}\ , \qquad
{\wtilde H}_{\ovl P}+{\wtilde H}_P=2\varepsilon {\wtilde {\cal T}}\ . 
\eeq
The form (\ref{11-1}) suggests us that the system under consideration is nothing but many-fermion 
system in two single-particle levels, ${\ovl P}$ and $P$ with the level distance $2|\varepsilon|$. 
If the relation (\ref{11-1}) is admitted, ${\wtilde H}$ represents the energy of the entire system. 
From this point of view, the state $\rket{\phi}$ is not the statistically mixed state, 
but the trial state of the time-dependent variation for ${\wtilde H}$ as the energy of the entire system. 
Therefore, the results obtained in this paper presents us the informations provided by $\rket{\phi}$ as a 
statistically pure state.

Next, we reexamine the correspondence (\ref{7-7}) and (\ref{7-8}). 
First, we notice the following: 
If the vacuum changes appropriately, fermion creation operator becomes annihilation operator, that is, 
if $\rrket{0}={\tilde c}^*\rket{0}\ ({\tilde c}\rket{0}=0)$, ${\tilde c}^*\rrket{0}=0$. 
In the case of boson operator, we can not find such a situation. 
If we note the above fact, the following correspondence may be also permitted: 
\beq\label{11-2}
({\hat b}, {\hat b}^*) \rightarrow (s_{\alpha}{\tilde c}_{\bar \alpha}^*, s_{\alpha}{\tilde c}_{\bar \alpha})\ , \qquad
({\hat a}, {\hat a}^*) \rightarrow ({\tilde c}_{\alpha}, {\tilde c}_{\alpha}^*)\ . 
\eeq
Then, for $({\hat T}-1/2)$, we have 
\beq\label{11-3}
{\hat T}-\frac{1}{2}\rightarrow 
{\maru {\cal T}}&=&
-\frac{1}{2}\sum_{\alpha}({\tilde c}_{\alpha}^*{\tilde c}_{\alpha}
-s_{\alpha}{\tilde c}_{\bar \alpha}s_{\alpha}{\tilde c}_{\bar \alpha}^*) \nonumber\\
&=&-\frac{1}{2}\sum_{\alpha}({\tilde c}_{\alpha}^*{\tilde c}_{\alpha}
+{\tilde c}_{\bar \alpha}^*{\tilde c}_{\bar \alpha})+\Omega_0\ .  
\eeq
The correspondence (\ref{11-2}) suggests us that the set $({\wtilde S}_{\pm,0})$ is replaced with 
the set $({\wtilde R}_{\pm,0})$. 
Then, another type of the pseudo $su(1,1)$-algebra can be defined in the form 
\beq\label{11-4}
& &{\maru {\cal T}}_+={\wtilde R}_+
\sqrt{\frac{\Omega_0+\frac{1}{2}+t+{\wtilde R}_0}{\Omega_0+\frac{1}{2}-t-{\wtilde R}_0+\epsilon}}\ , \quad
{\maru {\cal T}}_-=
\sqrt{\frac{\Omega_0+\frac{1}{2}+t+{\wtilde R}_0}{\Omega_0+\frac{1}{2}-t-{\wtilde R}_0+\epsilon}}{\wtilde R}_-\ , \nonumber\\
& &{\maru {\cal T}}_0=\Omega_0+\frac{1}{2}+{\wtilde R}_0\ . 
\eeq
The Hamiltonian in this case is expressed as 
\beq\label{11-5}
{\maru H}=2\varepsilon{\maru {\cal T}}-i\gamma\left({\maru {\cal T}}_+-{\maru {\cal T}}_-\right)\ . 
\eeq
It may be clear that the above does not correspond to the deformation of the Cooper-pair. 
It corresponds to the deformation of the density type fermion-pair. 
The Hamiltonian (\ref{11-5}) is applicable to the case where the single-particle energy of the level $P$ is equal to that 
of ${\ovl P}$.

Finally, we must mention two problems to be solved in the near future. 
By regarding the mixed-mode boson coherent state as the statistically mixed state, 
the expectation value of ${\hat H}_b=\omega {\hat b}^*{\hat b}$ is given by
\beq\label{11-6}
\langle {\hat H}_b\rangle
\sim \omega\cdot (2t-1)+\omega\cdot \frac{1}{e^{\omega \beta}-1}\ . \qquad (\beta=(k_B T)^{-1})
\eeq
The first and the second term represent the energy at the low temperature limit and the energy 
coming from the thermal fluctuation in the bose distribution, respectively \cite{5,6}. 
One of the future problems is to investigate the thermal effect such as shown in the relation (\ref{11-6}) by 
regarding the state $\rket{\phi}$ as the statistically mixed state for ${\wtilde H}_{\ovl P}
=\varepsilon\sum_{\alpha}{\tilde c}_{\bar \alpha}^*{\tilde c}_{\bar \alpha}$. 
In this case, our concern is to examine if the fermi distribution appears or not. 
Second problem is related to the Hamiltonians ${\wtilde H}_{\ovl P}$ and ${\wtilde H}_{P}$. 
They are on a level with ${\hat H}_b$ and ${\hat H}_a$. 
In Ref.\citen{5}, we can find some examples extended from ${\hat H}_b$ and ${\hat H}_a$. 
The future task is to investigate also the cases extended from ${\wtilde H}_{\ovl P}$ and ${\wtilde H}_P$. 
The above two are our future problems to be solved.

\section*{Acknowledgment}

One of the authors (Y.T.) 
is partially supported by the Grants-in-Aid of the Scientific Research 
(No.23540311) from the Ministry of Education, Culture, Sports, Science and 
Technology in Japan.


%\bibliographystyle{ptephy}
%\bibliography{sample}
%
% once the .bbl file has been generated then place the text in your article.

%\vfill\pagebreak

%\appendix

%\section{}

\end{document}